\documentclass[lettersize,journal]{IEEEtran}
\usepackage{balance}
\usepackage{amsmath,amsfonts}
\usepackage{algorithmic}
\usepackage{algorithm}
\usepackage{textcomp}
\usepackage{longtable}
\usepackage{float}
\usepackage{multirow}
\usepackage{threeparttable}
\usepackage{booktabs}
\usepackage{array}
\usepackage[caption=false,font=footnotesize,labelfont=rm,textfont=rm]{subfig}
\usepackage{textcomp}
\usepackage{stfloats}
\usepackage{url}
\usepackage{verbatim}
\usepackage{graphicx}
\usepackage{bm} 
\usepackage{xcolor}
\usepackage{cite}
\usepackage[numbers,sort&compress]{natbib}
\def\BibTeX{{\rm B\kern-.05em{\sc i\kern-.025em b}\kern-.08em
    T\kern-.1667em\lower.7ex\hbox{E}\kern-.125emX}}
\begin{document}

\title{GQE-Net: A Graph-based Quality Enhancement Network for Point Cloud Color Attribute}

\author{Jinrui Xing, Hui Yuan, \emph{Senior Member IEEE}, Raouf Hamzaoui, \emph{Senior Member IEEE}, \\Hao Liu, and Junhui Hou, \emph{Senior Member IEEE}
        % <-this % stops a space
\thanks{This work was supported in part by the National Natural Science Foundation of China under Grants 62222110 and 62172259, the Taishan Scholar Project of Shandong Province (tsqn202103001), the Natural Science Foundation of Shandong Province under Grant ZR2022ZD38, the Central Guidance Fund for Local Science and Technology Development of Shandong Province, under Grant YDZX2021002, the High-end Foreign Experts Recruitment Plan of the Chinese Ministry of Science and Technology under Grant G2023150003L, the Hong Kong Research
Grants Council under Grants 11202320 and 11218121, and the OPPO Research Fund.}% <-this % stops a space
\thanks{Jinrui Xing and Hui Yuan are with the School of Control Science and
Engineering, Shandong University, Ji'nan, 250061, China (e-mail: jinruixing@
mail.sdu.edu.cn; huiyuan@sdu.edu.cn). 

Raouf Hamzaoui is with the School of Engineering and Sustainable Development, De Montfort University, LE1 9BH Leicester, U.K. (e-mail: rhamzaoui@dmu.ac.uk). 

Hao Liu is with the School of Computer and Control Engineering, Yantai University, Yantai, 264005, China (e-mail: liuhaoxb@gmail.com).

Junhui Hou is with the Department of Computer Science, City University of Hong Kong, Hong Kong, China (e-mail: jh.hou@cityu.edu.hk). 

Jinrui Xing and Hui Yuan contributed equally to this work; Hui Yuan is the corresponding author.

This paper has supplementary downloadable material available at http://ieeexplore.ieee.org., provided by the author. The material includes a pdf file named Supplemental Materials. Contact jinruixing@mail.sdu.edu.cn for further questions about this work.
}
%\thanks{Manuscript received April 19, 2021; revised August 16, 2021.}
}

\markboth{Journal of \LaTeX\ Class Files,~Vol.~X, No.~X, February~2023}%
{Shell \MakeLowercase{\textit{et al.}}: GQE-Net: A Graph-based Quality Enhancement Network for Point Cloud Color Attribute}

% \IEEEpubid{0000--0000/00\$00.00~\copyright~2023 IEEE}

\maketitle

\begin{abstract}
In recent years, point clouds have become increasingly popular for representing three-dimensional (3D) visual objects and scenes. To efficiently store and transmit point clouds, compression methods have been developed, but they often result in a degradation of quality. To reduce color distortion in point clouds, we propose a graph-based quality enhancement network (GQE-Net) that uses geometry information as an auxiliary input and graph convolution blocks to extract local features efficiently. Specifically, we use a parallel-serial graph attention module with a multi-head graph attention mechanism to focus on important points or features and help them fuse together. Additionally, we design a feature refinement module that takes into account the normals and geometry distance between points. To work within the limitations of GPU memory capacity, the distorted point cloud is divided into overlap-allowed 3D patches, which are sent to GQE-Net for quality enhancement. To account for differences in data distribution among different color components, three models are trained for the three color components. Experimental results show that our method achieves state-of-the-art performance. For example, when implementing GQE-Net on a recent test model of the geometry-based point cloud compression (G-PCC) standard, $\bf{0.43}$ dB, $\bf{0.25}$ dB and $\bf{0.36}$ dB $\bf{\bm{{Bj \phi ntegaard}}}$ delta (BD)-peak-signal-to-noise ratio (PSNR), corresponding to $\bf{14.0\%}$, $\bf{9.3\%}$ and $\bf{14.5\%}$ BD-rate savings were achieved on dense point clouds for the Y, Cb, and Cr components, respectively.

The source code of our method is available at \textbf{https://github.com/xjr998/GQE-Net}.

\end{abstract}

\begin{IEEEkeywords}
point cloud, quality enhancement, graph neural network, G-PCC.
\end{IEEEkeywords}

\section{Introduction}
\IEEEPARstart{P}{oint} clouds are sets of 3D points given by their coordinates and attribute information such as color, reflectance, and normals. Point clouds are obtained from laser scans, multiview video cameras, or light field imaging systems \cite{guo2017real} \cite{perra2016analysis}. Due to their flexibility and powerful representation capability, they are increasingly being used in fields such as immersive communication, robotics, geographic information systems (GIS), and autonomous driving \cite{wang2019applications}. However, the large volume of data in point clouds can be a challenge for storage and transmission. Therefore, developing efficient compression technologies for point clouds is necessary.

To reduce resource consumption, the International Organization for Standardization (ISO) and the International Electrotechnical Commission (IEC) joint technical committee for ``Information technology'' (JTC1) / Work Group 7 (WG7) are currently developing two types of compression standards \cite{graziosi2020overview} for 3D point clouds: video-based point cloud compression (V-PCC) \cite{vpcc} and geometry-based point cloud compression (G-PCC) \cite{gpcc}. WG7 is also studying the potential of artificial intelligence technologies for point cloud compression, with the aim of developing a corresponding standard \cite{aipcc}. V-PCC projects the point cloud onto 2D planes and compresses these images using existing video coding standards, such as H.265/High Efficiency Video Coding (HEVC). G-PCC, on the other hand, is directly based on the 3D geometric structure. Compared to V-PCC, G-PCC has lower complexity and requires less memory use. However, the coding efficiency of G-PCC is usually lower than that of V-PCC when dealing with dense point clouds.

\begin{figure}[H]
\setlength{\belowcaptionskip}{-1.0em}
\captionsetup{}
\centering
\centerline{\includegraphics[width=6.5cm]{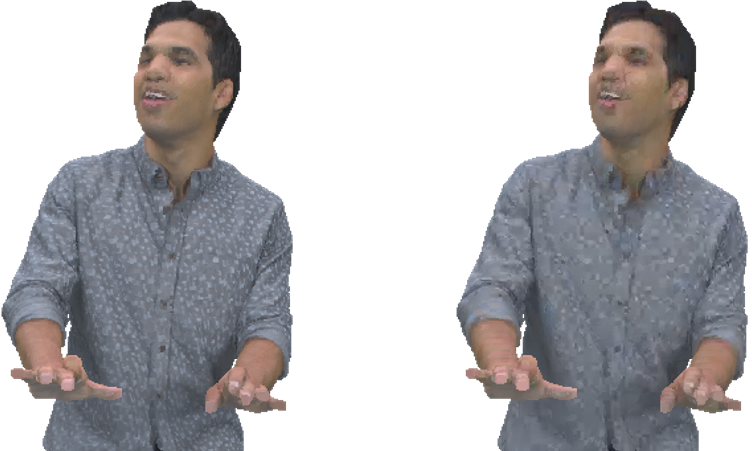}}
\caption{Visual quality of part of ``$loot\_vox10\_1200.ply$'', left: original point cloud, right: reconstructed point cloud with G-PCC TMC13v14.0 for quantization parameter $46$.} \label{o_r}
\end{figure}

Currently, many technologies for G-PCC and V-PCC are being developed and optimized, resulting in a higher compression efficiency. However, when the bandwidth is limited, the compression bit rate may have to be reduced by increasing the quantization step size. This can result in noticeable attribute distortions in the reconstructed point cloud and negatively affect the overall visual experience.

Both geometry and attribute compression can result in a reduction in quality. Geometry distortion is typically characterized by outliers and holes, while attribute distortion is mainly caused by blurring of textures and mismatches in color. Fig. 1 shows a comparison between an original point cloud and its reconstruction after compression using the lifting transform in a lossless geometry, lossy attribute configuration in G-PCC. The reconstructed point cloud is clearly different from the original point cloud due to the significant distortion in color. Therefore, it is essential to enhance the quality after compression.

In recent years, deep learning-based methods have had significant success in image processing. However, due to the irregular distribution of points, point clouds are difficult to process using conventional neural networks for 2D images and videos. To address this challenge, various approaches have been proposed. For example, projection-based methods aim to map the point cloud into multiple 2D images, which can then be processed by convolutional neural networks \cite{yang2020pbp,liu2021pqa,liu2022spu}. Another approach \cite{zhou2018voxelnet, roynard2018classification, kuang2020voxel} is to regularize the point cloud through voxelization and then apply a convolutional neural network with slight modifications to the voxels. However, the high complexity and memory consumption of these two approaches limit their applicability to some extent. As a result, many researchers have turned to working directly with 3D point clouds. Qi \emph{et al.} \cite{qi2017pointnet} proposed PointNet, which achieves permutation invariance of points and extracts features of the point cloud. Later, recognizing that each point in PointNet is learned independently and that the local features between points are not exploited, Qi \emph{et al.} \cite{qi2017pointnet++} proposed PointNet++ to capture local geometry information using a hierarchical structure. They also used 3D sparse convolutions to process cubes directly. Graham \emph{et al.} \cite{graham20183d} introduced new sparse convolutional operations to efficiently process spatially-sparse data. Su \emph{et al.} \cite{su2018splatnet} proposed sparse lattice networks with sparse bilateral convolutional layers. These layers apply convolutions only on occupied parts of the lattice, enabling hierarchical and spatially-aware feature learning. 

Using a graph-based neural network approach is a promising way to process point clouds. By creating a graph topological structure, points are closely related to each other, making it easy to extract local features. Building on previous successful graph-based processing techniques for point clouds \cite{wang2019dynamic,chen2021gapointnet,wang2019graph,shi2020point,te2018rgcnn,chen2019deep,
liang2019hierarchical}, we propose a graph neural network-based method for enhancing the quality of point clouds that have been distorted by compression. Our method splits a large point cloud into smaller, overlapping 3D patches that contain both geometry and color information. These patches are then fed into a neural network, called graph-based quality enhancement network (GQE-Net), for enhancement of the color attributes.The architecture of GQE-Net is shown in Fig. 2. As can be seen, GQE-Net includes special modules, such as a parallel-serial graph attention (PSGA) module and a feature refining (FR) module. Additionally, geometry coordinates are used to build graphs and calculate normal vectors. In our method, the Luma (Y) and Chroma (Cb and Cr) components of the color information are processed separately due to their distinct data distribution. Strategies are also implemented to handle overlapping and unprocessed points when merging the patches back into a single point cloud. In summary, this paper makes the following contributions:

\begin{itemize}
\item We introduce a parallel-serial graph attention module that uses a multi-head attention mechanism to focus on key points and points with high distortion. The module includes two parallel attention heads and one reinforcement head. 
\item We propose a feature refining  module that uses geometry information to determine the correlation between points. The module calculates normal vectors and incorporates them into the color features. Moreover, it assigns weights to the features based on the distance between points. 
\item Our experiments show that the proposed network significantly improves the quality of distorted point clouds and enhances coding efficiency. Moreover, the proposed network is robust in the sense that it can handle point clouds with various levels of distortion using a single pre-trained model.
\end{itemize}

The remainder of this paper is organized as follows. In Section II, we review related work in the areas of graph neural network-based point cloud processing, point cloud geometry denoising, and point cloud color quality enhancement. Section III provides a detailed description of our method, including the methodology of the proposed network and the structure of each module. In Section IV, we present experimental results and an ablation study to demonstrate the effectiveness of our method. Finally, Section V concludes the paper.

\begin{figure*}
\setlength{\belowcaptionskip}{-1.8em}
\captionsetup{font={small}}
\centering
\centerline{\includegraphics[width=16.8cm]{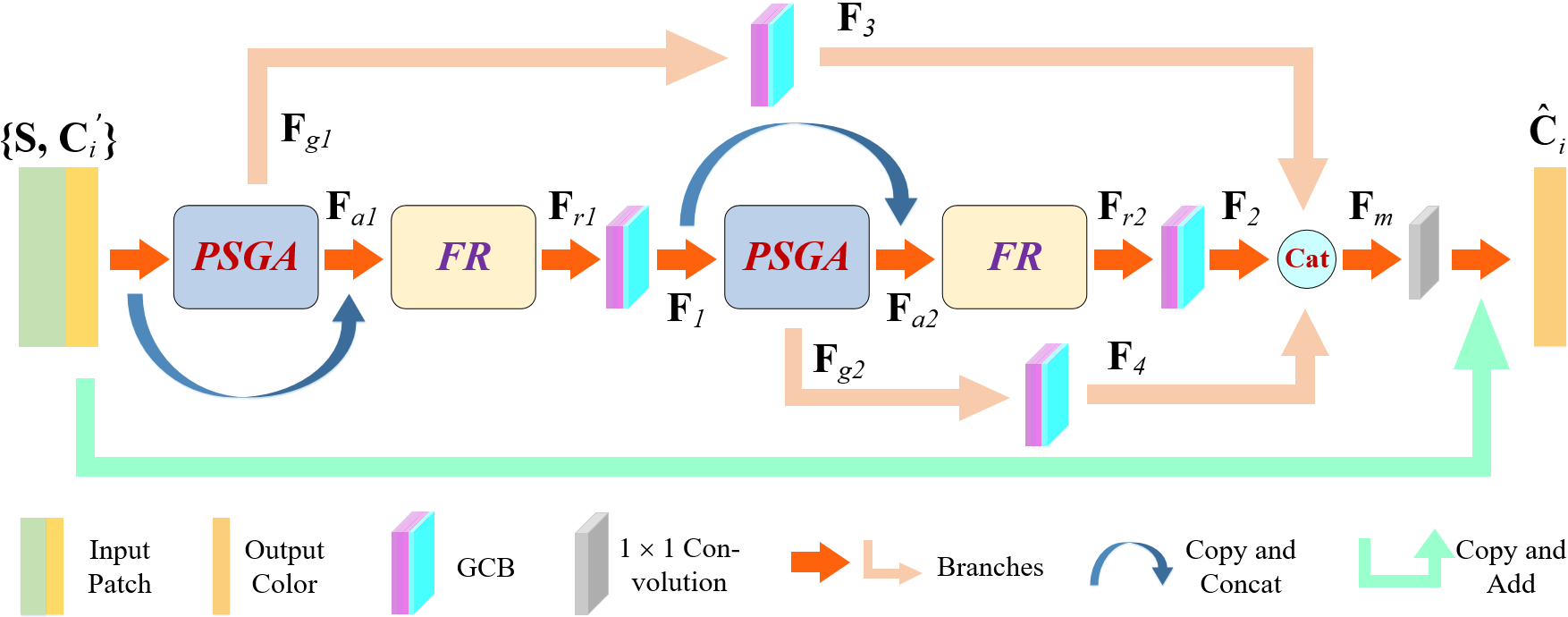}}
\caption{Structure of GQE-Net. The input are distorted color as well as auxiliary geometry information. The output is restored color. Different colored arrows indicate defferent branches or processes.} \label{GQE-Net}
\end{figure*}

\section{Related Works}
In this section, we begin by providing a brief overview of graph neural network-based point cloud processing methods, as our proposed method is also built on this framework. Next, we review methods for geometry denoising of point clouds, as they aid in the extraction of inter-point correlation by considering the geometry structure. Finally, we discuss several quality enhancement methods for point cloud color.
\subsection{Graph neural network-based point cloud processing}
Recently, graph-based neural networks have achieved great success in 3D point cloud processing. Notably, the EdgeConv module, a prominent graph convolution technique, was introduced by Wang \emph{et al.} \cite{wang2019dynamic} to facilitate high-level feature extraction from point clouds. This module uses a dynamic graph across network layers, significantly enhancing the capacity to capture intricate features. Additionally, the incorporation of graph-attention mechanisms plays an important role to focus on key points within the point cloud. Chen \emph{et al.} \cite{chen2021gapointnet} introduced a graph attention-based point network layer (GAPLayer) to learn local geometric representations by multiple single head graph attention mechanisms. Wang \emph{et al.} \cite{wang2019graph} proposed a graph attention convolution (GAC) that adapts its kernels to the structure of an object to capture structured features for fine-grained segmentation and avoid feature contamination between objects.
 
Many advancements in point cloud analysis have been achieved through the synergistic fusion of graph neural networks with classical models or mechanisms. This collaborative approach enhances the capability of these networks to tackle complex tasks effectively. Shi \emph{et al.} \cite{shi2020point} proposed a graph neural network for object detection. The network uses an auto-registration mechanism to reduce translation variance and a box merging and scoring operation to combine detections from multiple vertices accurately. Feng \emph{et al.} \cite{chen2019deep} proposed a deep auto-encoder with graph topology inference and filtering to achieve compact representations of unorganized 3D point clouds in an unsupervised manner. Liang \emph{et al.} \cite{liang2019hierarchical} proposed a hierarchical depth-wise graph convolutional neural network for point cloud semantic segmentation. A customized block called depthwise graph convolution (DGConv) was designed for local feature extraction. 

While these methods effectively extract correlation between points using the graph structure, they do not fully exploit the available information. The distances between points and the normals play a crucial role in determining the strength of correlation between two points. Moreover, as different points have varying importance, it is important to focus on key points and design more effective attention modules. In the field of quality enhancement, it is particularly important to capture points with high distortion or those located in key positions. Therefore, we design modules (PSGA and FR) that adaptively extract features and further exploit correlations between points.
\subsection{3D point cloud geometry denoising}
3D point cloud geometry denoising is crucial for various applications such as autonomous driving. The goal of point cloud denoising is to remove outliers or fill in holes to enable further processing or downstream applications. Within this context, the integration of graph Laplacian regularization (GLR) has proven effective in enhancing denoising capabilities \cite{dinesh2020point}\cite{zeng20193d}. Dinesh \emph{et al.} \cite{dinesh2020point} proposed a signal-dependent feature graph Laplacian regularizer which can be used as a signal prior. Zeng \emph{et al.} \cite{zeng20193d} proposed a discrete patch distance measure to quantify the similarity between two surface patches of the same size for graph construction. Given the huge data volume in point cloud processing, manifold-based methods hold promise for features projection and dimensionality reduction. Hu \emph{et al.} \cite{hu2021dynamic} proposed a method based on spatial-temporal graph representation that uses the temporal consistency between surface patches that correspond to the same underlying manifold. Xu \emph{et al.} \cite{xu2022tdnet} proposed a transformer-based end-to-end network with an encoder-decoder architecture. The encoder uses a transformer, while the decoder learns the latent manifold of each sampled point. To achieve optimal denoising, multi-stage or stepwise strategies are often applied. Chen \emph{et al.} \cite{chen2022repcd} proposed a task-specific point cloud denoising network, RePCDNet, which consists of four key modules based on recurrent neural networks (RNNs) for feature extraction and enhancement. Jia \emph{et al.} \cite{jia2021deep} proposed a two-step method for removing geometry artifacts and improving the compression efficiency of V-PCC. The first step is learning-based pseudo-motion compensation, aiming at denoising the artifacts. The second step takes advantage of the strong correlation between the near and far depth fields decomposed from geometry to further improve the results. Furthermore, within the specific context of V-PCC, the work in \cite{xing2022vpcc} introduced an adaptive denoising method, which applies the Wiener filter on the 2D geometry images.

The works reviewed in this section suggest that extracting effective geometry features is crucial for improving the quality of the point cloud. Techniques such as domain transformation, surface fitting, normal approximation, and distance measurement can be used to detect and eliminate noise in the point cloud. In addition, geometry features can provide important information to enhance color quality.
\subsection{3D point cloud color quality enhancement}
Most of the previous work in the field of point cloud quality enhancement is based on traditional algorithms. For example, Irfan and Magli \cite{irfan2021joint,irfan2021exploiting} make full use of graph transform or optimization to reduce color distortion. In \cite{irfan2021joint}, a point cloud quality enhancement method was proposed, which uses a spectral graph wavelet transform to jointly exploit both geometry and color in the graph spectral domain. The authors \cite{irfan2021exploiting} later proposed a graph-based optimization method that aims to remove both geometry and color distortion simultaneously. The method assumes that the color and location of a point are related to each other. Similarly based on the graph, Yamamoto \emph{et al.} \cite{yamamoto2016deblurring} proposed a deblurring method for point cloud attributes, which is inspired by multi-Wiener Stein’s unbiased risk estimate-linear expansion of thresholding deconvolution of images. The blurred textures are considered as graph signals that are decomposed into sub-bands after Wiener-like filtering, and then all filtered signals are added together with the coefficients obtained from the linear equation. Another work using Wiener filter for color quality enhancement is \cite{xing2022gpcc}, which is specifically designed for G-PCC. Optimal coefficients of the filter for each color component are calculated in the encoder and selectively written into the bit stream according to the rate-distortion cost. For V-PCC, the only color enhancement method we are aware of is the low-complexity color smoothing method \cite{color_smoothing} included as an optional mode in the V-PCC test model. In this method, the decoded points are split into 3D grids, and the color centroid of each grid is calculated and then smoothed with tri-linear filtering. 

The methods discussed above often offer straightforward yet effective ways to enhance the  color quality of point clouds. It becomes evident that the relationship between geometry and color is crucial for driving performance improvements. Nonetheless, these methods do have certain limitations. For instance, their objectives can sometimes be confined to specific tasks, such as G-PCC or V-PCC, resulting in a rather narrow scope. This has motivated the exploration of learning-based methods. Sheng \emph{et al.} \cite{sheng2022attribute} proposed a multi-scale graph attention network for removing attribute compression artifacts for G-PCC. This method uses Chebyshev graph convolutions to extract features of point cloud attributes and a multi-scale scheme to capture short and long-range correlations between the current point and its neighbors. However, the applicability of the method is restricted to G-PCC geometry lossless coding. Moreover, the need for separate models trained at various bit rates imposes limitations.

In summary, while current point cloud quality enhancement methods have shown promising results, there is still room for improvement, particularly within a point cloud coding system. Enhancing quality not only improves the reconstruction quality but also increases the efficiency of the coding system. As color plays a significant role in the overall subjective quality of point clouds and has an impact on subsequent tasks \cite{liu2021reduced}, we propose GQE-Net as a post-processing filter for a point cloud compression system to reduce the compression artifacts. 
\section{Proposed Method}
To reduce the artifacts caused by compression, we propose a graph-based quality enhancement network as a post-processing filter for point cloud compression. The proposed method is outlined in Fig. 3.

\begin{figure}[!ht]
\setlength{\belowcaptionskip}{-2.0em}
\captionsetup{}
\centering
\centerline{\includegraphics[width=8.5cm]{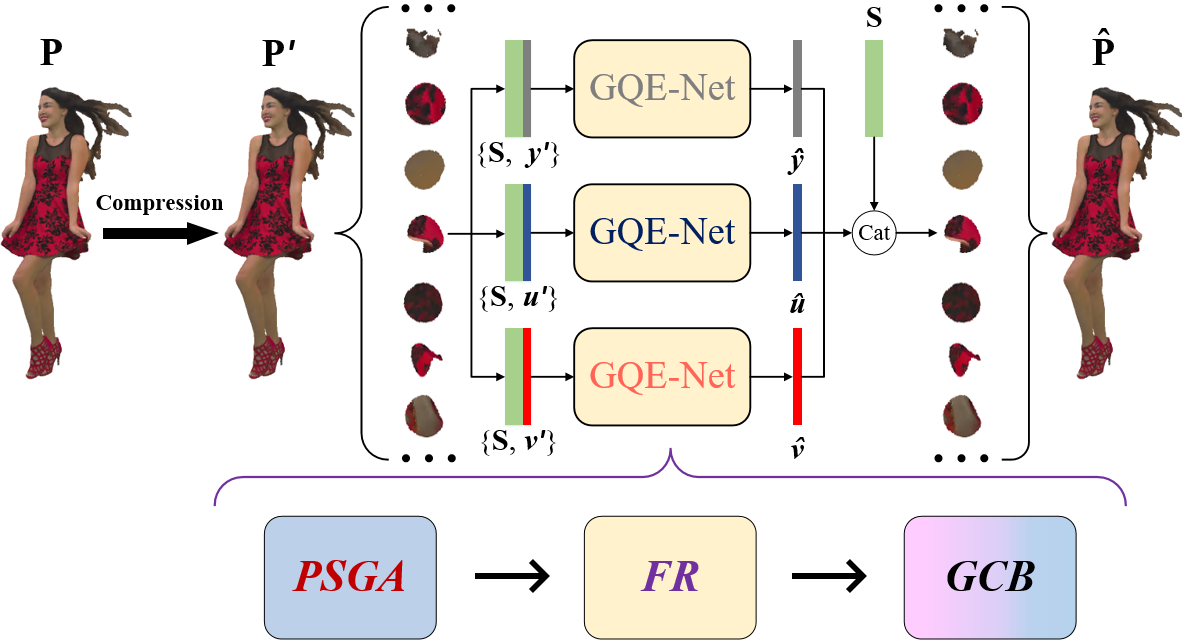}}
\caption{Overview of the proposed method.} \label{GQE-flow}
\end{figure}

There are four main modules in GQE-Net. The first module deals with graph construction, where the importance of local information is recognized for effective feature learning. To achieve this goal, we build a graph for each individual point. The second module, the Graph Convolution Block (GCB), facilitates efficient extraction of features from the graph. By integrating a maxpooling layer, GCB contributes to the selection of the most relevant features from various channels.

The third module, Parallel-Serial Graph Attention (PSGA), concentrates on salient points. A multi-head mechanism is used to enhance feature attention. To ensure coherence among the different heads and to further amplify the attention-related features, we adopt a structure that interconnects single heads in a parallel-serial configuration.

The fourth module introduces geometric information by computing normals, which are then integrated into the existing feature framework. Notably, we introduce a geometry-distance guided weighting matrix for each feature channel, mitigating the influence of points located farther away from the key point.

\subsection{Problem modelling}
To avoid color distortions caused by geometry losses, we assume that the compression of the geometry information is lossless, i.e., the coordinates of each point in the reconstructed point cloud remain unchanged after compression. We first convert the color space from RGB to YCbCr. We denote the original point cloud by $\bf{P} \in{ {\mathbb{R}}^{ \emph{N} \times \emph{6}}}$, where $N$ is the number of points, and $6$ corresponds to the 3D geometry coordinates and three color components. After decoding, the original point cloud is denoted by $\bf{P'} \in{ {\mathbb{R}}^{ \emph{N} \times \emph{6}}}$. By applying the proposed method to $\bf{P'}$, we obtain a quality-enhanced point cloud $\bf{\hat{P}} \in{ {\mathbb{R}}^{ \emph{N} \times \emph{6}}}$, as shown in Fig. 3. The aim of the proposed method is to minimize the difference between $\bf{\hat{P}}$ and $\bf{P}$.

\subsection{Patch generation and fusion}
Due to limitations in GPU memory capacity, it is challenging to input the entire point cloud directly into the neural network. As an alternative, we partition the point cloud into smaller 3D patches of fixed size $n$. These points are obtained as follows. Let $r$ be a parameter used to control the overlap between the 3D patches and let
\begin{subequations}
\setlength{\abovedisplayskip}{2.8pt} 
\setlength{\belowdisplayskip}{2.8pt}
\begin{gather}
m =\ \frac{N\times r}{n},  \tag{1}
\end{gather}
\end{subequations}
Using farthest point sampling (FPS), we select $m$ points from the point cloud. We call these $m$ points key points. Next, the k Nearest Neighbor (k-NN) algorithm \cite{peterson2009k} is applied to each key point to identify its $n-1$ closest neighbors. Each key point and its $n-1$ neighbors are grouped into a 3D patch represented as $\{\bm{\textbf{S}}, \bm{\textbf{C}}\}$, where $\bm{\textbf{S}} \in {{\mathbb{R}}^{n \times 3}}$ denotes the geometry coordinates and $\bm{\textbf{C}}=\{\bm{y}, \bm{u}, \bm{v}\} \in {\mathbb{R}}^{n \times 3}$ denotes the three color components of the $n$ points in the 3D patch. Thus, for each color component, the 3D patches have the same geometry but different color information. The architecture of GQE-Net for these three sets is identical, but with different parameters. For example, GQE-Net for the Luma component (denoted as Y) can be written as a function:
\begin{subequations}
\begin{gather}
\hat{\bm{y}} =\ \Psi(\bm{\textbf{S}},\  \bm{y'}\  | \ \Theta_{\bm{\textbf{Y}}}),  \tag{2}
\end{gather}
\end{subequations}
where $\Psi(\cdot)$ represents GQE-Net, $\Theta_{\bm{\textbf{Y}}}$ represents the learnable parameters for the Y component, $\bm{y'}$ represents the initial reconstructed Y component, and $\bm{\hat{\bm{y}}}$ represents the final, quality-enhanced Y component. The final color $\hat{\bm{\textbf{C}}} = \{ \hat{\bm{y}}, \hat{\bm{u}}, \hat{\bm{v}} \}$ of the patch is generated by combining the three quality-enhanced color components.

Using $\bm{\textbf{S}}$, we merge the 3D patches to create a high-quality point cloud, $\hat{\bm{\textbf{P}}}$. However, in this process, it is likely that some points will be selected multiple times across different patches, while others may not be selected at all. To address this problem, we use the following approach. For points that are processed in different patches, we take the average as the restored value. For points that are not included in any patch, we keep the same color. 
\subsection{Graph construction}
GQE-Net is a graph-based neural network, with most of its operations carried out on the graph topology. Given $n$ points with $L$-dimensional color features $\bm{\textbf{F}}=\{ \bm{f}_1,\bm{f}_2,…,\bm{f}_n\} \in {{\mathbb{R}}^{n \times L}}$ and 3D coordinates $\bm{\textbf{S}}=\{\bm{s}_1,\bm{s}_2,…,\bm{s}_n\} \in {{\mathbb{R}}^{n \times 3}}$, at the beginning, the $L$-dimensional feature is simply the corresponding color, i.e., $L = 1$. We construct a directed and self-loop graph $\bf{G}=(\bf{V}, \bf{E})$, where $\bf{V}$ represents the nodes and $\bf{E}$ represents the edges. To exploit the relationship between features of a node and its neighbors, while also simplifying the graph, each node is only connected to its $k-1$ closest neighboring nodes, as well as itself. These nodes are called the $k$ neighbors of the current node in the rest of the paper. The subsequent operations are carried out on the edge features $\bm{e}_{ij}=(\bm{f}_i,\bm{f}_{ij}-\bm{f}_i)\in {{\mathbb{R}}^{n \times 2L}}$, where $\bm{f}_i$ is the feature of the current node, $\bm{f}_{ij}$ is the feature of the $j^{th}$ connected node ($j=1,2,…,k$), and $\bm{f}_{i1}=\bm{f}_{i}$. By combining the edge features of all nodes, we obtain the final features $\bm{\textbf{P}}' \in {{\mathbb{R}}^{n \times k \times 2L}}$. Thus, for each node we calculate a $k \times 2L$ dimensional feature based on its $k$ nearest neighbors. This graph construction is shown in Fig. 4 and can be written as
\begin{subequations}
\begin{gather}
\bm{\textbf{F}}' = GC(\bm{\textbf{F}}),  \tag{3}
\end{gather}
\end{subequations}

\begin{figure}[!ht]
\setlength{\belowcaptionskip}{-2.0em}
\captionsetup{}
\centering
\centerline{\includegraphics[width=8.5cm]{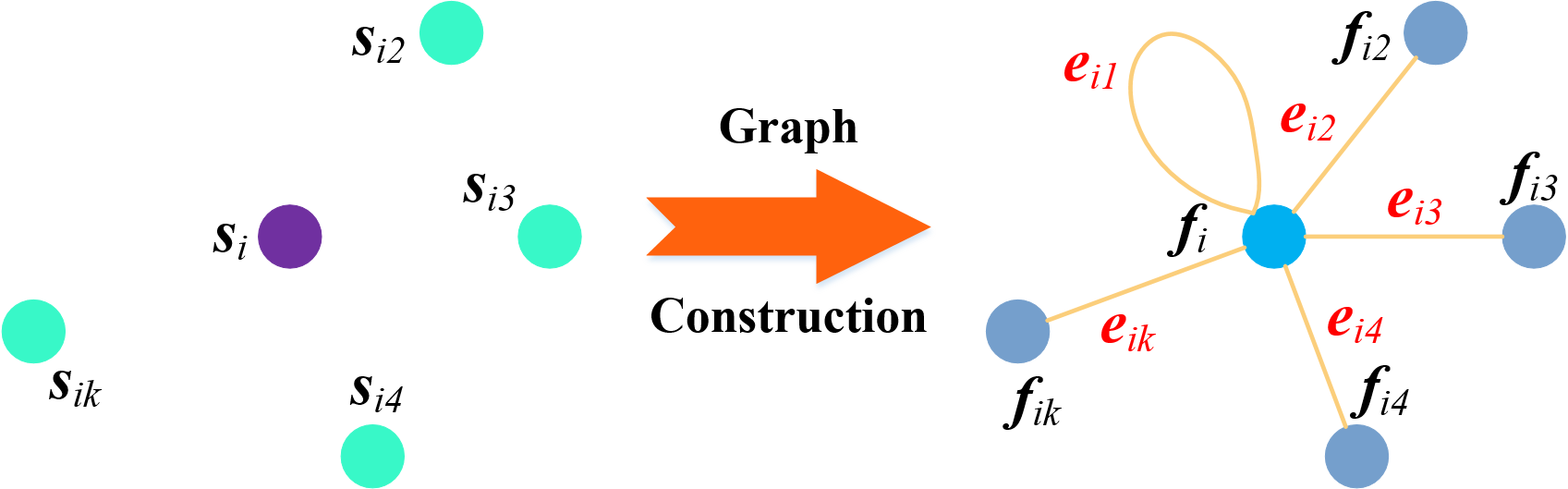}}
\caption{Graph construction.} \label{graph_new}
\end{figure}

\vspace{-0.2em}
\subsection{Graph convolution block}

The graph convolution block (GCB) is a pivotal element within GQE-Net, facilitating the efficient extraction of features from the graph (Fig. 5). By integrating a maxpooling layer, the GCB aids in the selection of the most pertinent features from various channels. The input edge feature, $\bm{\textbf{F}}_{g\_in}$, has a size of $n \times k \times L_1$, where $L_1$ is the number of feature channels. The input features are processed by three single blocks in series. Each single block contains a 2D convolution layer with a kernel size of $1 \times 1$, a batch normalization layer, and a Leaky ReLU for activation. After this process, the feature $\textbf{F}_{g\_mix}$, whose number of feature channels is $L_2$, can be obtained. At this point, the feature $\bm{\textbf{F}}_{g\_mix}$ is still an edge feature. To capture more abstract features from $\bm{\textbf{F}}_{g\_mix}$, a max pooling layer is used. It is worth noting that the max pooling layer selects the most important feature from the $k$ neighboring nodes at each channel. As a result, we can obtain the output feature $\textbf{F}_{o}$ with a size of $n \times L_2$. In other words, the neighboring information has been embedded into the $L_2$-dimensional feature, $\textbf{F}_{o}$.

\begin{figure}[!ht]
\setlength{\belowcaptionskip}{-2.0em}
\captionsetup{}
\centering
\centerline{\includegraphics[width=8.5cm]{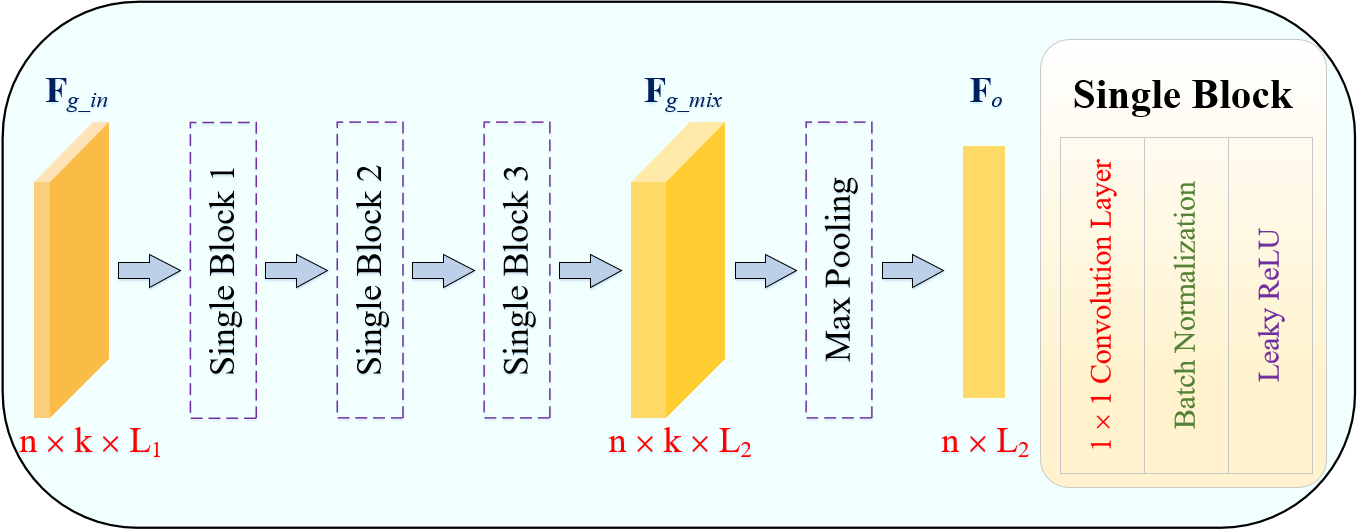}}
\caption{Graph convolution block structure.} \label{GCB}
\end{figure}

\vspace{-0.8em}
\subsection{PSGA module}
Points located in different areas may experience different levels of distortion after compression. It is important to focus on points with significant artifacts or those located in areas that significantly affect the subjective quality, such as high-frequency edges. To achieve this, we propose a PSGA module based on a multi-head mechanism \cite{chen2021gapointnet} \cite{vaswani2017attention}, as shown in Fig. 6 (a). Unlike existing methods, it adopts a parallel-serial structure. In the first layer, two single head attention (SHA) blocks ($\rm{SHA}_1$ and $\rm{SHA}_2$) are used in parallel to extract attention features separately. The output attention features from these two blocks are then combined and sent to a third SHA block ($\rm{SHA}_3$) for feature integration and reinforcement. This design is beneficial for exploring feature correlation and further integrating attention features.

The input to PSGA is $\textbf{F}$ with $L$ feature channels. The graph construction is done before each SHA block. The outputs of the parallel SHA blocks include an attention feature $(\textbf{F}_{ma} = \{\textbf{F}_{pa1},\ \textbf{F}_{pa2} \in {{\mathbb{R}}^{n \times 2 \times L}} \})$ obtained through graph-based self-attention on the input features, and a graph feature $(\textbf{F}_{mg} = \{\textbf{F}_{pg1},\ \textbf{F}_{pg2} \in {{\mathbb{R}}^{n \times k \times 2 \times L}} \})$ which contains the neighboring features of each point. For the two parallel heads, the graph features and the attention features are concatenated, respectively. A third SHA block $(\rm{SHA}_3)$ takes the concatenated attention features as input for fusion. Finally, $\textbf{F}_{ma}$ and the output attention feature of $\rm{SHA}_3$ are concatenated to get the final features $\textbf{F}_{a}$, while $\textbf{F}_{mg}$ and the output of $\rm{SHA}_3$ are also concatenated to get the updated graph features $\textbf{F}_{g}$.

\begin{figure}[!ht]
\centering
\setlength{\belowcaptionskip}{-2.0em}
\captionsetup{font=small}
\subfloat[]{\includegraphics[width=8.5cm]{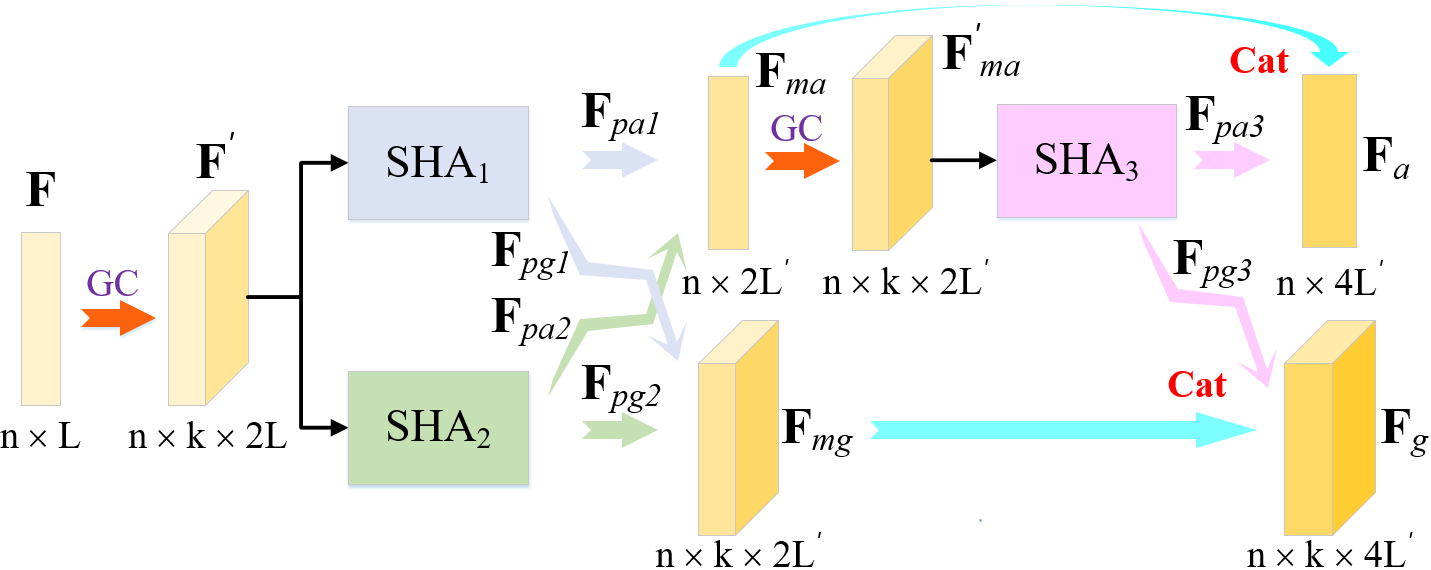}%
\label{}}
\hfil

\subfloat[]{\includegraphics[width=8.5cm]{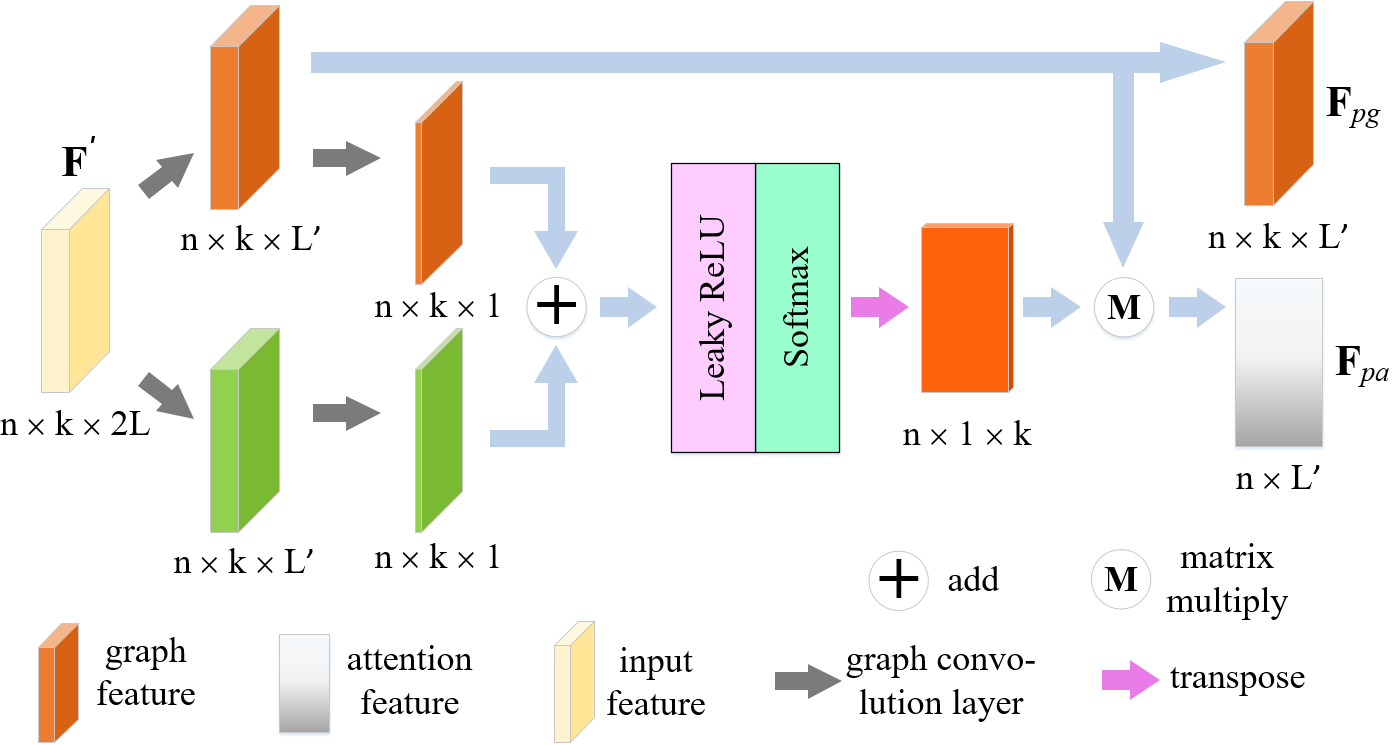}%
\label{fig_second_case}}
\caption{(a) PSGA module. ``GC'' means ``Graph Construction''. (b) Single head attention module.}
\label{}
\end{figure}

The structure of a single head attention is illustrated in Fig. 6 (b). We project the feature $\textbf{F}'$ into a higher-dimensional space, with a dimension of $L'$, using two branches. The output of one branch is $\textbf{F}_{pg}$. Specifically, we apply a graph convolution on $\textbf{F}'$ to obtain $\textbf{F}_{pg}$. In PSGA, the graph convolution layer uses a 2D convolution with a kernel size of $1\times 1$, followed by batch normalization and a Leaky ReLU activation function. Next, the graph features from both branches are compressed into the $n \times k \times 1$ dimension using a graph convolution layer without an activation function to obtain the most distinctive features of each neighboring point. We then add the features of the two branches and apply a Leaky ReLU activation function. The output, $\textbf{A} \in {{\mathbb{R}}^{n \times k }}$, is then normalized to weights ranging from $0$ to $1$. We use $softmax$ function to each row of $\textbf{A}$:
\begin{subequations}
\setlength{\abovedisplayskip}{2.5pt} 
\setlength{\belowdisplayskip}{2.5pt}
\begin{gather}
softmax(\textbf{A}_{i})_j = \frac{exp(\textbf{A}_{ij})}{\sum_{t=1}^k exp(\textbf{A}_{it})},  \tag{4}
\end{gather}
\end{subequations}
where $\textbf{A}_{i}$ is the $i^{th}$ row of $\textbf{A}$, and $\textbf{A}_{ij}$ is the $j^{th}$ feature value of $\textbf{A}_{i}$. The final attention feature, $\textbf{F}_{pa}$, is obtained by multiplying the graph feature $\textbf{F}_{pg}$ with the normalized attention weights. $\textbf{F}_{pa}$ provides information about the degree of distortion and importance of each point by taking into account the local information.

\begin{figure}
\setlength{\belowcaptionskip}{-2.0em}
\captionsetup{}
\centering
\centerline{\includegraphics[width=8.5cm]{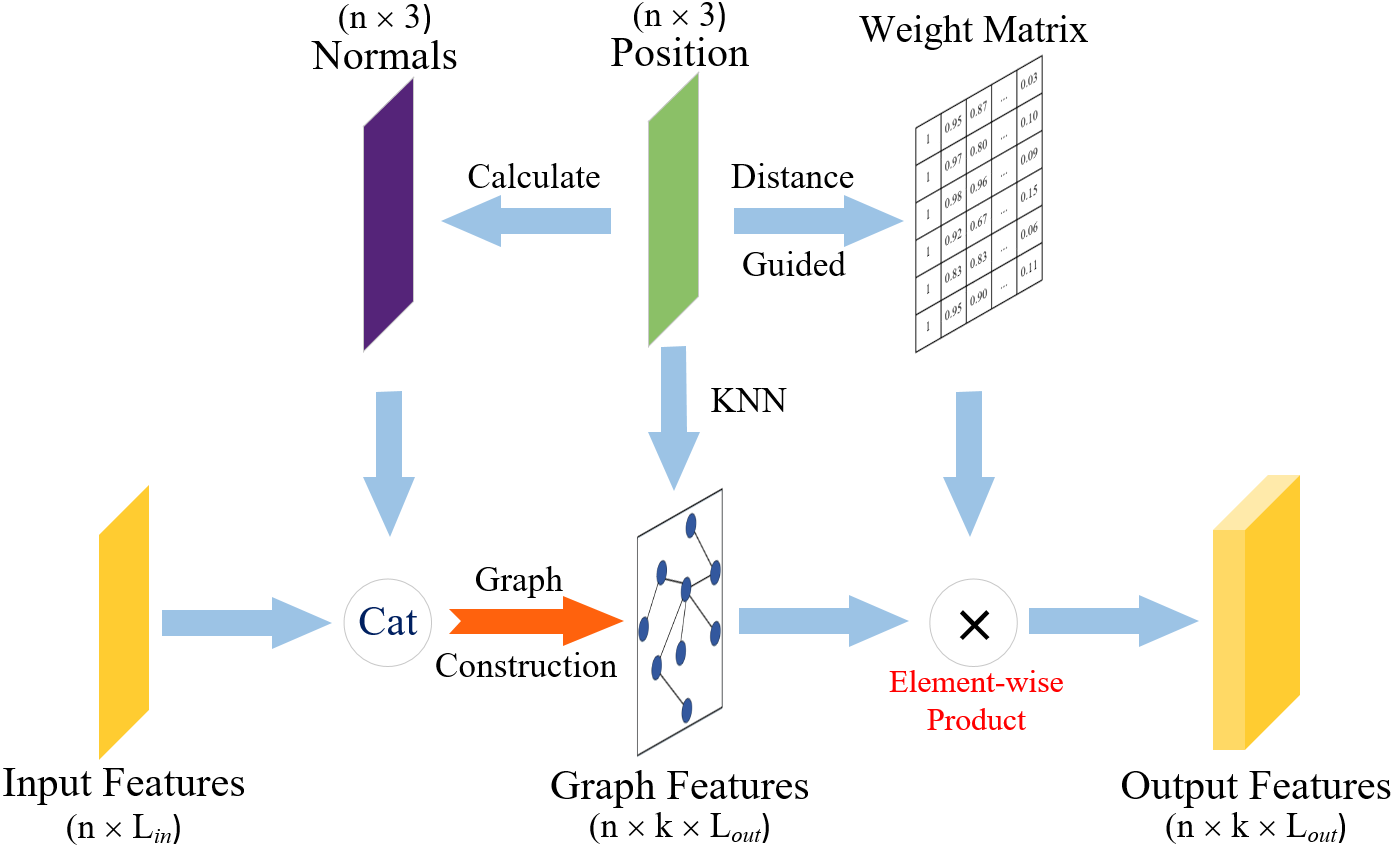}}
\caption{Feature refinement module.} \label{FR}
\end{figure}

\subsection{FR module}
The FR module is designed to effectively analyze and use the geometric information in the vicinity of each point, allowing for efficient leverage of the inter-correlation among points. The structure of the FR module is illustrated in Fig. 7. Specifically, the normal vector of each point is calculated and integrated with the input features. After constructing the graph, a weight matrix is also computed based on the distance between each point and its neighboring points, which is then multiplied by the graph feature.

The normal gives a measure of the angle of the tangent plane on which a point is located. The normals of points can reflect changes in the curvature of the surface. Both the coordinates and the normal contribute to the correlation between points. For example, if the normal discrepancy between two neighboring points is large, their color may differ greatly because they are not on the same plane and do not have similar local features. Therefore, we incorporate the normal of each point into the features, increasing the dimension of the features from $L_{in}$ to $L_{out}=L_{in}+3$.

The computation of the normal vector is achieved through principal component analysis (PCA) \cite{abdi2010principal}, a method that uses the k nearest neighbors of a given point to approximate the underlying plane. By extracting a specific direction from this plane, the distribution of all points projected onto this direction becomes the most concentrated, effectively revealing the least significant vector in PCA. This calculated normal vector subsequently enriches the feature set, enabling a more comprehensive characterization of each point.

Then, we also extract local correlation based on the distances between points. Since points that are farther away from the current point may have less structural similarity, we aim to reduce the influence of features from distant neighboring points. To do this, we first calculate the distance $d_{ij}$, $i \in [1, n]$, $j \in [1, k]$ between the $i^{th}$ node and each of its $k$ neighboring nodes in the graph. This defines a weight matrix $\textbf{W} = (w_{ij}) \in {\mathbb{R}}^{n \times k}$, where
\begin{subequations}
\setlength{\abovedisplayskip}{2.5pt} 
\setlength{\belowdisplayskip}{2.5pt}
\begin{gather}
w_{ij} = 2 \times (1-sigmoid(d_{ij})),  \tag{5}
\end{gather}
\end{subequations}
and
\begin{subequations}
\setlength{\abovedisplayskip}{2.5pt} 
\setlength{\belowdisplayskip}{2.5pt}
\begin{gather}
sigmoid(d_{ij}) = \frac{1}{1+exp(-d_{ij})}.  \tag{6}
\end{gather}
\end{subequations}
Note that $w_{ij}$ is close to $1$ when $d_{ij}$ is close to $0$, while $w_{ij}$ is close to $0$ when $d_{ij}$ is large. Since the number of channels in the current graph features is $L'$, we replicate matrix $\textbf{W}$ $L'$ times to generate a tensor $\textbf{W}'$ with dimension $n \times k \times L'$ to deal with each channel of the graph features. Finally, we calculate an element-wise product between $\textbf{W}'$ and the graph features to generate the output features.
\vspace{-0.7em}
\subsection{Architecture of GQE-Net}
Fig. 2 shows the architecture of GQE-Net. The input of the network is divided into two parts: distorted color ${\textbf{C}_i}'$ and geometry coordinates $\textbf{S}$. The PSGA module is applied to the input color to focus on important points or points with large distortions. The output of the PSGA module is graph features $\textbf{F}_{g1}$ and attention features $\textbf{F}_{a1}$:
\begin{subequations}
\setlength{\abovedisplayskip}{2.5pt} 
\setlength{\belowdisplayskip}{2.5pt}
\begin{gather}
(\textbf{F}_{a1}, \textbf{F}_{g1}) = PSGA({\textbf{C}_i}', \textbf{S}).  \tag{7}
\end{gather}
\end{subequations}
The number of feature channels of $\textbf{F}_{a1}$ and $\textbf{F}_{g1}$ is $64$. In the FR module, the attention features $\textbf{F}_{a1}$ are fused with the input color ${\textbf{C}_i}'$. The normals and distance are taken into account, and the neighbors are weighted by the geometry distance. The output of the FR module is the refined features 
\begin{subequations}
\setlength{\abovedisplayskip}{2.5pt} 
\setlength{\belowdisplayskip}{2.5pt}
\begin{gather}
\textbf{F}_{r1}=FR(cat({\textbf{C}_i}', \textbf{F}_{a1}), \textbf{S}),  \tag{8}
\end{gather}
\end{subequations}
where $cat(\cdot)$ represents concatenation. To effectively fuse and extract features, the proposed GCB is applied to the graph feature $\textbf{F}_{r1}$:
\begin{subequations}
\setlength{\abovedisplayskip}{2.5pt} 
\setlength{\belowdisplayskip}{2.5pt}
\begin{gather}
\textbf{F}_{1}=GCB(\textbf{F}_{r1}),  \tag{9}
\end{gather}
\end{subequations}
where $GCB(\cdot)$ denotes the graph convolution block. This operation changes the number of feature channels to $64$. To further extract features point-by-point, another PSGA module is applied:
\begin{subequations}
\setlength{\abovedisplayskip}{2.5pt} 
\setlength{\belowdisplayskip}{2.5pt}
\begin{gather}
(\textbf{F}_{a2}, \textbf{F}_{g2}) = PSGA(\textbf{F}_1, \textbf{S}),  \tag{10}
\end{gather}
\end{subequations}
where the number of feature channels for $\textbf{F}_{a2}$ and $\textbf{F}_{g2}$ is $256$. The normals and neighbors are also combined for further feature refinement:
\begin{subequations}
\setlength{\abovedisplayskip}{2.5pt} 
\setlength{\belowdisplayskip}{2.5pt}
\begin{gather}
\textbf{F}_{r2}=FR(cat(\textbf{F}_1, \textbf{F}_{a2}), \textbf{S}),  \tag{11}
\end{gather}
\end{subequations}
where $\textbf{F}_{r2}$ aggregates two features ($\textbf{F}_{1}$ and $\textbf{F}_{a2}$) and incorporates distance and normal information, resulting in $643$ channels. Then, another GCB is applied for feature fusion and to select the most important neighbor in each feature channel. 
\begin{subequations}
\setlength{\abovedisplayskip}{2.5pt} 
\setlength{\belowdisplayskip}{2.5pt}
\begin{gather}
\textbf{F}_{2}=GCB(\textbf{F}_{r2}).  \tag{12}
\end{gather}
\end{subequations}
This results in $256$ feature channels for $\textbf{F}_{2}$. Graph features are crucial for obtaining local information of each point. Therefore, we use the graph features generated by the PSGAs and process them through GCBs to generate $\textbf{F}_{3}$ and $\textbf{F}_{4}$: 
\begin{subequations}
\setlength{\abovedisplayskip}{2.5pt} 
\setlength{\belowdisplayskip}{2.5pt}
\begin{gather}
\textbf{F}_{3}=GCB(\textbf{F}_{g1}),\  \textbf{F}_{4}=GCB(\textbf{F}_{g2}).  \tag{13}
\end{gather}
\end{subequations}
The features obtained through this process have incorporated information from the neighborhood through graph convolution and also emphasized important details. Moreover, the features are fused through skip connections to enhance the representation capabilities of the network. In this way, the features from different layers can be fully exploited:
\begin{subequations}
\setlength{\abovedisplayskip}{2.5pt} 
\setlength{\belowdisplayskip}{2.5pt}
\begin{gather}
\textbf{F}_{m}=cat(\textbf{F}_{2}, \textbf{F}_{3}, \textbf{F}_{4}).  \tag{14}
\end{gather}
\end{subequations}
Finally, by using an $1 \times 1$ convolution layer without batch normalization and activation, the mixed features are restored, and residual learning is also used to speed up convergence and ensure the performance of the proposed network:
\begin{subequations}
\setlength{\abovedisplayskip}{2.5pt} 
\setlength{\belowdisplayskip}{2.5pt}
\begin{gather}
\hat{\textbf{C}}_i = {\textbf{C}_i}' + Conv(\textbf{F}_m),     \tag{15}
\end{gather}
\end{subequations}
where $Conv(\cdot)$ is a convolution layer with kernel size of $1 \times 1$.

\subsection{Loss function}
The loss function needs to reflect the distance between the quality-enhanced patch and the ground truth for each color component. Therefore, we use the mean squared error (MSE) as the loss function of GQE-Net:
\begin{subequations}
\setlength{\abovedisplayskip}{2.5pt} 
\setlength{\belowdisplayskip}{2.5pt}
\begin{gather}
l_i = \frac{1}{n}\sum_{j=1}^n {|| \textbf{C}_{ij} - \hat{\textbf{C}}_{ij}||}^2,     \tag{16}
\end{gather}
\end{subequations}
where $l_i$ is the loss of the $i^{th}$ color component, $\textbf{C}$ and $\hat{\textbf{C}}$ represent the true color and the restored color, respectively. Note that the loss is calculated in the YCbCr color space.

\section{Experimental Results}
We conducted a series of experiments to assess the objective and subjective quality of the point clouds when our method is used. We also compared the coding efficiency before and after incorporating our method into the coding system. Additionally, we carried out several ablation studies to study the impact of each component of GQE-Net on the overall performance. 

\subsection{Experimental configuration}
We evaluated the performance of our method on point clouds encoded with G-PCC Test Model Category 13 version 14.0 (TMC13v14.0) using a computer with an Intel i7-7820X CPU and 64GB of memory. The encoding process was conducted under the Common Test Condition (CTC) C1 \cite{schwarz2018common}, which involves lossless geometry compression and lossy attribute compression. We used the lifting transform \cite{gpcc} in the coding procedure.

We trained two GQE-Net models, one for dense point clouds and one for sparse point clouds. To train each model, we used point clouds compressed at various bit rates. Training was carried out on an NVIDIA GeForce RTX3090 GPU, using PyTorch v1.10.  The experimental details, results, and analysis for each case are presented in the following sections. 
\subsubsection{GQE-Net for dense point clouds}
\ 
\newline
\indent{To train our first GQE-Net model, we first selected point clouds with color information from the Waterloo point cloud sub-dataset (WPCSD) \cite{liu2021pqa}. The training data set is shown in the supplemental materials. As can be seen, the point clouds have rich color and texture. During training, we set the number of points $n$ in a patch to $2048$, and the overlapping ratio $r$ to $2$. In the k-NN-based graph construction, we set $k$ to $20$. We collected a total of $52164$ patches for training. We ran a total of $180$ epochs, during which we started with an initial learning rate of $0.0016$ and decreased it by $0.25$ every $60$ epochs. The batch size was $12$. We used the Adam optimizer \cite{kingma2014adam} with a momentum of $0.9$.}

We assessed the performance of GQE-Net on $14$ point clouds from the Cat\_A dataset. We selected several representative point clouds provided by MPEG \cite{schwarz2018emerging}. The test data set is shown in the supplemental materials. Each point cloud was compressed with G-PCC using the quantization parameters (QP) $51$, $46$, $40$, $34$, $28$, $22$, which correspond to the six bit rates R01, R02, ..., R06. During the test, the number of points in a patch was set to $2048$, and the overlapping ratio was set to $2$. This ensures that the percentage of unprocessed points is less than $0.1\%$.

\subsubsection{GQE-Net for sparse point clouds}
\ 
\newline
\indent{To train our second GQE-Net model, we selected point clouds of the ``building'' type from the MPEG Cat\_B dataset \cite{schwarz2018emerging}. For testing, we used a different set of point clouds from the same dataset. The training data set and the test data set are shown in the supplemental materials. As in Section IV-A-1), we encoded the point clouds with G-PCC at the six bit rates R01, R02, ..., R06 and set the number of points in each patch to $2048$, with overlapping ratio $2$.}

\subsection{Objective quality evaluation}
To evaluate both the objective quality and coding efficiency, we used the commonly used BD-PSNR and BD-rate \cite{bjontegaard2001calculation} metrics. The BD-rate measures the average bit rate reduction in bits per input point (bpip) at the same PSNR. A negative BD-rate indicates better performance compared to the test model. In addition to calculating the PSNR for each color component, we used YCbCr-PSNR \cite{torlig2018novel} to evaluate the overall color quality gains brought by our method to the point cloud. The results for Cat\_A and Cat\_B are shown in Table I and Table II, respectively. More results can be found in the supplemental materials.

\begin{table}[!htbp]
\scriptsize
\renewcommand{\arraystretch}{1}
\captionsetup{font={small}}
\caption{Rate-distortion comparison with G-PCC TMC13v14.0 on Cat\_A.}
\setlength{\belowcaptionskip}{-1.0em}
\centering
{\begin{tabular}{cccccccc}
\toprule
\centering
\multirow{2}{*}{\textbf{Sequence}}& \multicolumn{4}{c}{\textbf{BD-PSNR [dB]}}  & \multicolumn{3}{c}{\textbf{BD-rate [\%]}}    \\
  &\textbf{Luma}& \textbf{Cb}\ & \textbf{Cr}& \textbf{YCbCr} &\textbf{Luma}& \textbf{Cb} &  \textbf{Cr} \\
\midrule
basketball & 0.381 & 0.171 & 0.413 & 0.359 & -13.5  & -8.5 &  -20.4 \\
dancer & 0.399 & 0.181 & 0.415 & 0.374  & -13.5 & -8.3 & -18.6 \\
exercise & 0.255 & 0.141 & 0.281 & 0.244  & -11.2 &  -8.2 & \textbf{-21.6} \\
longdress & 0.449 & 0.375 & 0.374 & 0.431  & -12.4 &  -10.7 &  -12.0 \\
loot & 0.472 & 0.428 & 0.479 & 0.467  & -13.9  &  -11.7 & -16.7 \\
model & 0.375 & 0.194 & 0.438 & 0.360 &  -12.5 &  -7.7 & -15.8 \\
queen & 0.242 & \textbf{0.417} & 0.489 & 0.289 &  -8.0 & \textbf{-12.2} &  -13.8 \\
redandblack & 0.449 & 0.242 & \textbf{0.498} & 0.429 & -14.1 & -8.6 & -14.5 \\
soldier & 0.550 & 0.375 & 0.480 & \textbf{0.519} & -15.2 & -10.0 & -15.7 \\
Andrew & \textbf{0.584} & 0.179 & 0.120 & 0.476 &  \textbf{-20.5} & -8.4 & -6.1 \\
David & 0.335 & 0.169 & 0.266 & 0.305 & -12.2 & -8.8 & -15.9 \\
Phil & 0.558 & 0.165 & 0.159 & 0.459 & -17.8 & -7.0 & -6.3 \\
Ricardo & 0.390 & 0.240 & 0.299 & 0.360 & -14.2 & -10.1 & -14.1 \\
Sarah & 0.535 & 0.227 & 0.249 & 0.461 & -17.4 &  -9.8 & -11.3 \\
\midrule

\textbf{Average} & \textbf{0.427} & \textbf{0.253} & \textbf{0.355} &\textbf{0.395} &\textbf{-14.0} & \textbf{-9.3} & \textbf{-14.5} \\ 
\bottomrule
\end{tabular}}
\end{table}

From Table I, we can see that for Cat\_A point clouds GQE-Net achieved an average improvement of $0.43$ dB, $0.25$ dB and $0.36$ dB for the Y, Cb, and Cr components, respectively, corresponding to -$14.0\%$, -$9.3\%$, -$14.5\%$ BD-rates. The largest PSNR gains were $1.36$ dB (Luma component of Andrew at R05), $0.83$ dB (Cb component of soldier at R06), and $0.93$ dB (Cr component of soldier at R06). We observed that the performance gains were limited at the high bitrates. This is due to two reasons. First, since the quality of the point clouds at high bitrates is already high, the room for improvement is smaller than at low bitrates. Second, since we trained GQE-Net with point clouds compressed at various bitrates, the training process is likely to have geared the network towards minimizing large distortions. Such large distortions typically correspond to low bitrates. Fig. 8 compares the rate-PSNR curves for Cat\_A point clouds. It can be seen that the coding efficiency can be greatly improved by incorporating the proposed method into the coding system, particularly at medium and high bit rates.

\begin{table}[!htbp]
\scriptsize
\renewcommand{\arraystretch}{1}
\captionsetup{font={small}}
\caption{Rate-distortion comparison with G-PCC TMC13v14.0 on Cat\_B.}
\setlength{\belowcaptionskip}{-1.0em}
\centering
{\begin{tabular}{cccccccc}
\toprule
\centering
\multirow{2}{*}{\textbf{Sequence}}& \multicolumn{4}{c}{\textbf{BD-PSNR [dB]}}  & \multicolumn{3}{c}{\textbf{BD-rate [\%]}}    \\
  &\textbf{Luma}& \textbf{Cb}\ & \textbf{Cr}& \textbf{YCbCr} &\textbf{Luma}& \textbf{Cb} &  \textbf{Cr} \\
\midrule
arco\_valentino & -0.15 & 0.091 & \textbf{0.086} & -0.09 & -0.2  & -4.4 &  \textbf{-5.0} \\
egyptian\_mask & \textbf{0.062} & 0.133 & 0.070 & \textbf{0.072}  & \textbf{-2.9} & \textbf{-5.4} & -3.8 \\
facade\_00009 & -0.09 & 0.070 & 0.034 & -0.05  & -0.5 &  -2.3 & -2.0 \\
staue\_klimt & 0.002 & \textbf{0.105} & -0.03 & 0.011  & -1.7 &  -3.9 &  -0.7 \\
\midrule
\textbf{Average} & \textbf{-0.04} & \textbf{0.100} & \textbf{0.038} &\textbf{-0.02} &\textbf{-1.3} & \textbf{-4.0} & \textbf{-2.8} \\ 
\textbf{Average\_low} & \textbf{0.119} & \textbf{0.103} & \textbf{0.071} &\textbf{0.111} &\textbf{-3.1} & \textbf{-4.6} & \textbf{-4.9} \\ 
\bottomrule
\end{tabular}}
\begin{tablenotes}
\scriptsize
\item[*]$*$ \textbf{Average\_low} shows the average performance at the low bitrates (R01, R02, R03, R04).
\end{tablenotes}
\end{table}

\vspace{-1em}
Table II shows the performance of our second GQE-Net model. The average BD-rates for the Y, Cb, Cr components were -$1.3\%$, -$4.0\%$ and -$2.8\%$, respectively. This is because the relationship between the points in these sparse point clouds is more challenging to capture. Moreover, the correlation between points is also relatively low, making it difficult to improve the quality of the current point using its neighbors. Furthermore, the inherent noise present in Cat\_B point clouds may also have a negative impact on performance. If we consider the low bitrates only (i.e., R01, R02, R03, R04), the BD-rates can improve, achieving -$3.1\%$, -$4.6\%$ and -$4.9\%$, for Y, Cb, Cr, respectively.

\begin{figure*}[!ht]
\centering
\captionsetup{font=small}
\subfloat{\includegraphics[width=3cm]{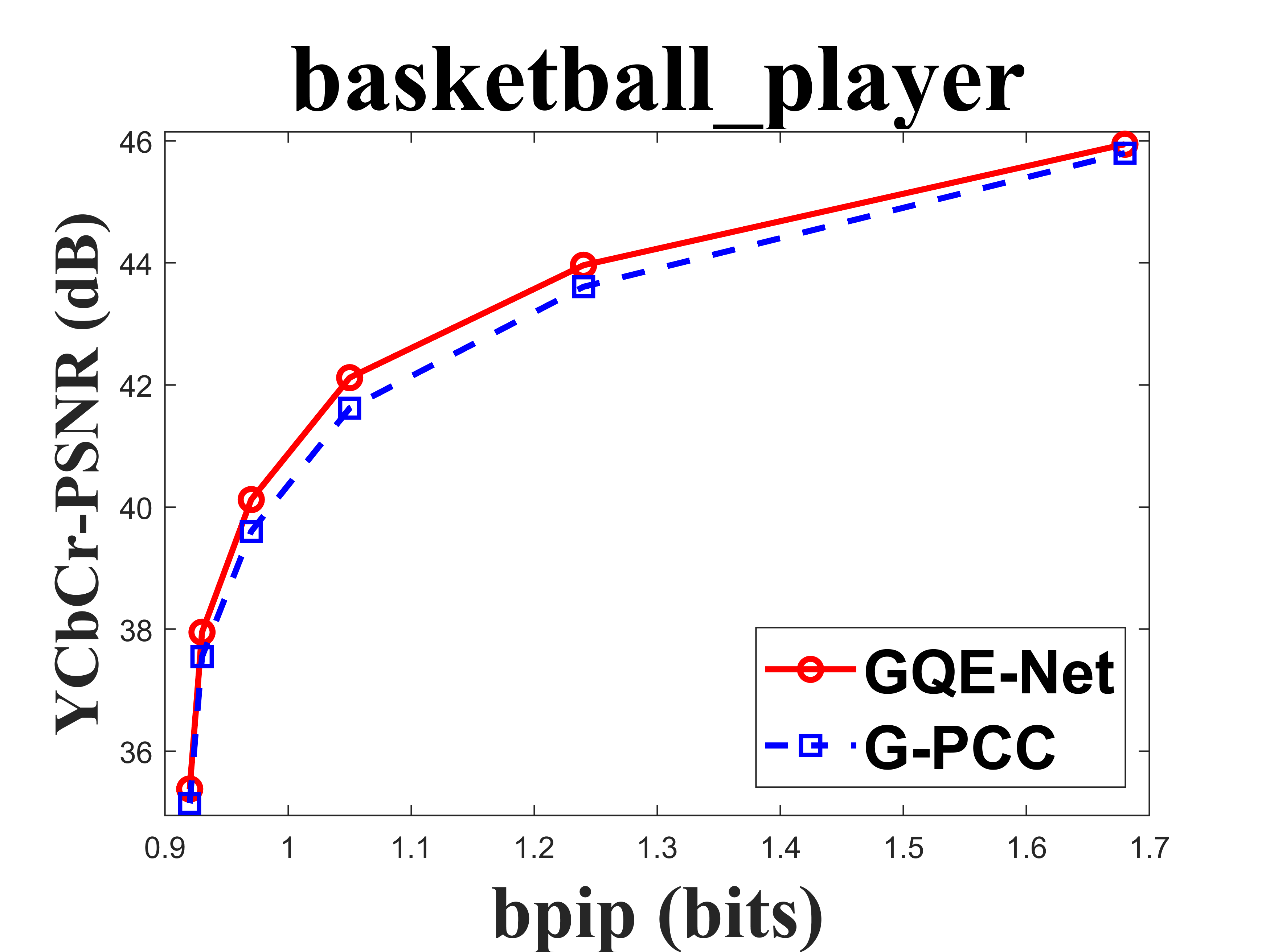}%
\label{}}
\subfloat{\includegraphics[width=3cm]{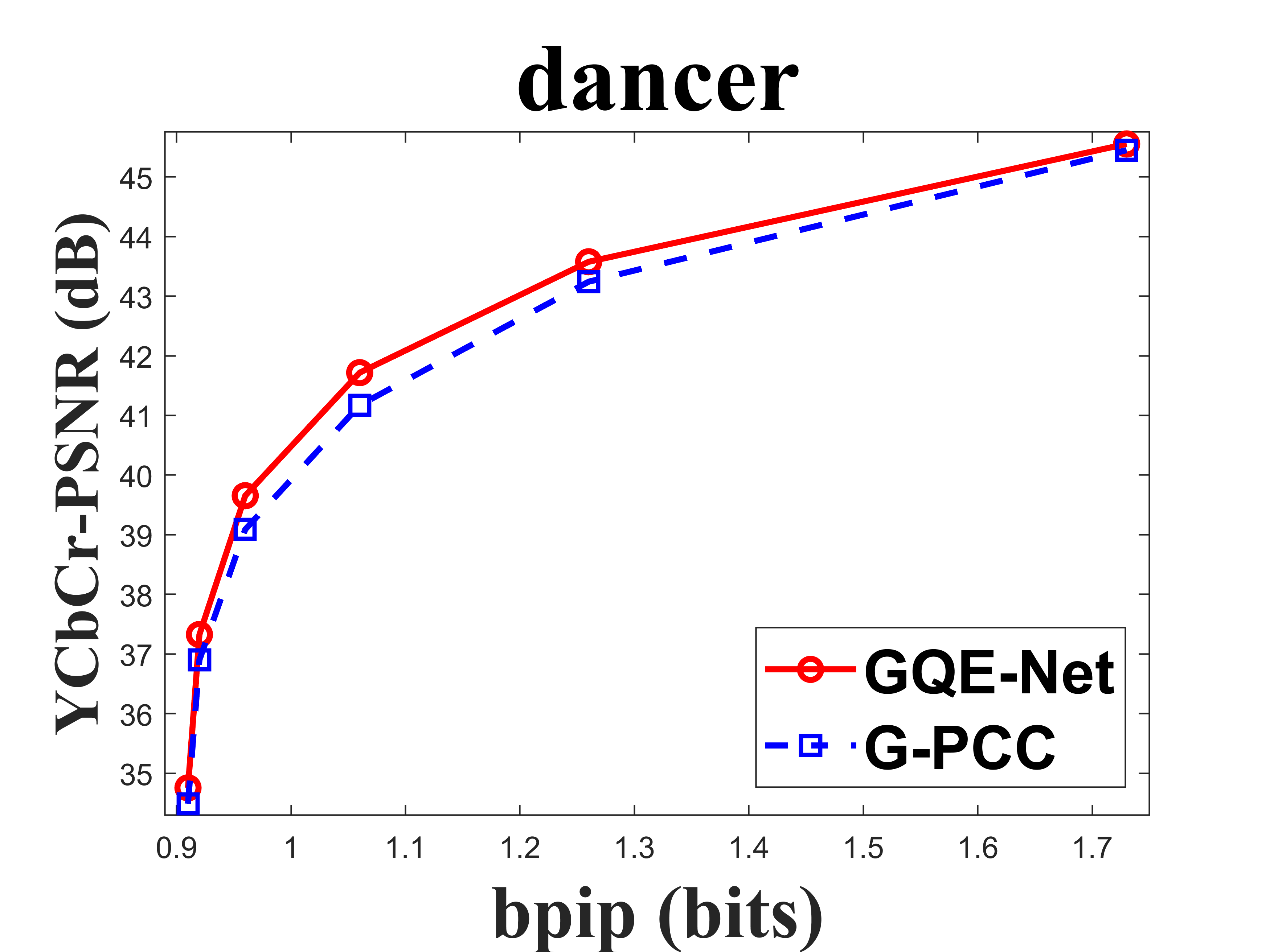}%
\label{}}
\subfloat{\includegraphics[width=3cm]{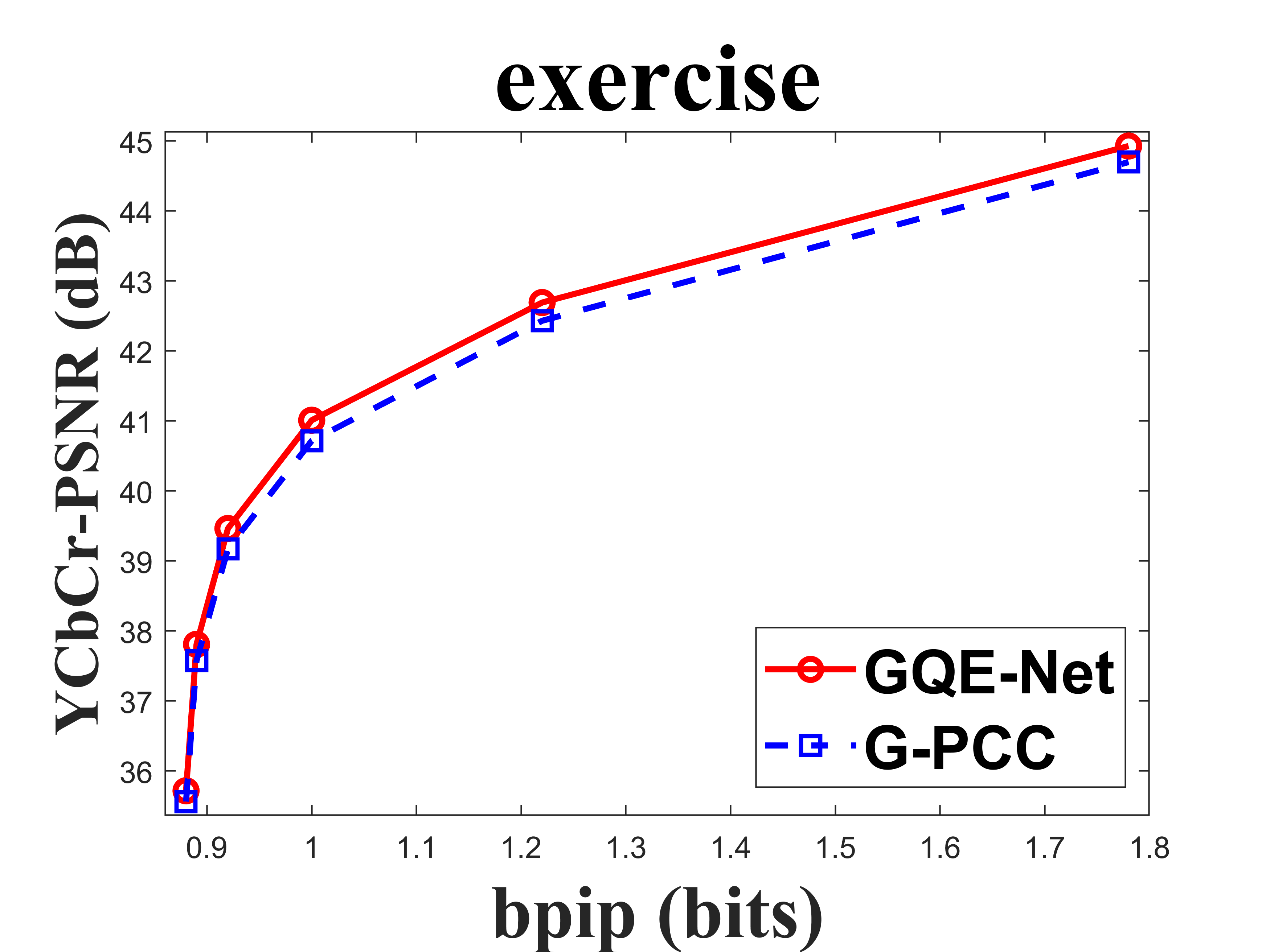}%
\label{}}
\subfloat{\includegraphics[width=3cm]{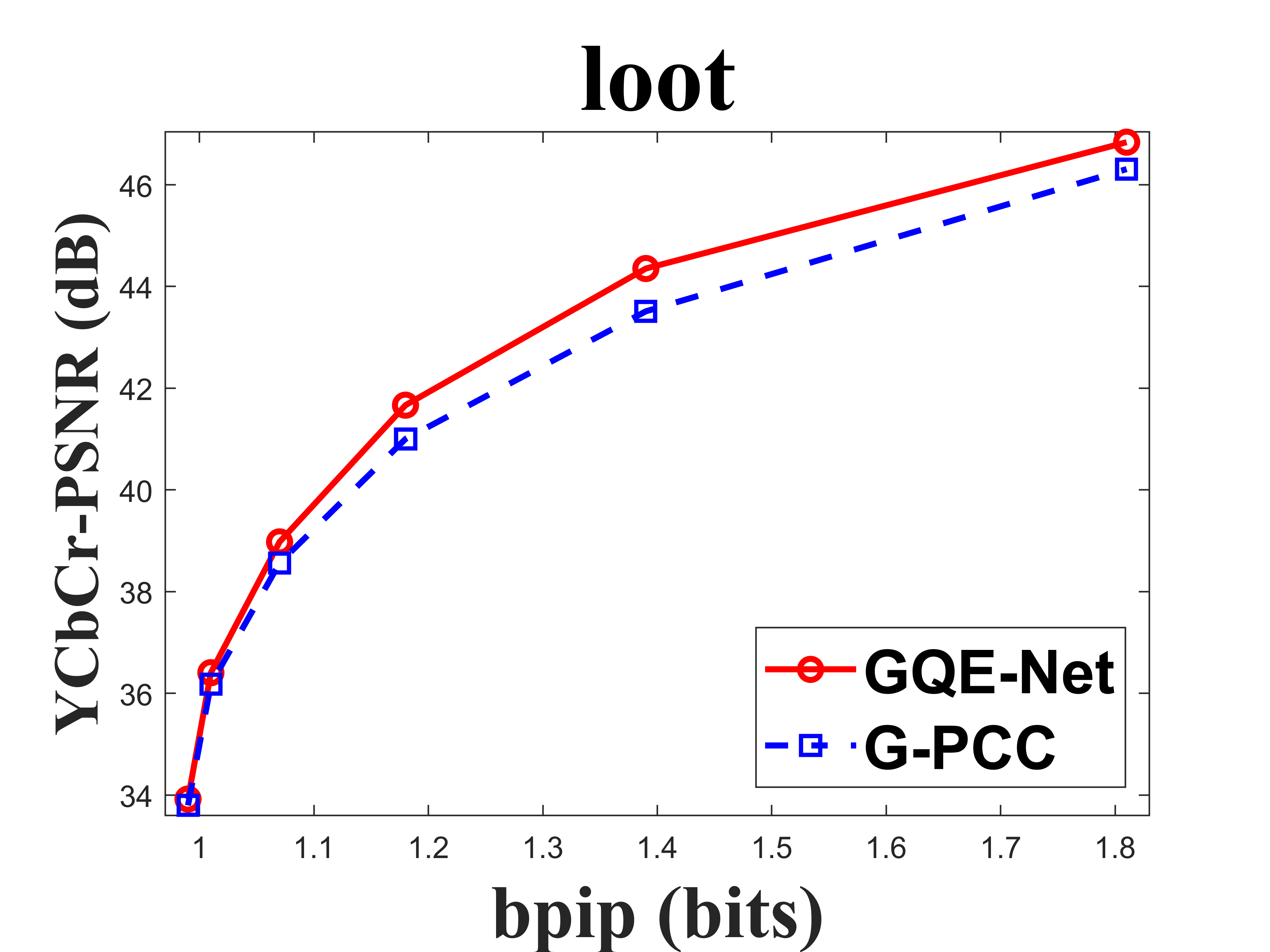}%
\label{}}
\subfloat{\includegraphics[width=3cm]{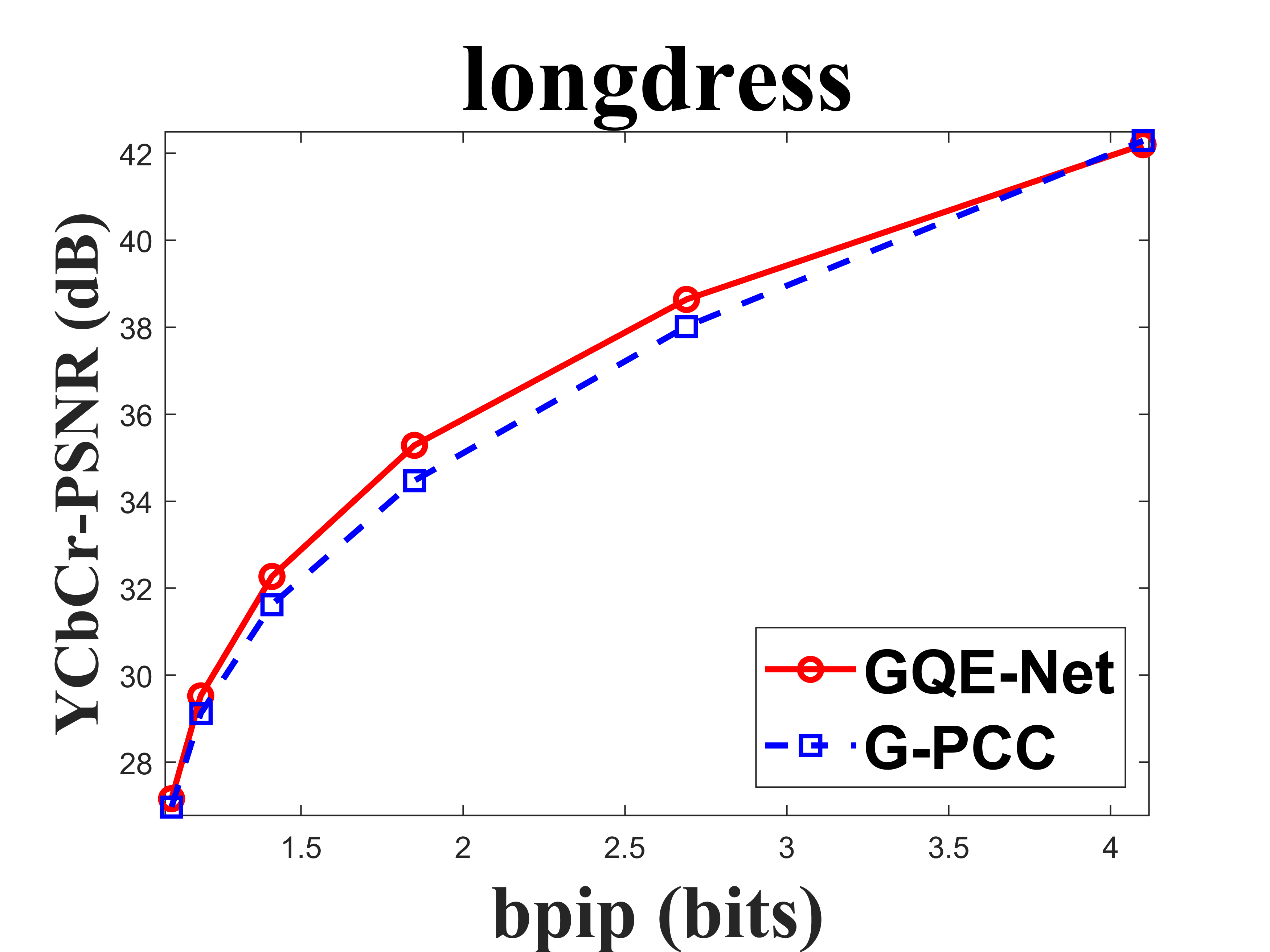}%
\label{}}
\subfloat{\includegraphics[width=3cm]{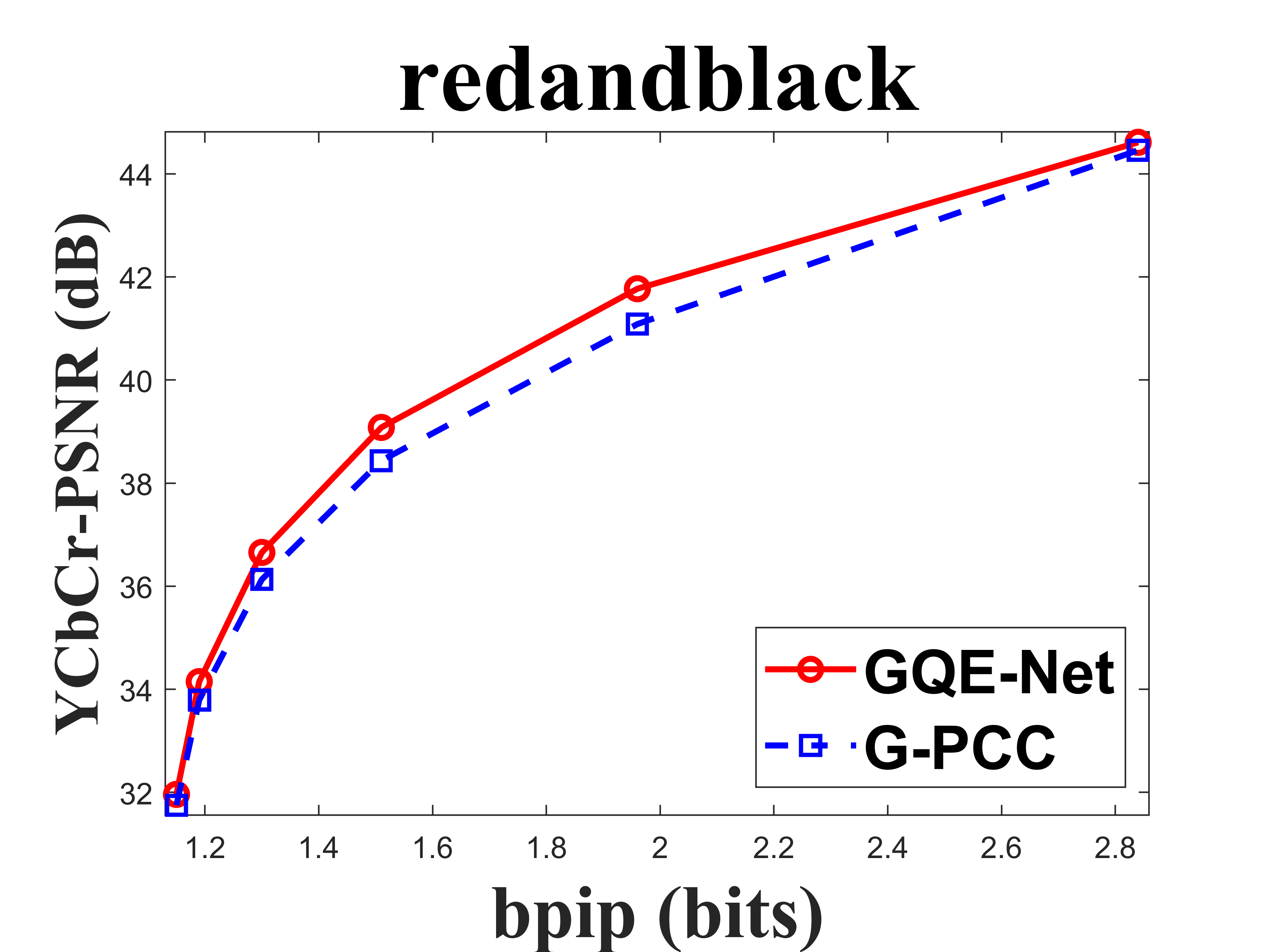}%
\label{}}

\vspace{-0.8em}
\quad
\subfloat{\includegraphics[width=3cm]{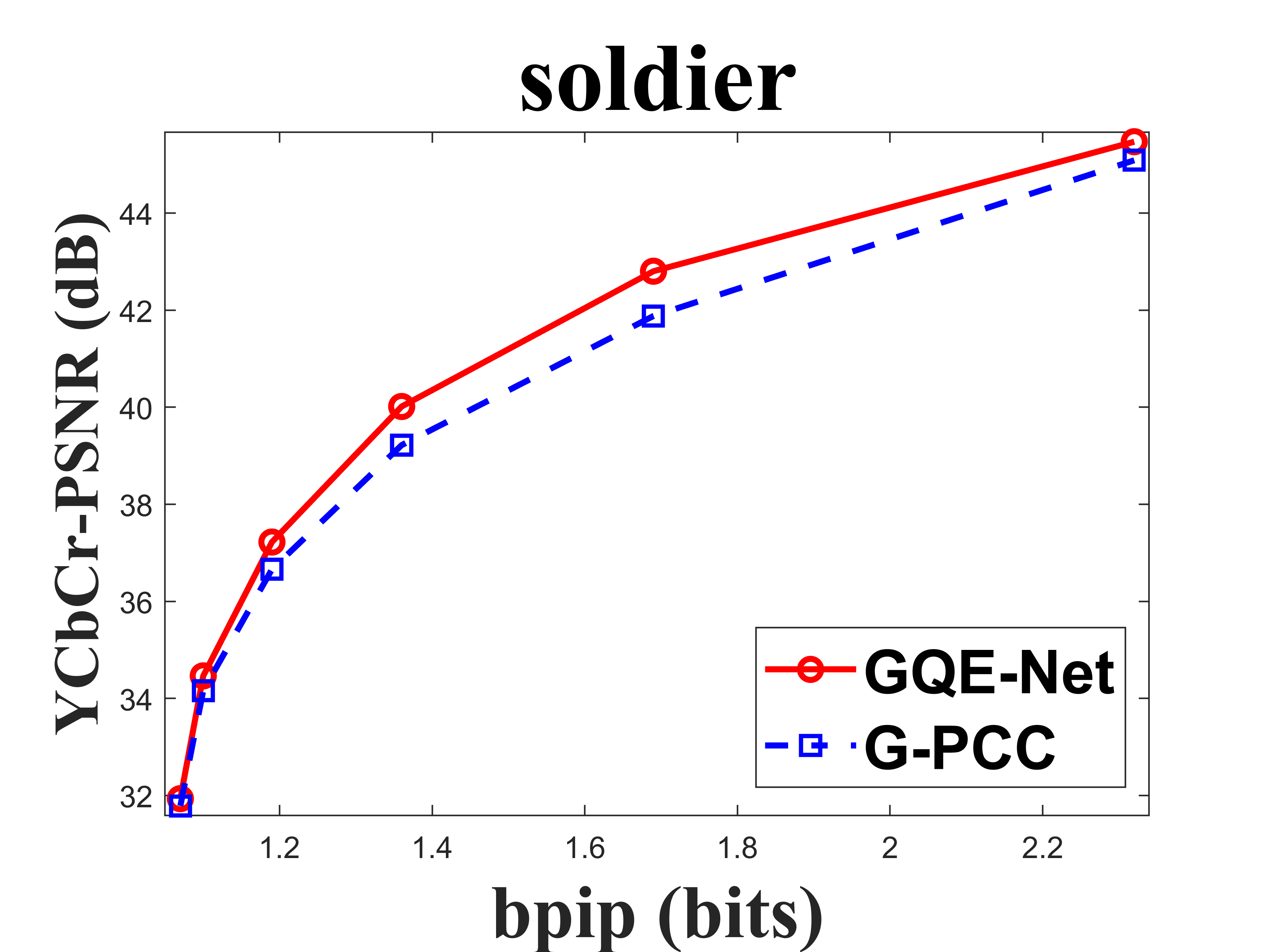}%
\label{}}
\subfloat{\includegraphics[width=3cm]{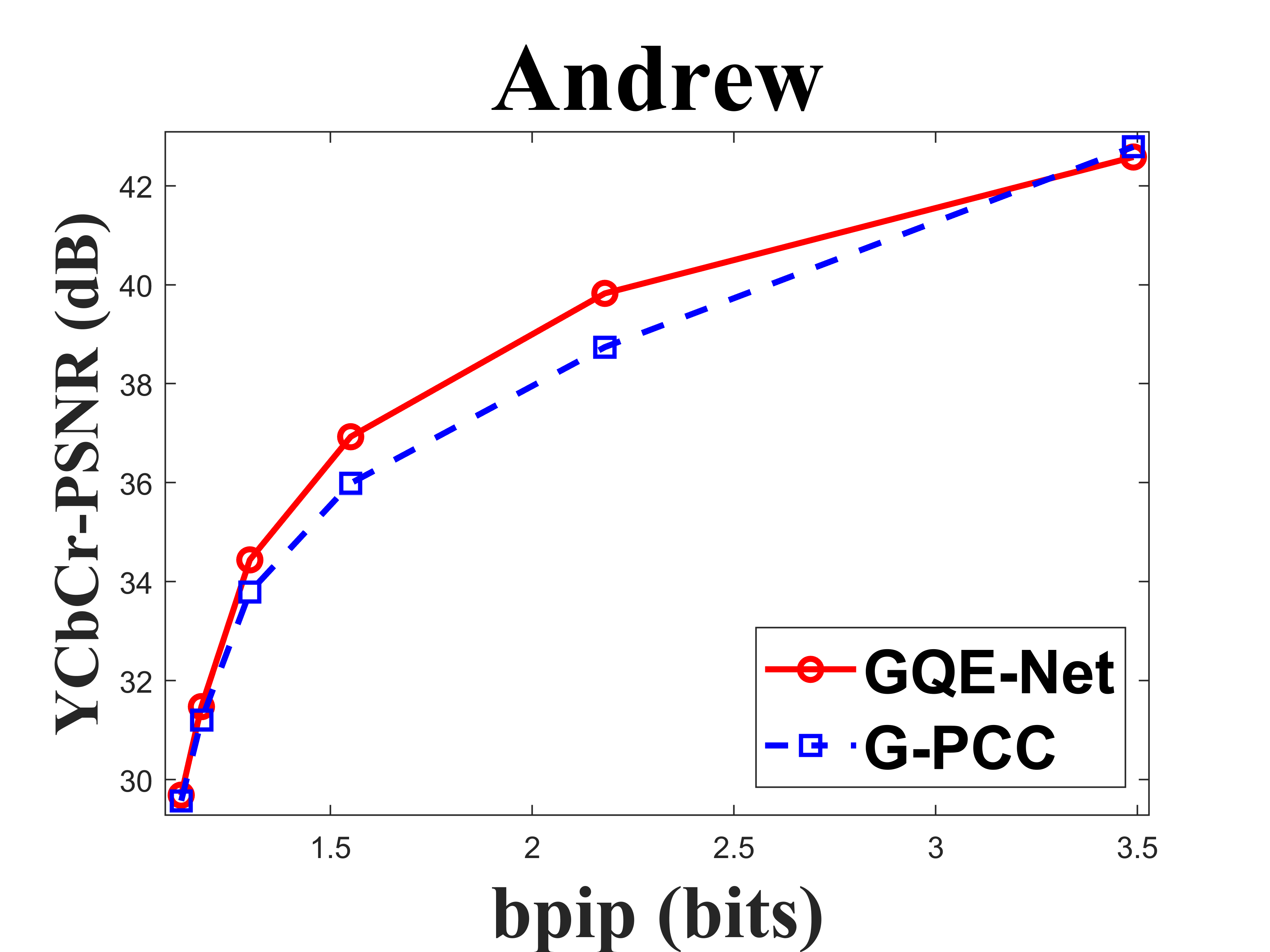}%
\label{}}
\subfloat{\includegraphics[width=3cm]{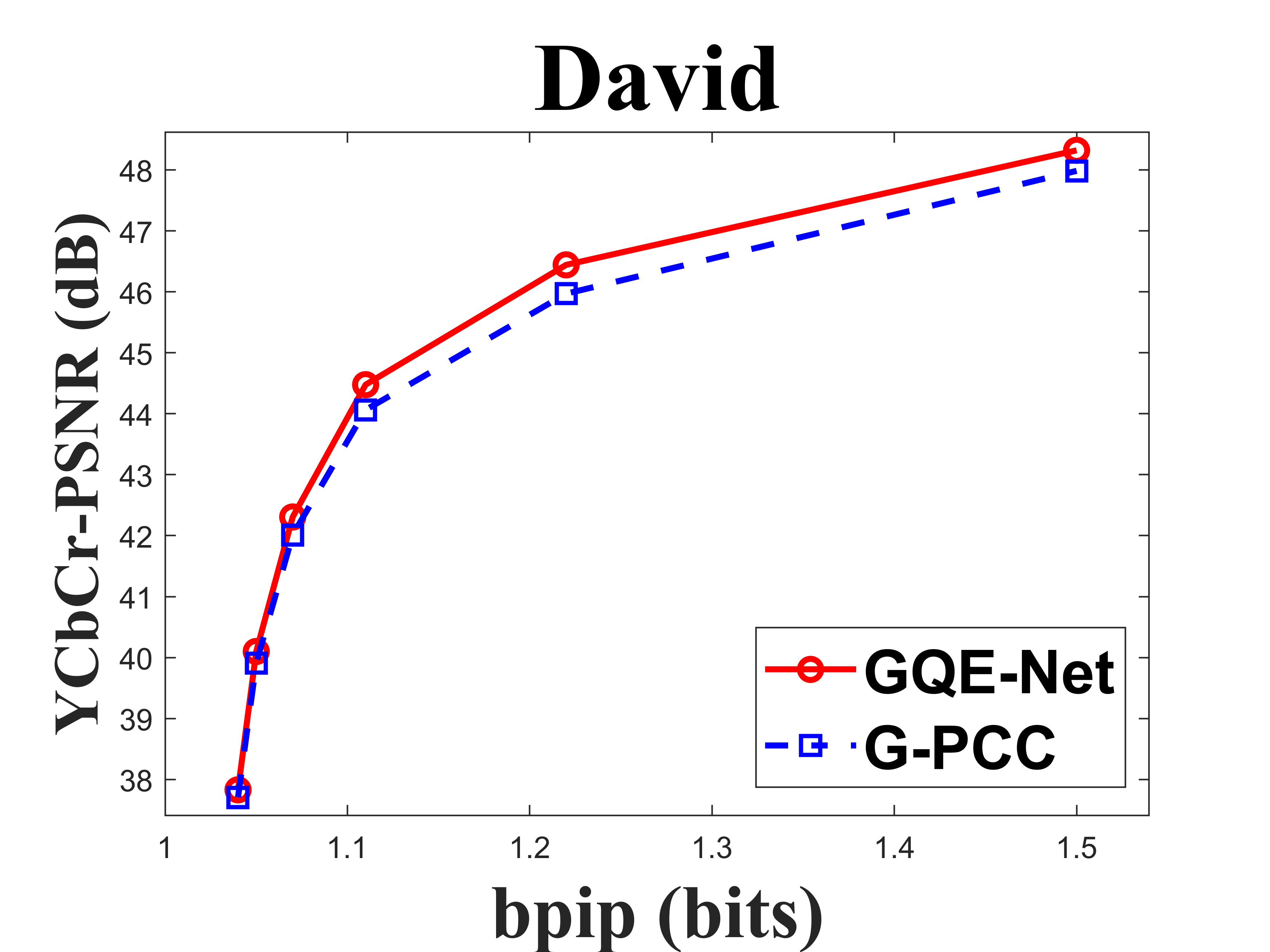}%
\label{}}
\subfloat{\includegraphics[width=3cm]{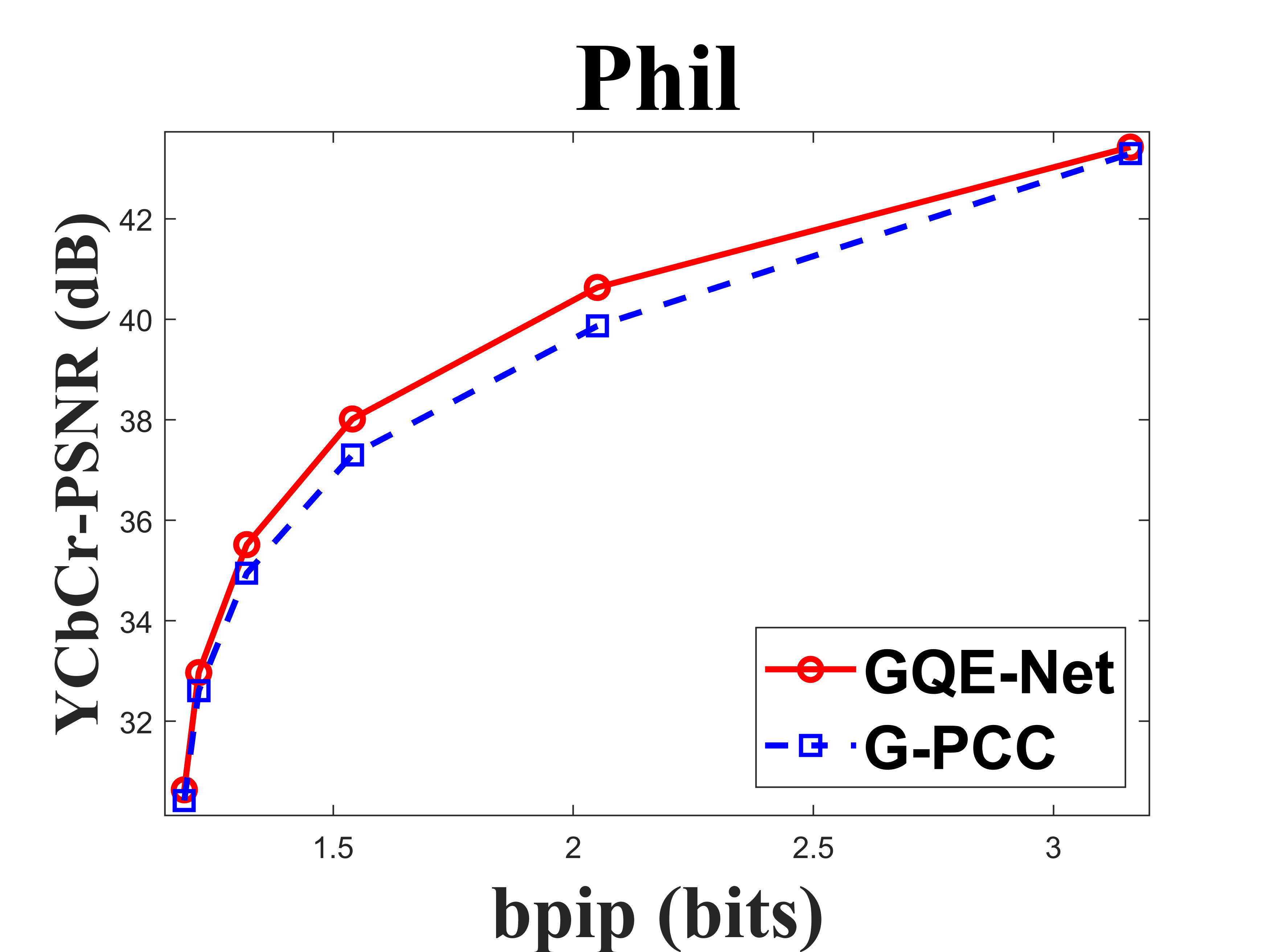}%
\label{}}
\subfloat{\includegraphics[width=3cm]{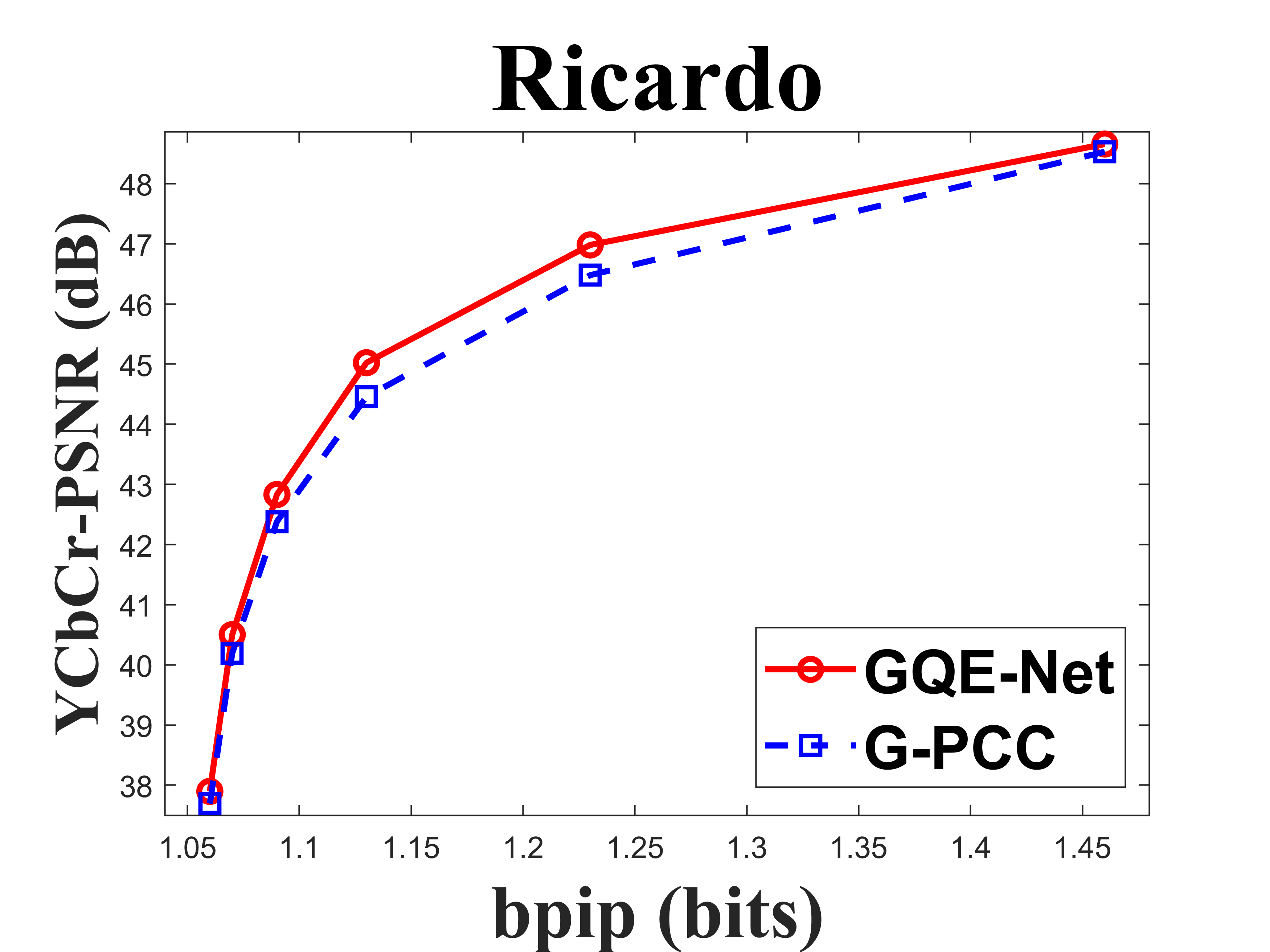}%
\label{}}
\subfloat{\includegraphics[width=3cm]{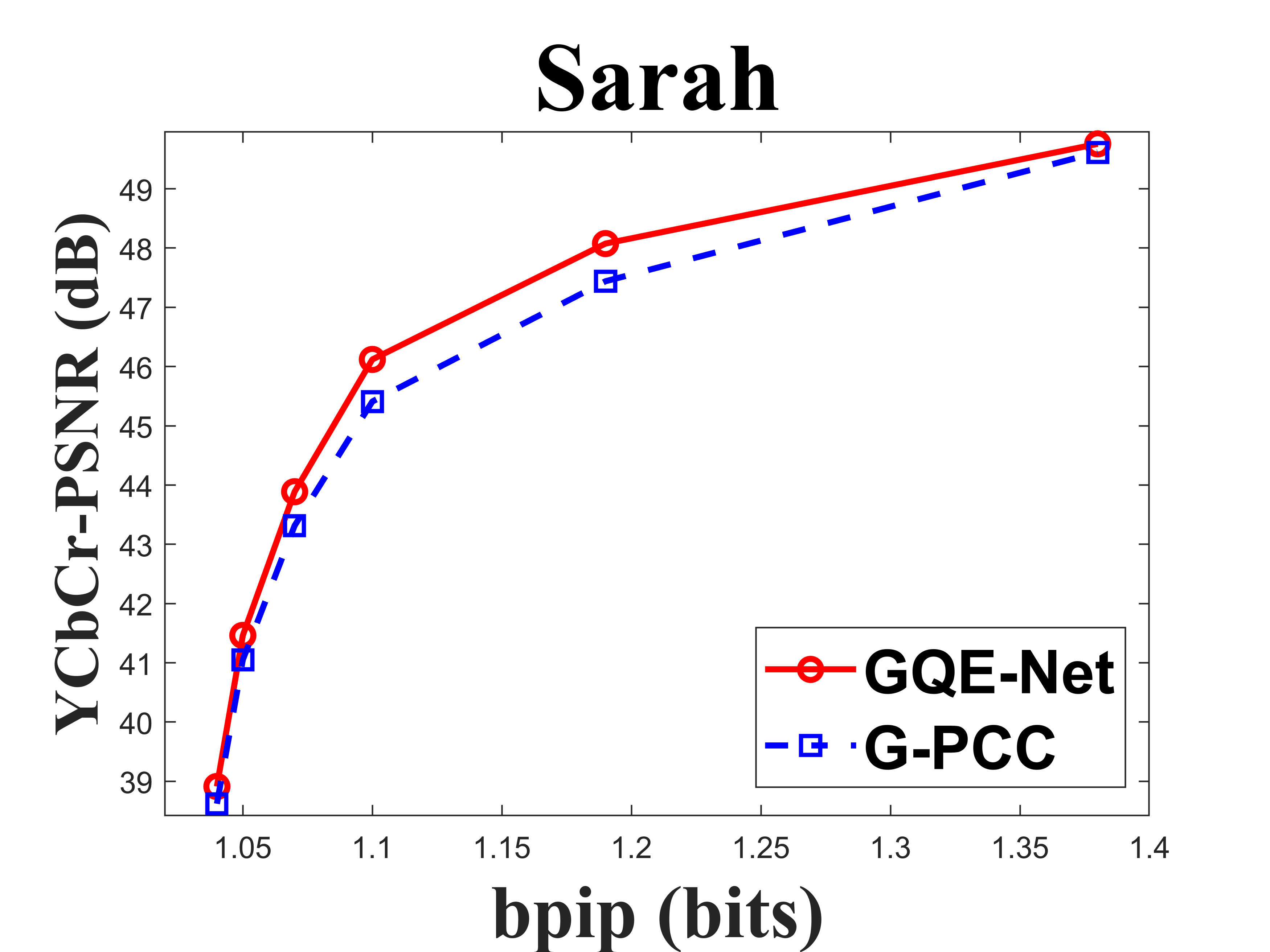}%
\label{}}
\caption{Comparison of rate-PSNR curves.}
\label{lift_main}
\end{figure*}

\begin{figure*}[!ht]
\setlength{\belowcaptionskip}{-1.8em}
\captionsetup{font={small}}
\centering
\centerline{\includegraphics[width=17.6cm]{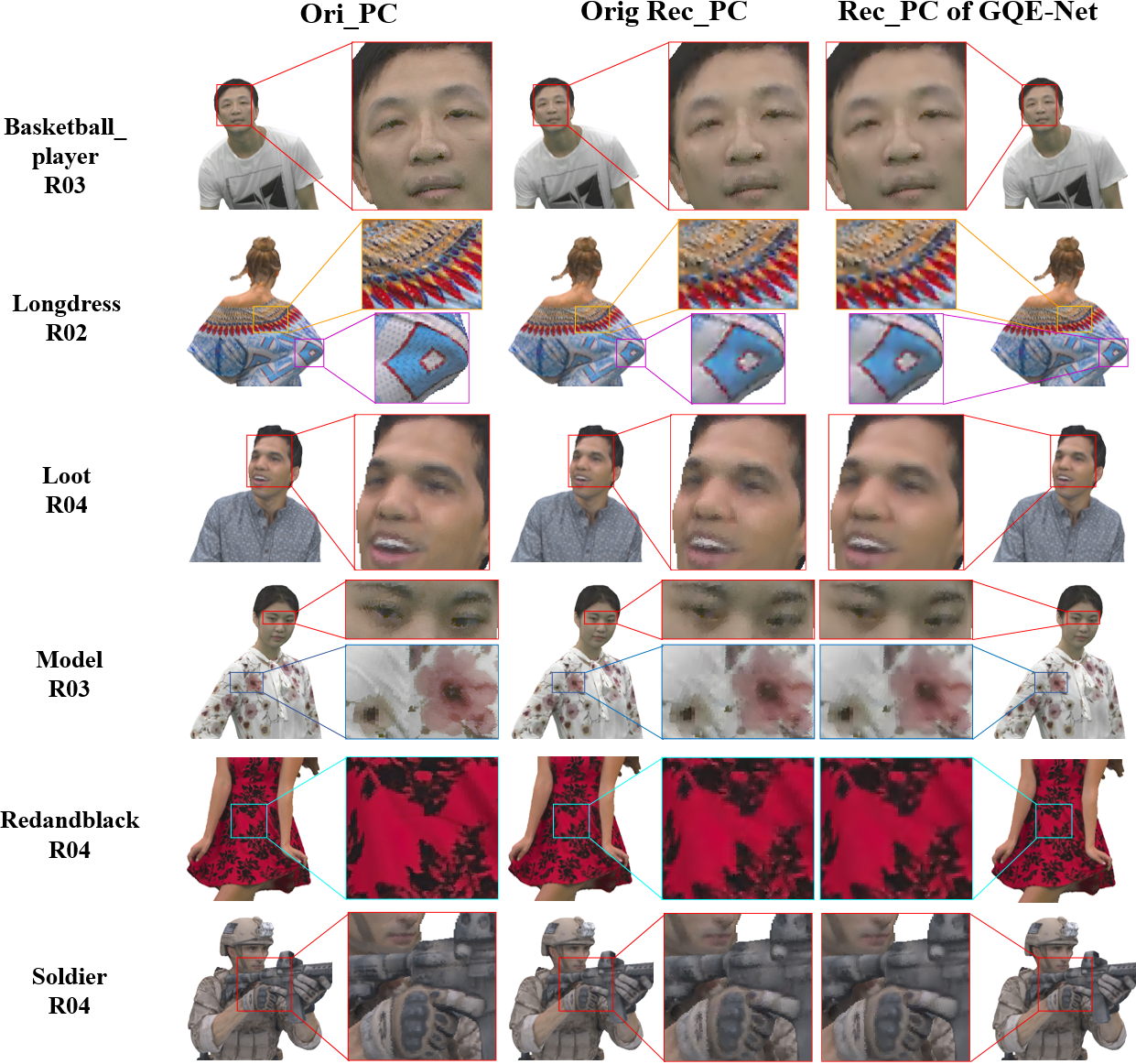}}
\vspace{-0.3em}
\caption{Visual comparison. ``Ori\_PC'' denotes the original point cloud, ``Orig Rec\_PC'' denotes the point clouds reconstructed by G-PCC TMC13v14.0. ``Rec\_PC of GQE-Net'' denotes the point clouds processed by GQE-Net.} \label{subjective}
\end{figure*}

\subsection{Subjective quality evaluation}
Fig. 9 compares the original point clouds (ground truth), the point clouds reconstructed without any enhancement, and the point clouds processed by GQE-Net. The texture of the point clouds processed by GQE-Net is clearer and the color transitions are smoother, resulting in a better overall visual experience. 

\begin{figure*}[!ht]
\centering
\captionsetup{font=small}
\subfloat{\includegraphics[width=3cm]{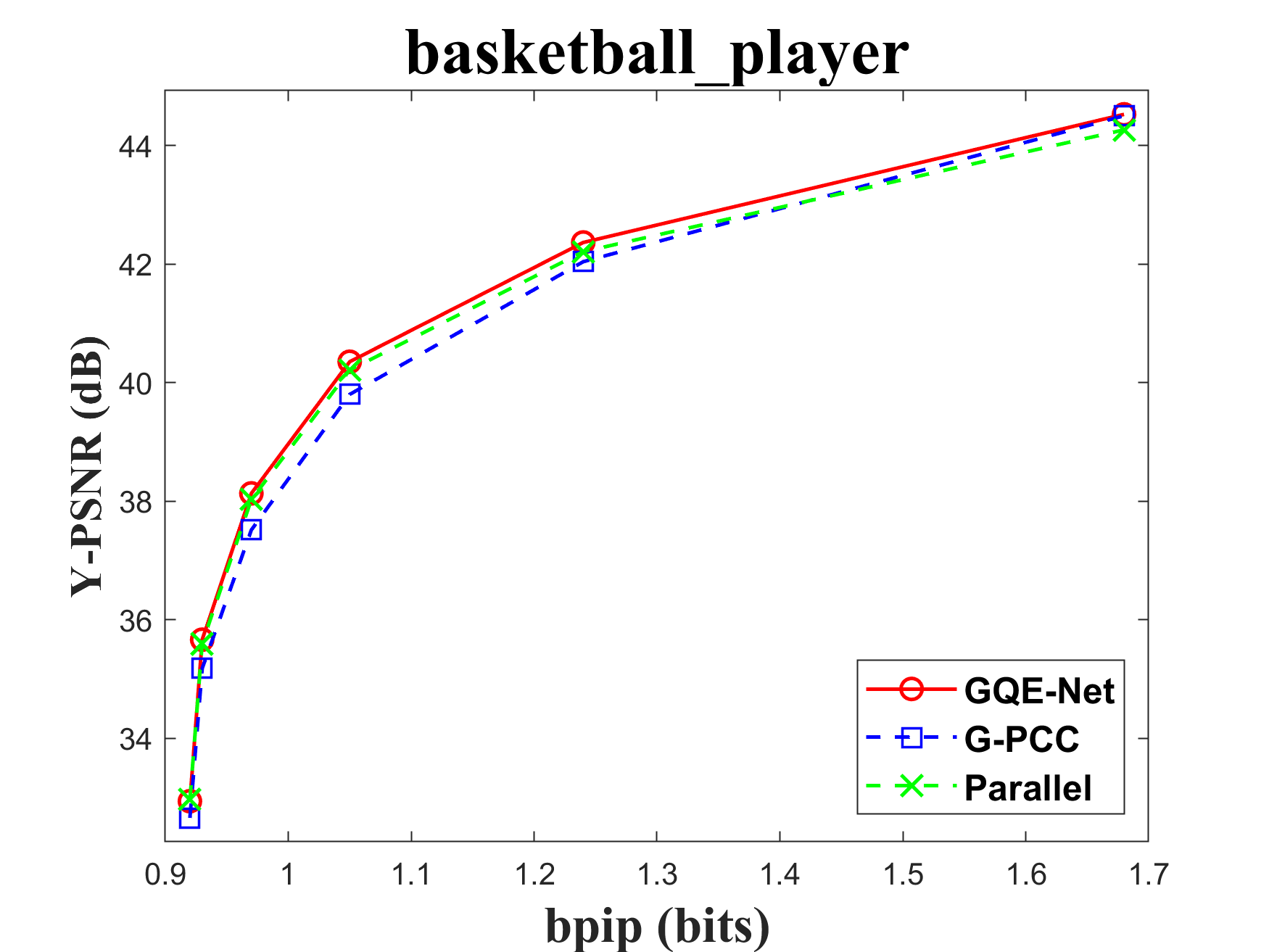}%
\label{}}
\subfloat{\includegraphics[width=3cm]{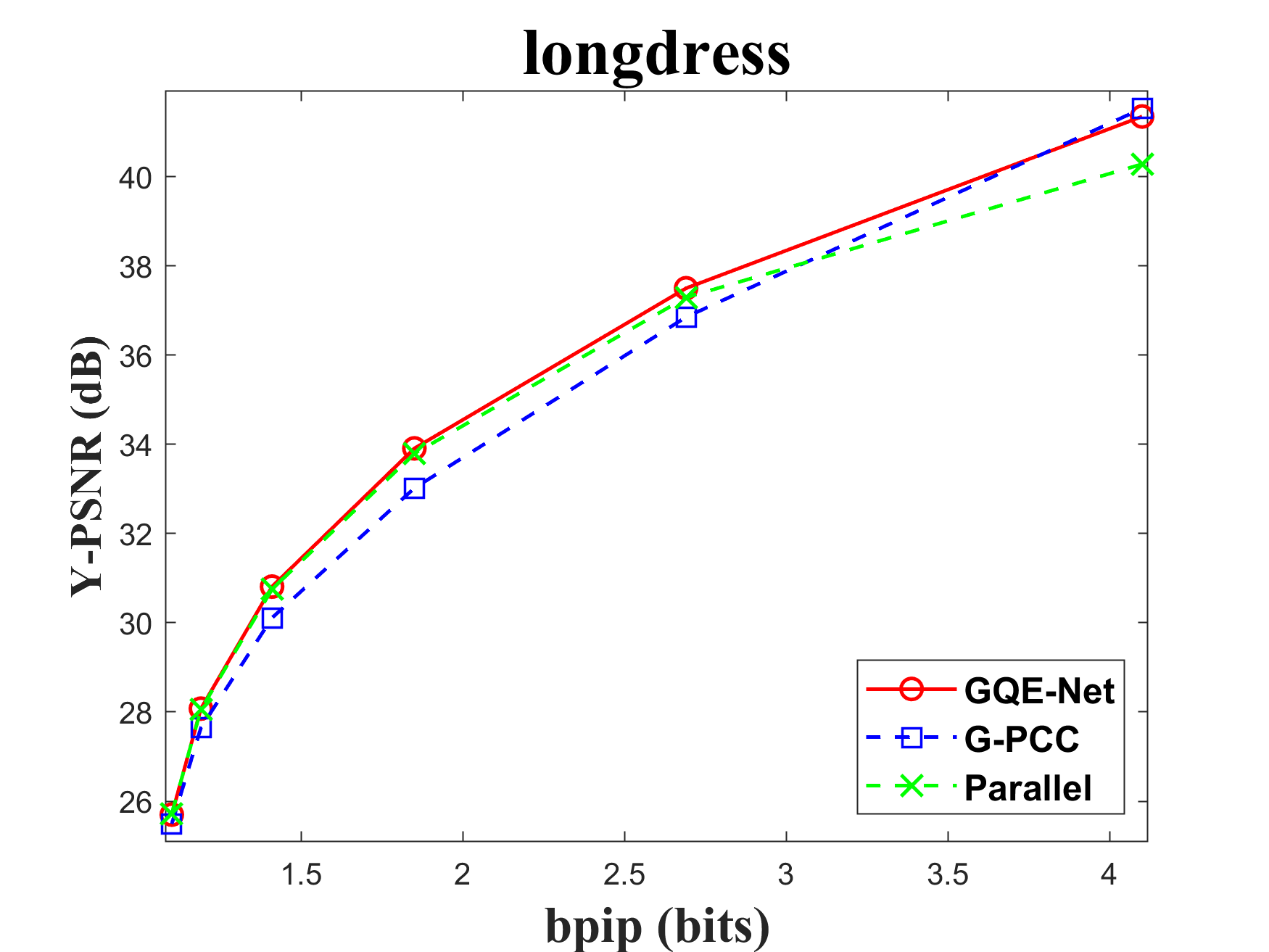}%
\label{}}
\subfloat{\includegraphics[width=3cm]{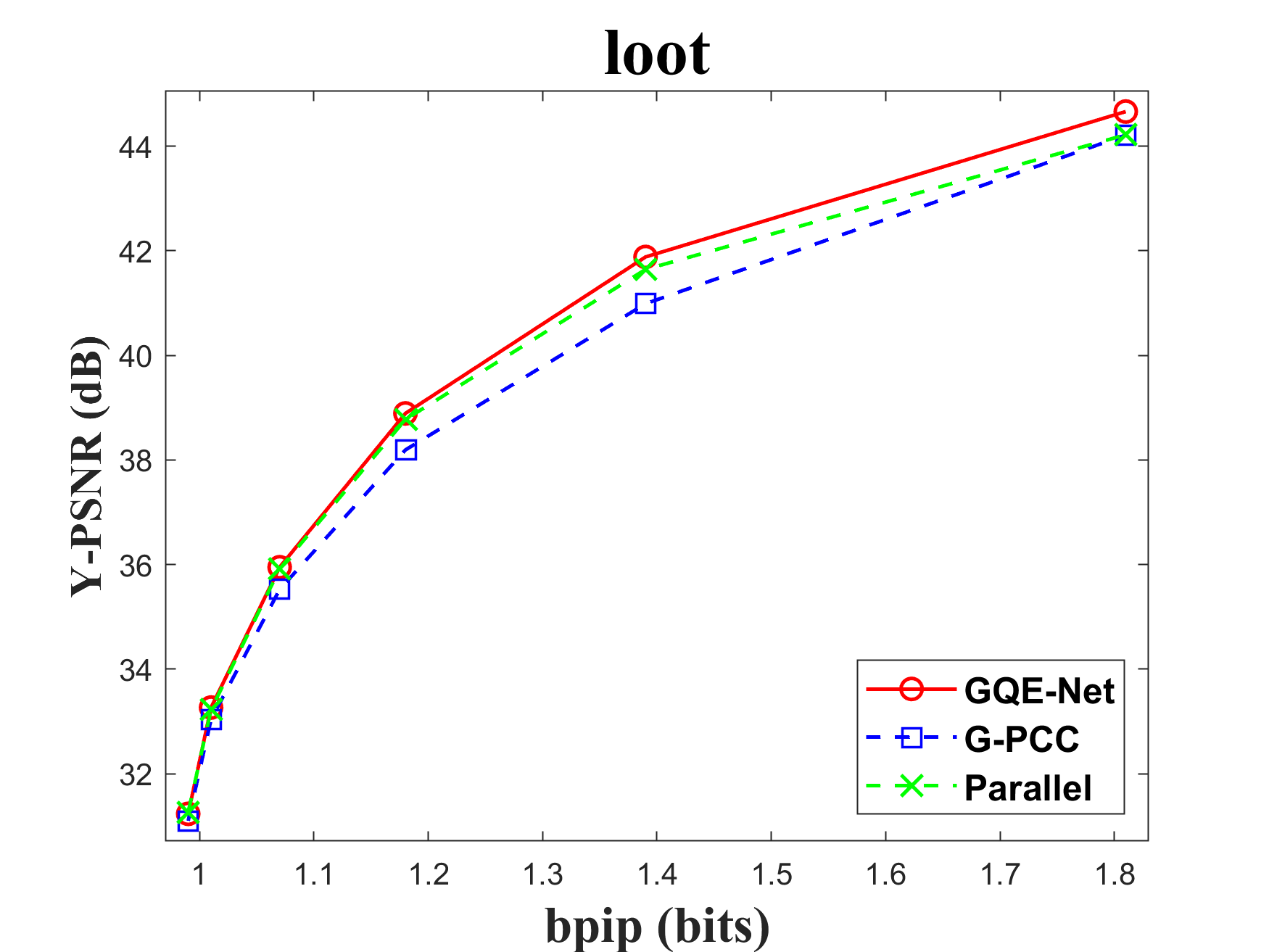}%
\label{}}
\subfloat{\includegraphics[width=3cm]{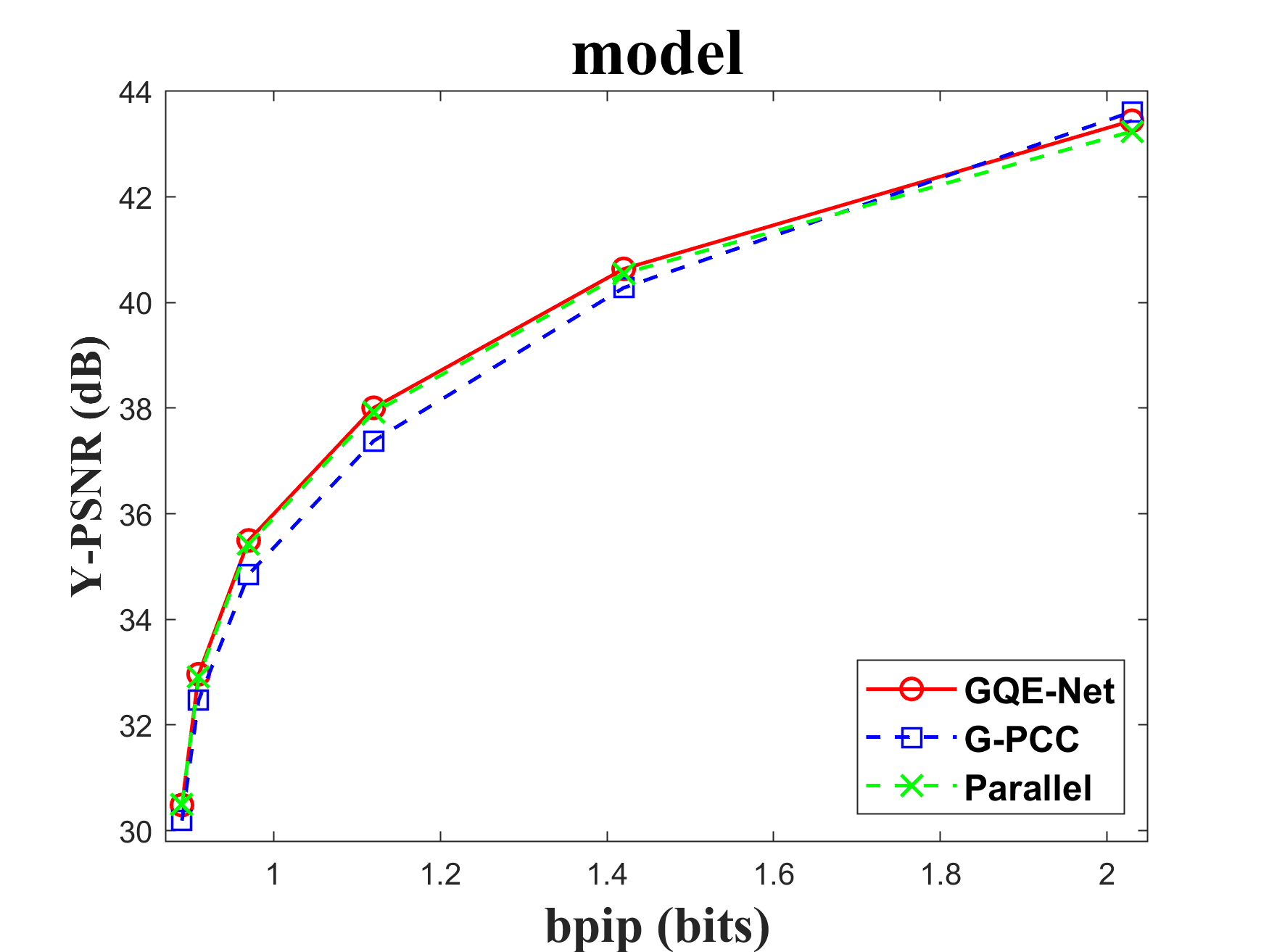}%
\label{}}
\subfloat{\includegraphics[width=3cm]{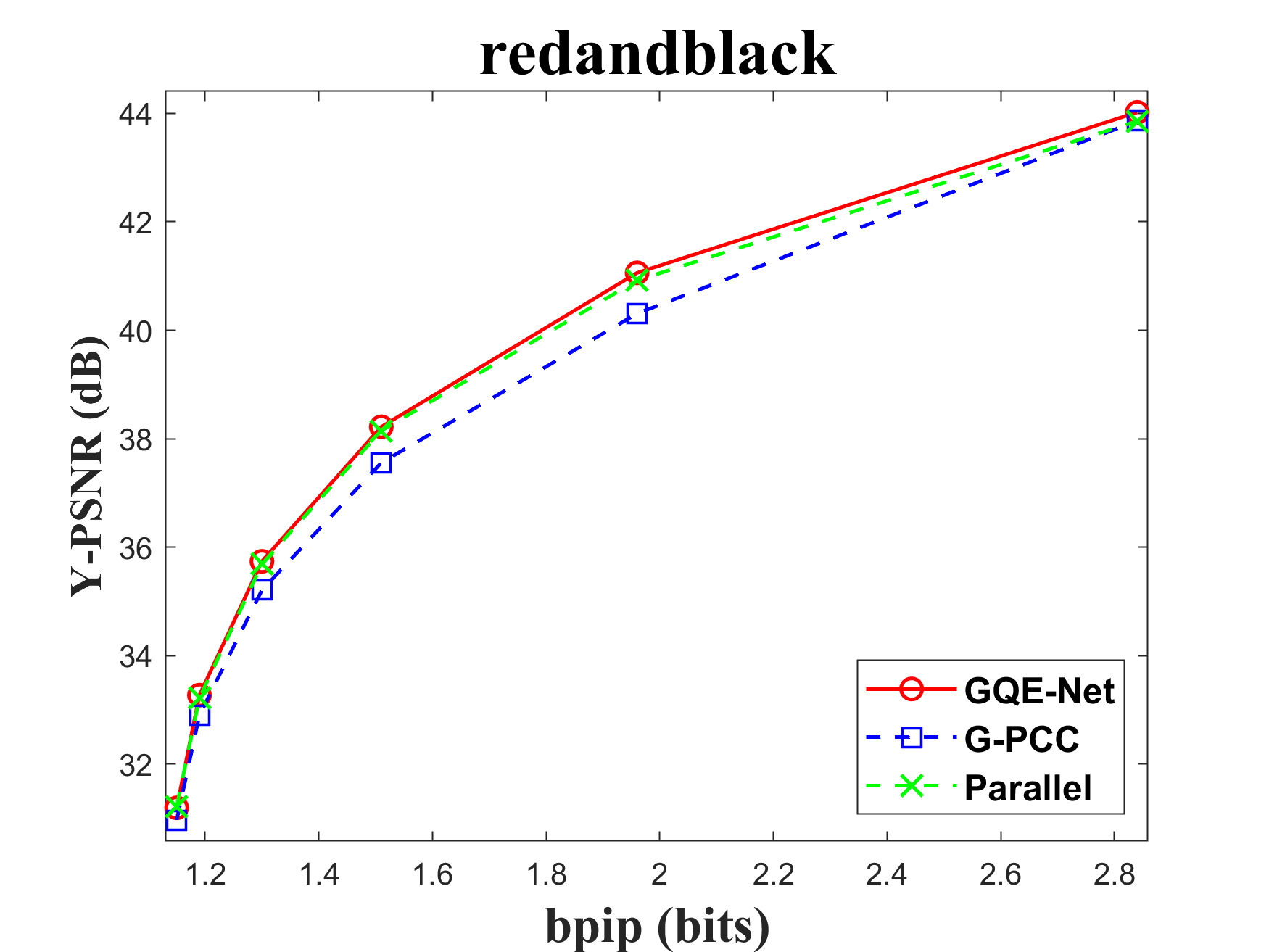}%
\label{}}
\subfloat{\includegraphics[width=3cm]{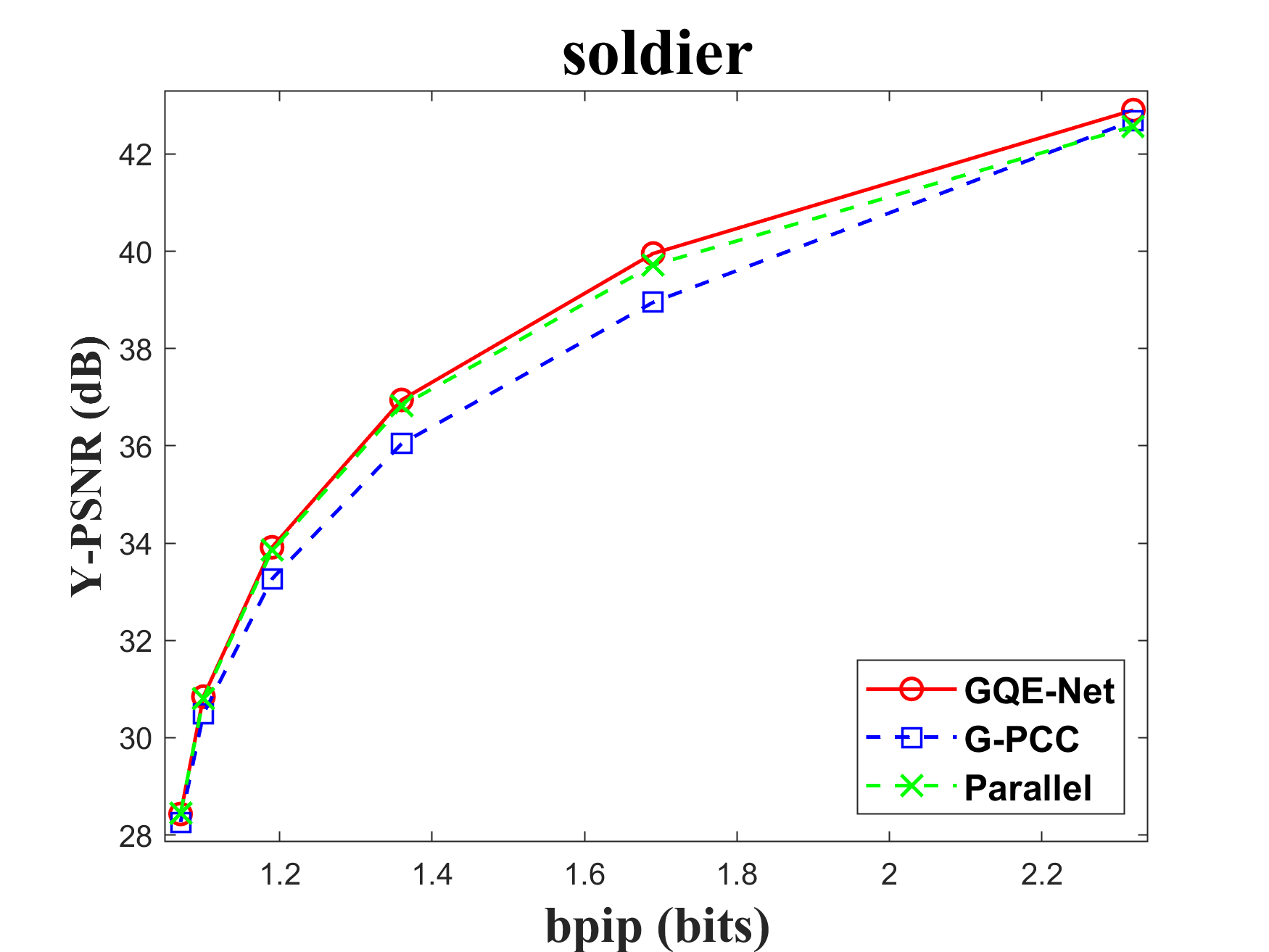}%
\label{}}
\caption{Efficiency of the PSGA module.}
\label{PSGA}
\end{figure*}

\begin{figure*}[!ht]
\centering
\captionsetup{font=small}
\subfloat{\includegraphics[width=3cm]{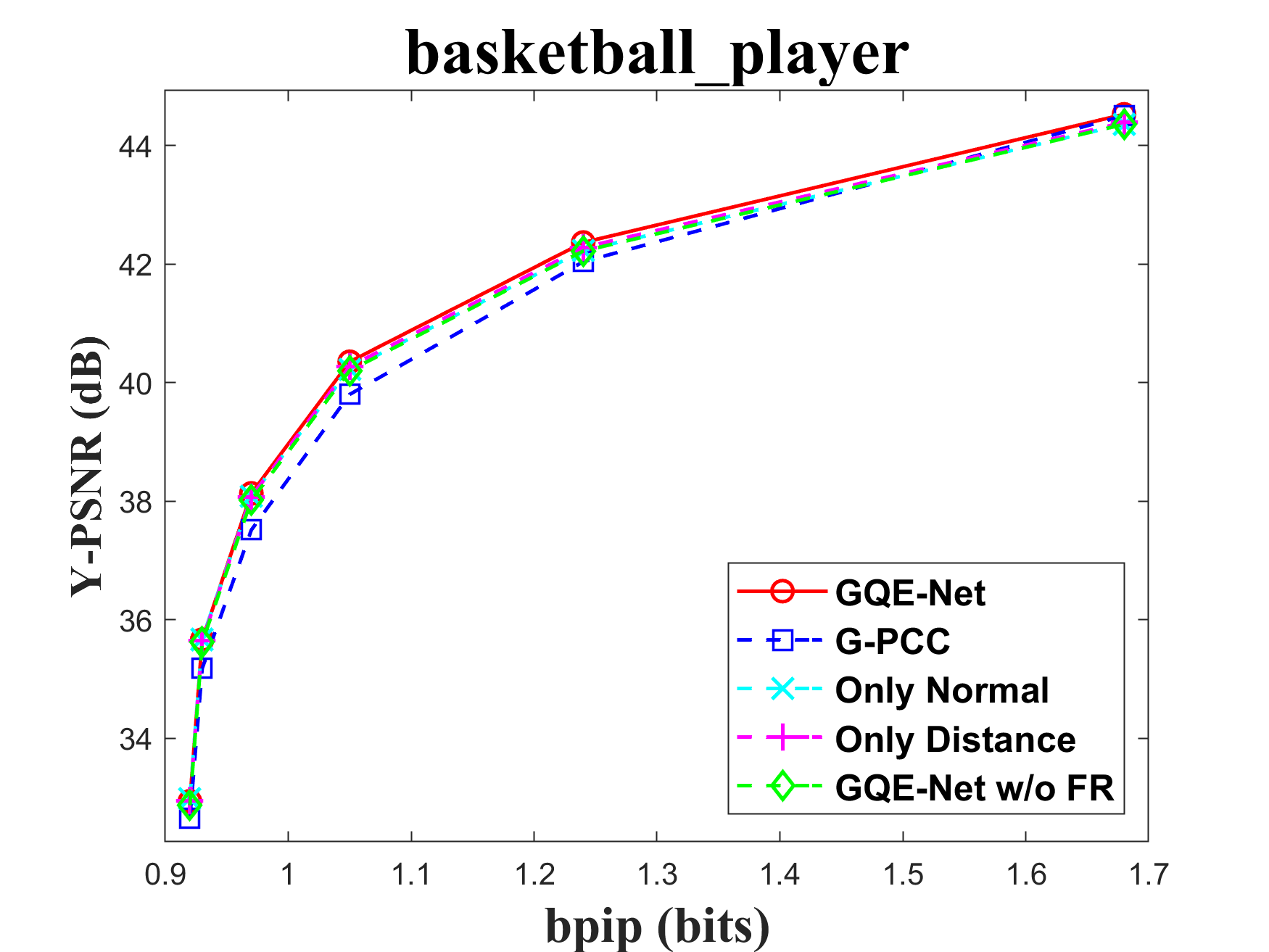}%
\label{}}
\subfloat{\includegraphics[width=3cm]{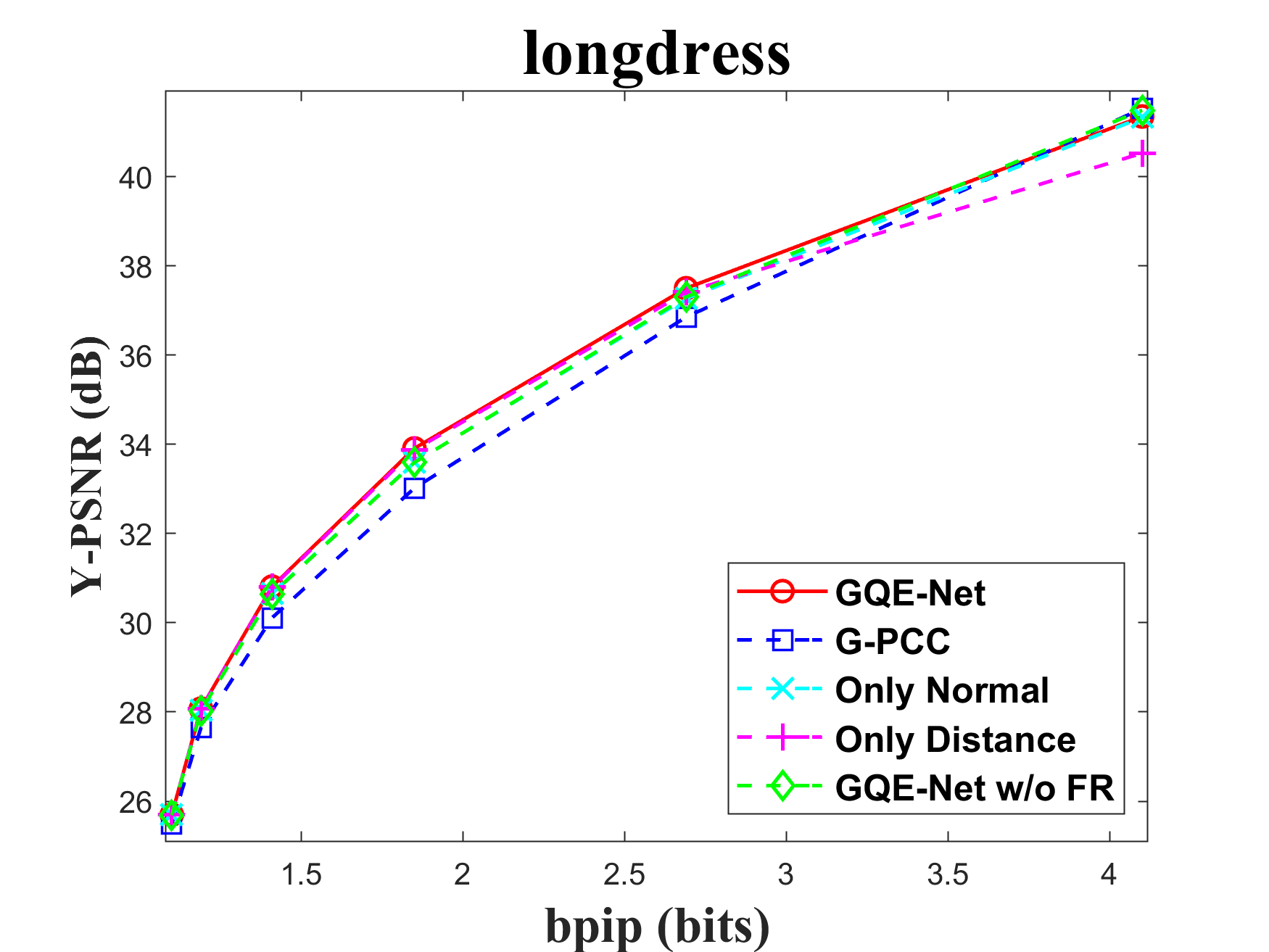}%
\label{}}
\subfloat{\includegraphics[width=3cm]{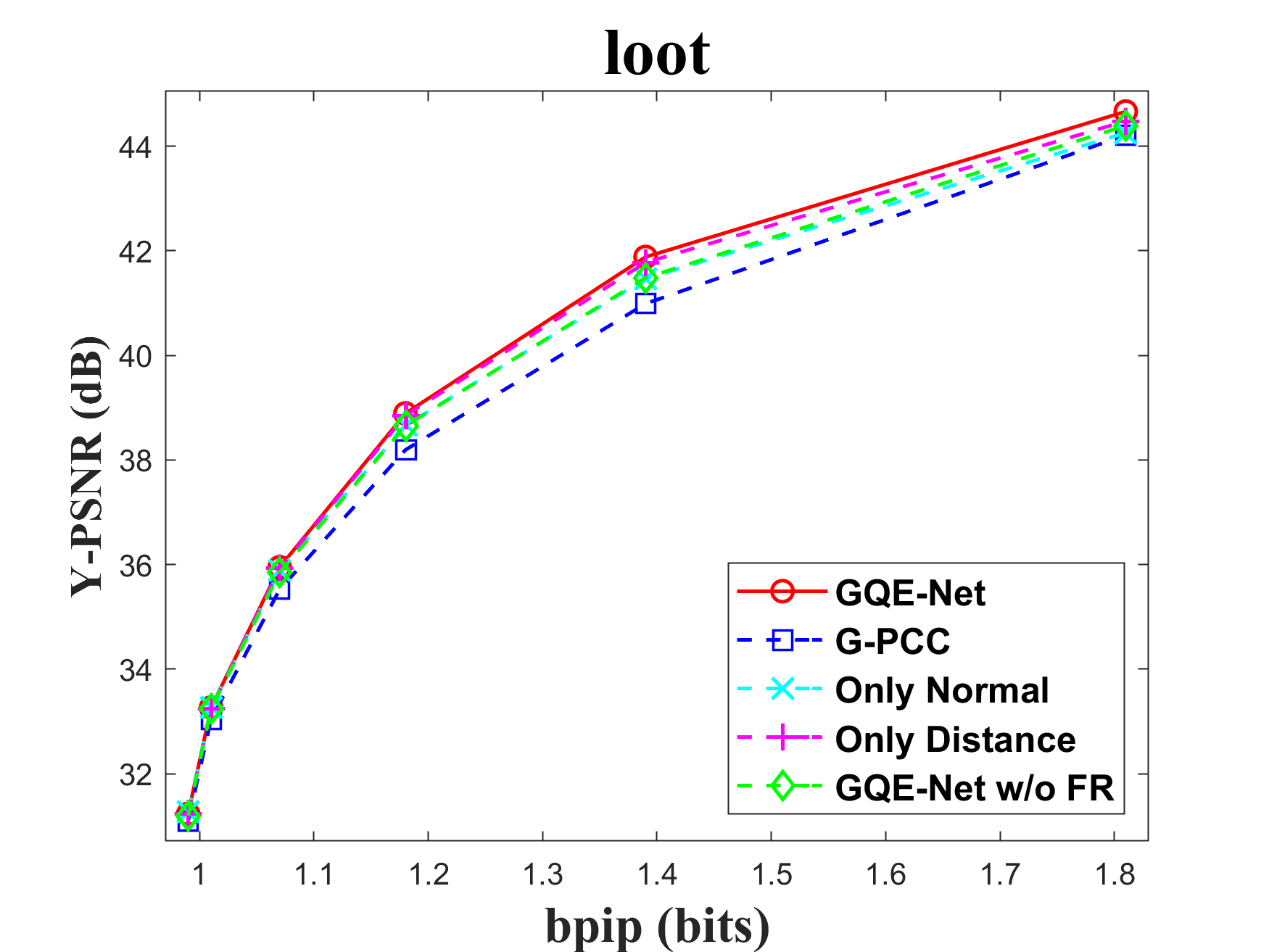}%
\label{}}
\subfloat{\includegraphics[width=3cm]{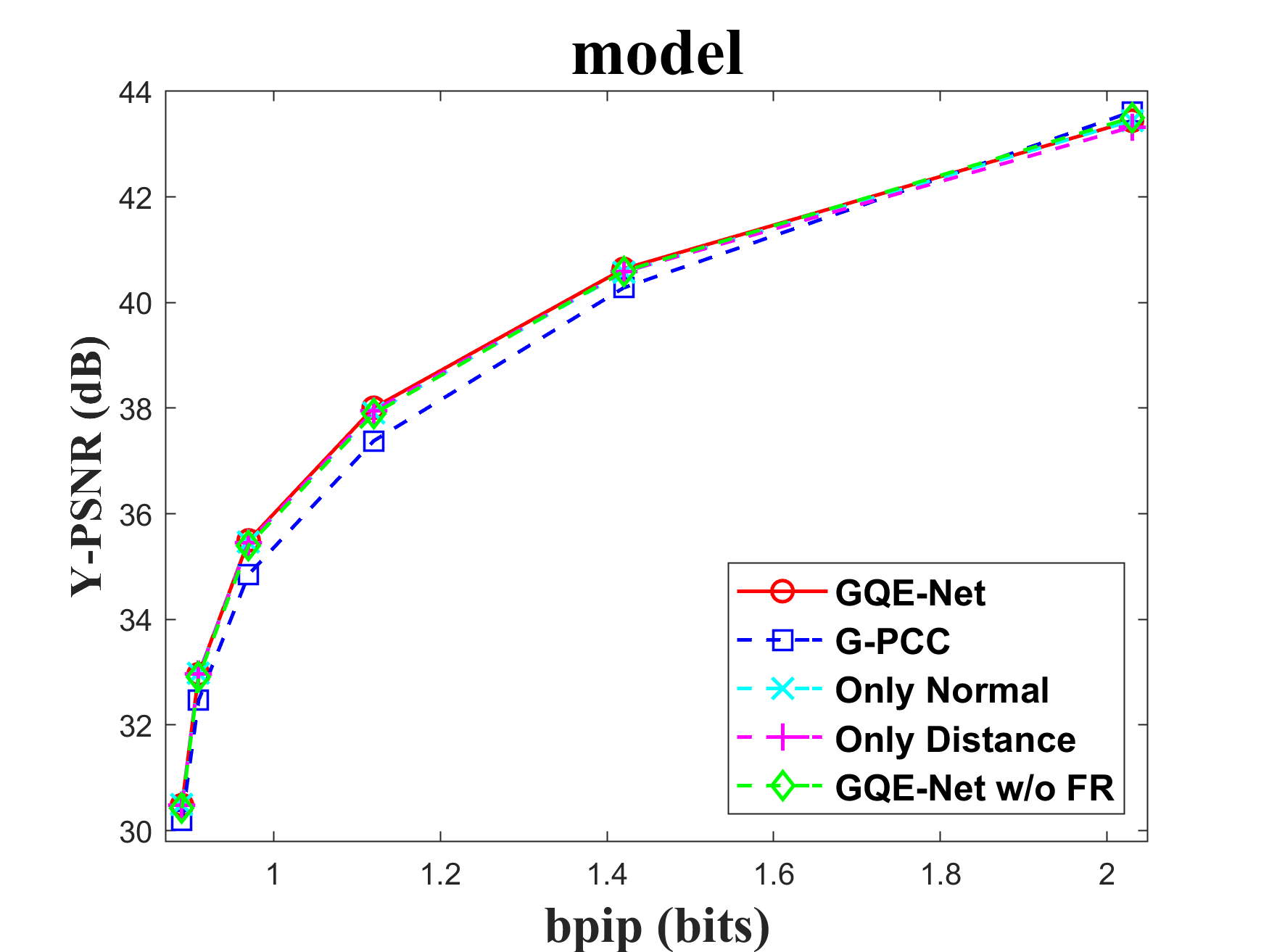}%
\label{}}
\subfloat{\includegraphics[width=3cm]{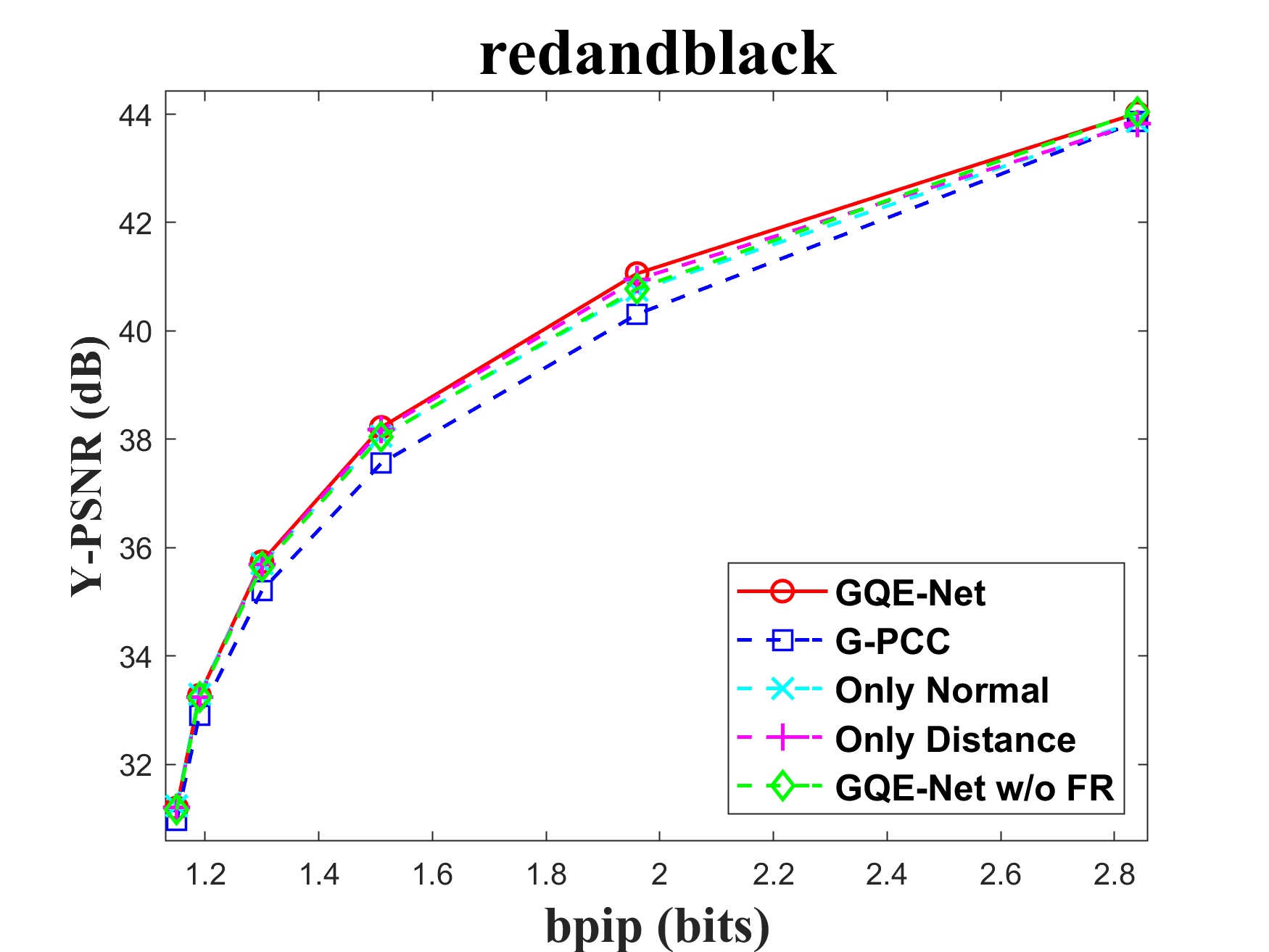}%
\label{}}
\subfloat{\includegraphics[width=3cm]{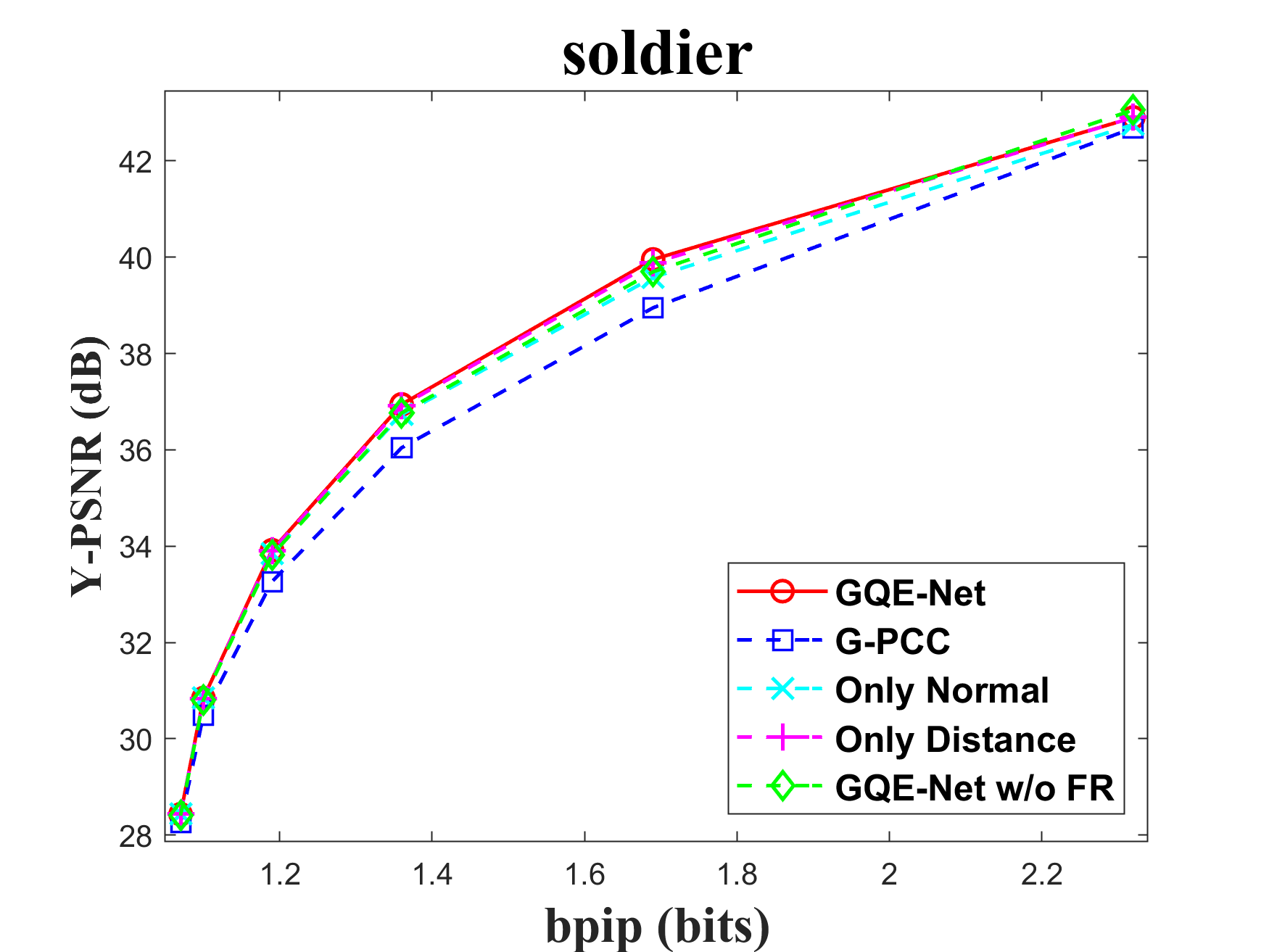}%
\label{}}
\caption{Efficiency of the FR module.}
\label{FR}
\end{figure*}

\subsection{Results for RAHT}
To further demonstrate the effectiveness of GQE-Net, we applied the trained GQE-Net models directly to the point clouds in Cat\_A encoded under the G-PCC lossless geometry, lossy attribute and \textbf{RAHT} configuration. The BD-PSNR and BD-rate results are presented in Table III. More results are available in the supplemental materials. 

It can be seen that competitive results can be achieved at all bit rates. Compared with G-PCC TMC13v14.0, GQE-Net achieved $0.28$ dB, $0.26$ dB, and $0.36$ dB BD-PSNR, corresponding to -$9.5\%$, -$11.0\%$, and -$13.9\%$ BD-rates. Moreover, the PSNR gain was up to $0.84$ dB (Cr component of redandblack at R04). We believe that the performance could be improved if GQE-Net is fine-tuned or retrained based on point clouds compressed with RAHT. 

\begin{table}[!htbp]
\scriptsize
\renewcommand{\arraystretch}{1}
\captionsetup{font={small}}
\caption{Rate-distortion comparison with G-PCC TMC13v14.0 when RAHT is used.}
\setlength{\belowcaptionskip}{-1.0em}
\centering
{\begin{tabular}{cccccccc}
\toprule
\centering
\multirow{2}{*}{\textbf{Sequence}}& \multicolumn{4}{c}{\textbf{BD-PSNR [dB]}}  & \multicolumn{3}{c}{\textbf{BD-rate [\%]}}    \\
  &\textbf{Luma}& \textbf{Cb}\ & \textbf{Cr}& \textbf{YCbCr} &\textbf{Luma}& \textbf{Cb} &  \textbf{Cr} \\
\midrule
basketball & 0.298 & 0.180 & 0.402 & 0.297 & -9.8  & -11.1 &  -18.0  \\
dancer & 0.399 & 0.181 & 0.415 & 0.374  & -10.6 & -12.7 & -17.9 \\
exercise & 0.255 & 0.141 & 0.281 & 0.244  & -9.5 &  -13.3 & \textbf{-22.0} \\
longdress & 0.180 & 0.320 & 0.354 & 0.219  & -6.7 &  -8.8 &  -10.2 \\
loot & 0.327 & 0.383 & \textbf{0.477} & 0.353  & -9.2  &  -13.4 & -15.8 \\
model & 0.297 & 0.194 & 0.418 & 0.299 &  -10.0 &  -9.8 & -15.2 \\
queen & 0.242 & \textbf{0.417} & 0.439 & 0.289 &  -6.9 & -12.2 &  -12.8 \\
redandblack & 0.320 & 0.228 & 0.344 & 0.312 & -8.7 & -7.9 & -9.8 \\
soldier & 0.343 & 0.353 & 0.423 & 0.354 & -9.6 & -13.3 & -15.9 \\
Andrew & 0.209 & 0.160 & 0.119 & 0.192 & -10.8 & -9.0 & -6.9 \\
David & 0.315 & 0.225 & 0.316 & 0.304 & -10.6 & -11.6 & -15.3 \\
Phil & 0.245 & 0.151 & 0.192 & 0.227 & -9.3 & -7.3 & -7.9 \\
Ricardo & 0.306 & 0.300 & 0.351 & 0.314 & -10.0 & \textbf{-13.6} & -14.2 \\
Sarah & \textbf{0.405} & 0.295 & 0.378 & \textbf{0.388} & \textbf{-11.3} &  -9.9 & -12.6 \\
\midrule

\textbf{Average} & \textbf{0.284} & \textbf{0.255} & \textbf{0.355} &\textbf{0.290} &\textbf{-9.5} & \textbf{-11.0} & \textbf{-13.9} \\ 
\bottomrule
\end{tabular}}
\end{table}

\begin{table}[!htbp]
\scriptsize
\renewcommand{\arraystretch}{1}
\captionsetup{font={small}}
\caption{Rate-distortion comparison with V-PCC TMC2v18.0.}
\setlength{\belowcaptionskip}{-1.0em}
\centering
{\begin{tabular}{cccccccc}
\toprule
\centering
\multirow{2}{*}{\textbf{Sequence}}& \multicolumn{4}{c}{\textbf{BD-PSNR [dB]}}  & \multicolumn{3}{c}{\textbf{BD-rate [\%]}}    \\
  &\textbf{Luma}& \textbf{Cb}\ & \textbf{Cr}& \textbf{YCbCr} &\textbf{Luma}& \textbf{Cb} &  \textbf{Cr} \\
\midrule
basketball & 0.119 & 0.121 & 0.239 & 0.134 & -4.9  & -10.3 &  -11.2  \\
dancer & 0.142 & 0.144 & 0.260 & 0.157  & -5.6 & -11.4 & -12.6 \\
exercise & 0.100 & 0.135 & 0.206 & 0.118  & -4.8 &  -13.7 & -15.0 \\
longdress & 0.139 & 0.151 & 0.1614 & 0.143  & -5.2 &  -6.4 &  -7.4 \\
loot & 0.220 & 0.443 & \textbf{0.426} & 0.273  & -6.1  & \textbf{-19.6} & \textbf{-18.4} \\
model & 0.099 & 0.126 & 0.235 & 0.119 &  -4.9 &  -9.0 & -10.3 \\
queen & \textbf{0.310} & 0.205 & 0.150 & 0.277 &  \textbf{-15.1} & -11.0 &  -9.0 \\
redandblack & 0.228 & 0.204 & 0.241 & 0.227 & -6.8 & -11.3 & -8.7 \\
soldier & 0.282 & 0.342 & 0.332 & \textbf{0.296} & -8.2 & -16.1 & -15.3 \\
Andrew & 0.135 & 0.065 & 0.008 & 0.111 & -6.3 & -5.8 & -1.9 \\
David & 0.118 & 0.300 & 0.270 & 0.159 & -3.5 & -17.0 & -14.5 \\
Phil & 0.184 & 0.156 & 0.079 & 0.167 & -6.8 & -16.5 & -15.8 \\
Ricardo & 0.243 & 0.386 & 0.354 & 0.274 & -9.8 & -11.1 & -18.0 \\
Sarah & 0.201 & \textbf{0.451} & 0.403 & 0.2744 & -5.6 &  -15.6 & -15.5 \\
\midrule

\textbf{Average} & \textbf{0.180} & \textbf{0.230} & \textbf{0.240} & \textbf{0.194} &\textbf{-6.5} & \textbf{-12.4} & \textbf{-11.5} \\ 
\bottomrule
\end{tabular}}
\end{table}

\subsection{Application to V-PCC}
To further demonstrate the effectiveness of GQE-Net, we used it to improve the quality of point clouds in Cat\_A, which were compressed with V-PCC TCM2v18.0 under the lossy geometry and lossy attribute configuration. The quantization parameters for the geometry coding were set to $32$, $28$, $24$, $20$, and $16$. Since the coding structure of V-PCC is different from that of G-PCC, we retrained GQE-Net using the WPCSD dataset. The results, shown in Table IV, indicate that an average BD-PSNR of $0.18$ dB, $0.23$ dB, and $0.24$ dB can be achieved for the Y, Cb, and Cr components, respectively, corresponding to $6.5\%$, $12.4\%$ and $11.5\%$ BD-rate savings. The performance of GQE-Net for V-PCC was slightly reduced compared to G-PCC. This is because the geometry of the reconstructed point cloud is lossy, making the FR module unable to extract the correct distance and normal information. A comprehensive comparison of rate-distortion and rate-PSNR performance with V-PCC is provided in the supplemental materials.

\begin{table}
\scriptsize
\renewcommand{\arraystretch}{1}
\captionsetup{font={small}}
\caption{Rate-distortion comparison with G-PCC TMC13v14.0 when a lossy geometry, lossy attribute configuration is used.}
\setlength{\belowcaptionskip}{-1.0em}
\centering
{\begin{tabular}{cccccccc}
\toprule
\centering
\multirow{2}{*}{\textbf{Sequence}}& \multicolumn{4}{c}{\textbf{BD-PSNR [dB]}}  & \multicolumn{3}{c}{\textbf{BD-rate [\%]}}    \\
  &\textbf{Luma}& \textbf{Cb}\ & \textbf{Cr}& \textbf{YCbCr} &\textbf{Luma}& \textbf{Cb} &  \textbf{Cr} \\
\midrule
basketball & \textbf{0.352} & 0.042 & 0.121 & \textbf{0.285} & \textbf{-13.4}  & -2.6 &  -6.4  \\
dancer & 0.265 & 0.063 & 0.079 & 0.217  & -12.1 & -4.3 & -4.8 \\
exercise & 0.175 & 0.056 & 0.060 & 0.146  & -9.4 &  -4.2 & -5.1 \\
longdress & 0.124 & 0.094 & 0.079 & 0.115  & -5.8 &  -5.4 &  -5.9 \\
loot & 0.127 & 0.252 & 0.140 & 0.144  & -6.1  &  -9.7 & -6.0 \\
model & 0.282 & 0.061 & 0.100 & 0.231 &  -10.8 &  -3.2 & -4.9 \\
queen & 0.182 & \textbf{0.339} & -0.012 & 0.177 &  -8.6 & \textbf{-12.2} &  -3.8 \\
redandblack & 0.202 & 0.095 & \textbf{0.161} & 0.184 & -9.4 & -5.3 & \textbf{-7.5} \\
soldier & 0.141 & 0.137 & 0.073 & 0.132 & -6.5 & -7.0 & -4.5 \\
Andrew & 0.134 & 0.026 & -0.014 & 0.102 & -6.3 & -2.9 & -0.5 \\
David & 0.044 & 0.103 & 0.044 & 0.051 & -3.0 & -6.3 & -1.7 \\
Phil & 0.214 & 0.031 & 0.006 & 0.165 & -10.4 & -2.8 & -1.7 \\
Ricardo & 0.108 & 0.143 & 0.059 & 0.107 & -5.1 & -6.9 & -3.5 \\
Sarah & 0.180 & 0.152 & 0.078 & 0.164 & -6.1 &  -6.4 & -3.8 \\
\midrule

\textbf{Average} & \textbf{0.181} & \textbf{0.114} & \textbf{0.070} &\textbf{0.159} &\textbf{-8.1} & \textbf{-5.7} & \textbf{-4.3} \\ 
\bottomrule
\end{tabular}}
\end{table}

\subsection{Results for lossy geometry and lossy attribute condition}
Table V compares GQE-Net to G-PCC TMC13 for a lossy geometry, lossy attribute condition. More results are provided in the supplemental materials. The results show that GQE-Net remains capable of enhancing performance, although to a lesser extent compared to the lossless geometry, lossy attribute condition.

\subsection{Ablation study}
In this section, we evaluate the effectiveness of the PSGA module and FR module. We also study the effect of patch overlap on the performance of GQE-Net.

\begin{table*}[!ht]
\scriptsize
\captionsetup{font={small}}
%\caption{Rate-distortion performance comparison ($\Delta$PSNR ($\rm{dB}$) and BD-Rate (\%)) in the ablation study.}
\caption{Study of PSGA and FR modules. {\upshape \textbf{Time}} gives the processing time for one patch. {\upshape \textbf{Paras}} is the number of learnable parameters in the network.}
\setlength{\belowcaptionskip}{-1.0em}
\centering
{\begin{tabular}{ccccccccccc}
\toprule
\centering
\multirow{2}{*}{\textbf{Sequence}} & \multicolumn{2}{c}{\textbf{GQE-Net w/o FR}} & \multicolumn{2}{c}{\textbf{Only Normal}} & \multicolumn{2}{c}{\textbf{Only Distance}} & \multicolumn{2}{c}{\textbf{Parallel}} & \multicolumn{2}{c}{\textbf{GQE-Net}}\\
 & BD-PSNR & BD-rate & BD-PSNR & BD-rate & BD-PSNR & BD-rate & BD-PSNR & BD-rate & BD-PSNR & BD-rate \\
\midrule
basketball & 0.266 & -10.2 & 0.312 & -11.4 & 0.320 & -11.8 & 0.260 & -10.2 & 0.381 & -13.5 \\
dancer & 0.310 & -10.9 & 0.337 & -11.6 & 0.353 & -12.4 & 0.301 & -10.9 & 0.399 & -13.5 \\
exercise & 0.182 & -8.5 & 0.205 & -9.1 & 0.213 & -9.7 & 0.180 & -8.2 & 0.255 & -11.2 \\
longdress & 0.348 & -9.3 & 0.325 & -9.3 & 0.290 & -11.7 & 0.206 & -10.6 & 0.449 & -12.4 \\
loot & 0.296 & -9.4 & 0.292 & -9.7 & 0.411 & -12.7 & 0.332 & -11.1 & 0.472 & -13.9 \\
model & 0.325 & -10.8 & 0.348 & -11.5 & 0.329 & -11.5 & 0.292 & -10.7 & 0.375 & -12.5 \\
queen & 0.213 & -7.1 & 0.196 & -6.9 & 0.112 & -5.4 & 0.128 & -5.8 & 0.242 & -8.0 \\
redandblack & 0.350 & -10.8 & 0.347 & -11.1 & 0.380 & -12.6 & 0.373 & -12.1 & 0.449 & -14.1\\
soldier & 0.475 & -12.7 & 0.418 & -12.0 & 0.531 & -14.7 & 0.423 & -12.8 & 0.550 & -15.2 \\
Andrew & 0.505 & -15.6 & 0.532 & -16.6 & 0.534 & -19.6 & 0.452 & -17.7 & 0.584 & -20.5 \\
David & 0.109 & -4.9 & 0.152 & -6.7 & 0.295 & -11.5 & 0.143 & -6.5 & 0.335 & -12.2 \\
Phil & 0.520 & -15.8 & 0.519 & -16.2 & 0.544 & -17.7 & 0.478 & -16.5 & 0.558 & -17.8 \\
Ricardo & 0.276 & -10.3 & 0.298 & -11.2 & 0.340 & -13.1 & 0.148 & -7.4 & 0.390 & -14.2 \\
Sarah & 0.423 & -13.6 & 0.424 & -14.0 & 0.471 & -15.8 & 0.148 & -7.7 & 0.535 & -17.4 \\
\textbf{Average} & 0.328 & -10.7 & 0.336 & -11.2 & 0.365 & -12.8 & 0.276 & -10.6 & \textbf{0.427} & \textbf{-14.0} \\
\midrule
\textbf{Time [ms]} & \multicolumn{2}{c}{25.82} & \multicolumn{2}{c}{25.94} & \multicolumn{2}{c}{26.52} & \multicolumn{2}{c}{25.80} & \multicolumn{2}{c}{26.67} \\ 
\textbf{Paras} & \multicolumn{2}{c}{476,186} & \multicolumn{2}{c}{477,338} & \multicolumn{2}{c}{476,186} & \multicolumn{2}{c}{440,610} & \multicolumn{2}{c}{477,338} \\ 
\bottomrule
\end{tabular}}

\end{table*}

\begin{table*}[!ht]
\scriptsize
\captionsetup{font={small}}
\caption{Effect of patch overlap. {\upshape Gen [s], Pro [s]} and {\upshape Fus [s]} denote the patch generation time, GQE-Net processing time, and patch fusion time, respectively. {\upshape $\bm{r}$} is the overlapping ratio. ``{\upshape \textbf{Sequential selection}}" is the patch segmentation method used in \cite{sheng2021deep}.}
\setlength{\belowcaptionskip}{-1.0em}
\centering
{\begin{tabular}{ccccccccc}
\toprule
\centering
\multirow{2}{*}{\textbf{Sequence}} & \multicolumn{2}{c}{\textbf{Sequential selection}} & \multicolumn{2}{c}{\bm{$r = 0.5$}} & \multicolumn{2}{c}{\bm{$r = 1$}} & \multicolumn{2}{c}{\textbf{GQE-Net (\bm{$r =2 $})}}\\
 & BD-PSNR & BD-rate & BD-PSNR & BD-rate & BD-PSNR & BD-rate & BD-PSNR & BD-rate \\
\midrule
basketball & 0.377 & -13.5 & 0.161 & -6.0 & 0.293 & -10.7 & \textbf{0.381} &\textbf{-13.5} \\
dancer & \textbf{0.411} & \textbf{-13.9} & 0.167 & -5.9 & 0.307 & -10.7 & 0.399 & -13.5 \\
exercise & 0.233 & -10.6 & 0.117 & -5.3 & 0.203 & -9.0 & \textbf{0.255} & \textbf{-11.2} \\
longdress  & 0.261 & -11.1 & 0.119 & -5.4 & 0.251 & -9.9 & \textbf{0.449} & \textbf{-12.4} \\
loot & 0.398 & -12.5 & 0.200 & -6.4 & 0.368 & -11.3 & \textbf{0.472} & \textbf{-13.9} \\
model & \textbf{0.387} & \textbf{-12.6} & 0.151 & -5.4 & 0.286 & -9.8 & 0.375 & -12.5 \\
queen & 0.177 & -6.9 & 0.097 & -3.4 & 0.174 & -5.9 & \textbf{0.242} & \textbf{-8.0} \\
redandblack & 0.392 & -12.8 & 0.183 & -6.2 & 0.345 & -11.3 &\textbf{0.449} & \textbf{-14.1}\\
soldier & 0.496 & -14.2 & 0.220 & -6.6 & 0.416 & -12.0 & \textbf{0.550} & \textbf{-15.2} \\
Andrew & \textbf{0.598} & -19.9 & 0.210 & -8.6 & 0.419 & -16.1 & 0.584 & \textbf{-20.5} \\
David & 0.309 & -11.5 & 0.117 & -4.6 & 0.230 & -8.9 & \textbf{0.335} & \textbf{-12.2} \\
Phil & \textbf{0.641} & \textbf{-19.5} & 0.208 & -7.2 & 0.392 & -13.1 & 0.558 & -17.8 \\
Ricardo & \textbf{0.426} & \textbf{-15.3} & 0.132 & -5.1 & 0.264 & -10.1 & 0.390 & -14.2 \\
Sarah & \textbf{0.542} & \textbf{-17.6} & 0.200 & -7.1 & 0.363 & -12.4 & 0.535 & -17.4 \\
\textbf{Average} & 0.403 & -13.7 & 0.163 & -5.9 & 0.308 & -10.8 & \textbf{0.427} & \textbf{-14.0} \\
\midrule
\textbf{Gen [s]} & \multicolumn{2}{c}{\textbf{5.45}} & \multicolumn{2}{c}{9.45} & \multicolumn{2}{c}{14.83} & \multicolumn{2}{c}{29.67} \\
\textbf{Pro [s]} & \multicolumn{2}{c}{138.72} & \multicolumn{2}{c}{\textbf{59.50}} & \multicolumn{2}{c}{130.19} & \multicolumn{2}{c}{291.29} \\
\textbf{Fus [s]} & \multicolumn{2}{c}{\textbf{4.93}} & \multicolumn{2}{c}{22.17} & \multicolumn{2}{c}{31.36} & \multicolumn{2}{c}{43.98} \\
\bottomrule
\end{tabular}}

\end{table*}

\subsubsection{Influence of PSGA}
\ 
\newline
\indent{The proposed parallel-serial multi-head mechanism is designed to enhance the fusion and reinforcement of the attention features. Compared to a standard multi-head attention module, it establishes a stronger relationship and interaction between each single head attention layer. To evaluate the performance of the PSGA module, we replaced the parallel-serial structure with four parallel single heads \cite{chen2021gapointnet} (referred to as \textbf{Parallel}) in which each head operates independently. In this test, we only evaluated the performance on the Y component.}

The results are shown in Fig. 10. Since we used a four-head attention mechanism in parallel, the number of learnable parameters in this module remained almost unchanged. However, we can see that the parallel-serial structure (used in GQE-Net) led to the best performance, particularly at high bit rates. The detailed statistics are given in Table VI.

\begin{table*}[!ht]
\scriptsize
\captionsetup{font={small}}
\caption{BD-rate comparison between MSGAT \cite{sheng2022attribute} , GQE-Net\_1, and GQE-Net\_2.}
\setlength{\belowcaptionskip}{-1.5em}
\centering
{\begin{tabular}{cccc|ccc|ccc}
\toprule
\centering
\multirow{2}{*}{\textbf{Sequence}} & \multicolumn{3}{c}{\textbf{MSGAT}} & \multicolumn{3}{c}{\textbf{GQE-Net\_1}} & \multicolumn{3}{c}{\textbf{GQE-Net\_2}} \\
 & Y [\%] & Cb [\%] & Cr [\%] &  Y [\%] & Cb [\%] & Cr [\%] & Y [\%] & Cb [\%] & Cr [\%] \\
\midrule
Longdress & -12.73 & 0.15 & -2.73 & -14.7 & \textbf{-8.7} & \textbf{-13.1} & \textbf{-15.9} & -6.5 & -7.1\\
Redandblack & -8.008 & 1.06 & -3.04 & -13.2 & -5.9 & \textbf{-14.8} & \textbf{-17.6} & \textbf{-8.8} & -4.2\\
Soldier & -11.72 & -1.28 & -1.45 & -13.4 & -5.5 & \textbf{-10.7} & \textbf{-16.3} & \textbf{-12.6} & -6.1\\
Dancer & -13.18 & -2.26 & -1.71 & -16.2 & -5.1 & \textbf{-12.9} & \textbf{-23.0} & \textbf{-13.5} & -9.3\\
Model & -13.23 & -3.87 & -4.40 & -15.9 & -5.3 & \textbf{-10.7} & \textbf{-20.6} & \textbf{-12.4} & -9.6\\
Andrew & -7.02 & -4.37 & \textbf{-5.66} & -16.3 & -5.3 & -5.2 & \textbf{-19.9} & \textbf{-6.3} & -3.8\\
David & -8.33 & -3.37 & -2.12 & -9.0 & -5.9 & \textbf{-10.8} & \textbf{-12.7} & \textbf{-11.2} & -5.1\\
Phil & -13.73 & -0.53 & 0.68 & -16.3 & -4.4 & -3.9 & \textbf{-25.2} & \textbf{-6.7} & \textbf{-4.7}\\
Sarah & -12.75 & -3.52 & -3.16 & -15.2 & -7.6 & \textbf{-7.8} & \textbf{-16.3} & \textbf{-9.4} & -3.8\\
Ricardo & -8.93 & -3.05 & -4.55 & -12.7 & -5.4 & \textbf{-8.8} &\textbf{ -17.4} & \textbf{-14.7} & -6.9\\

\midrule
\textbf{Average} & \textcolor{red}{-10.97} & \textcolor{red}{-2.10} & \textcolor{red}{-2.81} & \textcolor{red}{-14.3} & \textcolor{red}{-5.9} & \textcolor{red}{ \textbf{-9.9}} & \textcolor{red}{\textbf{-18.5}} & \textcolor{red}{\textbf{-10.2}} &\textcolor{red}{ -6.1}\\
\bottomrule
\end{tabular}}
\end{table*}

\begin{table*}[!ht]
\scriptsize
\renewcommand{\arraystretch}{1.02}
\captionsetup{font={small}}
\caption{BD-rate and time complexity comparison between the Wiener filter-based method \cite{xing2022gpcc} and GQE-Net. {\upshape Gen, Pro} and {\upshape Fus} give the patch generation time, GQE-Net processing time, and patch fusion time, respectively.}
\setlength{\belowcaptionskip}{-1.5em}
\centering
{\begin{tabular}{ccc|ccccc|cccccc}
\toprule
\centering
\multirow{2}{*}{\textbf{Sequence}} & \multicolumn{2}{c}{\textbf{G-PCC lifting}}  & \multicolumn{5}{c}{\textbf{\cite{xing2022gpcc}}} & \multicolumn{6}{c}{\textbf{GQE-Net}} \\
 & Enc [s] & Dec [s] & Y [\%] & Cb [\%] & Cr [\%] & Enc [s] & Dec [s] &  Y [\%] & Cb [\%] & Cr [\%] & Gen [s] & Pro [s] & Fus [s]  \\
\midrule
basketball & 2.85 & 1.62 & -3.3  & -2.2 &  0.5 & 9.59 & 7.56  & \textbf{-13.5}  & \textbf{-8.5} &  \textbf{-20.4} & 72.47  &  550.90  &  121.16  \\
dancer & 2.51 & 1.42  &  -4.1 & -2.7 & -0.9  & 8.31 & 6.67 & \textbf{-13.5} & \textbf{-8.3} & \textbf{-18.6} & 115.68  & 489.89  &  105.01 \\
exercise & 2.57  &  1.54  & -3.1 &  -2.8 & -0.4 & 7.67  &  6.26 & \textbf{-11.2} &  \textbf{-8.2} & \textbf{-21.6} & 105.91  &  472.03 &  106.02 \\
longdress & 0.79 & 0.48   & -7.5 &  -4.9  &  -6.3 & 2.67  &  2.03   & \textbf{-12.4} &  \textbf{-10.7} &  \textbf{-12.0} &  4.72 &  166.61 & 38.55  \\
loot & 0.74 & 0.45   & -6.9  &  0.2 & 0.1  & 2.50  &  1.93 & \textbf{-13.9}  &  \textbf{-11.7} & \textbf{-16.7} & 4.14   &  157.71  &  38.25 \\
model &  2.65  &  1.62 & -3.6 & -1.5 & -0.3  &  8.15  &  6.27  & \textbf{-12.5} &  \textbf{-7.7} & \textbf{-15.8} & 58.81  &  508.13 & 116.69  \\
queen & 0.87 & 0.52  &  -1.7 & -0.7 &  -0.3  &  3.22 & 2.51  &  \textbf{-8.0} & \textbf{-12.2} &  \textbf{-13.8} & 17.39  & 185.46  &   41.05\\
redandblack & 0.70 & 0.43  & -5.3 & -0.9 & -8.7 &  2.28  &  1.90 & \textbf{-14.1} & \textbf{-8.6} & \textbf{-14.5} & 3.67  &  147.80 &  34.39 \\
soldier & 1.03 & 0.67  & -8.1 & 0.6 & 0.5 &  4.16  &  3.04 & \textbf{-15.2} & \textbf{-10.0} & \textbf{-15.7} & 7.51  & 216.59  & 50.17  \\
Andrew & 1.53  &  1.06 & -7.2 & -2.7 & -3.0  & 3.92  &  3.27  &  \textbf{-20.5} & \textbf{-8.4} & \textbf{-6.1} & 30.38  & 242.67  & 52.52  \\
David &  1.37 &  0.83  &  -5.5 & 0.5 & 0.6  &  4.02  &  3.22  &  \textbf{-12.2} & \textbf{-8.8} & \textbf{-15.9} &  29.86 & 245.34  &  53.46 \\
Phil &  1.56 &  1.07  & -6.5 & -2.6 & -1.8  & 4.48  &  3.40 &  \textbf{-17.8} & \textbf{-7.0} & \textbf{-6.3} &  36.61 & 273.72  & 59.65  \\
Ricardo & 0.98 & 0.55  &  -2.8 & 0.2 & 0.2 & 2.75  &  2.31 &  \textbf{-14.2} & \textbf{-10.1} & \textbf{-14.1} &  15.33 & 174.58  & 37.38  \\
Sarah &  1.03  &  0.70  & -4.5 & 0.9 & 0.7 & 3.36  & 2.73  & \textbf{-17.4} &  \textbf{-9.8} & \textbf{-11.3} & 20.54  & 205.13  &  43.42 \\
\midrule
arco\_valentino & 5.94 & 3.88   &  \textbf{-3.2} & 1.7 & -1.3 & 5.65  &  4.38 &  -1.3  & \textbf{-6.2} &  \textbf{-9.2} &  19.17 & 300.56  & 57.51  \\
egyptian\_mask & 1.09 & 0.76   & 0.0 & 0.2 & 0.3  & 0.91  & 0.17  & \textbf{-4.7} & \textbf{-4.4} & \textbf{-3.7} & 0.54  &  49.95 & 10.90  \\
facade\_00009 & 6.10 & 4.27   &\textbf{ -7.3} &  -1.9 & -0.2  &  5.33  &  4.26 & -3.6 &  \textbf{-2.7} & \textbf{-3.8} &  20.82 &  304.85 & 66.92  \\
staue\_klimt & 2.03 & 1.42  & -2.6 &  -1.6 &  0.1  &  1.54 &  1.20 & \textbf{-2.9} &  \textbf{-5.0} & \textbf{-3.1} & 1.71  &  99.04 & 22.18  \\

\midrule
\textbf{Average} &  \textcolor{red}{2.02} & \textcolor{red}{1.18}  & \textcolor{red}{-4.6} & \textcolor{red}{-1.1} & \textcolor{red}{-1.1}  & \textcolor{red}{4.44} & \textcolor{red}{3.48}  & \textcolor{red}{\textbf{-11.6}} & \textcolor{red}{\textbf{-8.3}} & \textcolor{red}{ \textbf{-12.4}} & \textcolor{red}{31.40} & \textcolor{red}{266.44}  & \textcolor{red}{58.62} \\
%\textbf{Average} &  2.02 & 1.18  & -4.6 & -1.1 & -1.1  & 4.44 & 3.48 & \textbf{-11.6} & \textbf{-8.3} & \textbf{-12.4} & 31.40 & 266.44  & 58.62 \\

\bottomrule
\end{tabular}}

\end{table*}

\subsubsection{Influence of FR module}
\ 
\newline
\indent{The FR module is crucial for integrating geometry information into color features. To assess the impact of the FR module, we tested the performance of GQE-Net both with and without it. There are two key components in this module: normal calculation and distance-based feature weighting. Normals aid in identifying the similarity between the central point and its neighbors, while the weighting matrix reinforces features that are geometrically close to each other and weakens features that are geometrically far apart. We tested versions of GQE-Net that fully use the FR module, only use normal information (denoted as \textbf{Only Normal}), only use feature weighting (denoted as \textbf{Only Distance}), and use none of them (denoted as \textbf{GQE-Net w/o FR}). The corresponding rate-PSNR curves are compared in Fig. 11. We can observe that the GQE-Net version that fully uses the FR module leads to the best performance for all tested point clouds, demonstrating the efficiency of the FR module. Similar conclusions can also be seen in Table VI.}

In Table VI, we also compare time and space complexity of GQE-Net in each case. We evaluate time complexity by the processing time of one patch and space complexity by the number of learnable parameters in the network. It can be seen that with a small increase in complexity, the FR and PSGA modules provided significant performance gains for GQE-Net.

\subsubsection{Study of patch overlap}
\ 
\newline
\indent{The way in which the point cloud is partitioned into patches has a significant impact on both performance and efficiency. In this section, we study the effect of the overlapping ratio $r$. We also compare our patch generation approach to an alternative method \cite{sheng2021deep} that leads to non-overlapped patches. This method sequentially partitions the point cloud into non-overlapping patches of the same size.}

Table VII shows the results for the proposed patch generation method for $r = 0.5$, $1$, and $2$. It also shows the results for the method used in \cite{sheng2021deep} when the number of points in each patch was set to $2048$.

Doubling the overlapping ratio improved the rate-distortion performance as it decreased the number of unprocessed points (about $50\%$ for $r = 0.5$, $10\%$ for $r = 1$, and $0.01\%$ for $r = 2$). At the same time, it significantly increased the time complexity. For this reason, a further increase in the overlapping ratio is not recommended. 

The time complexity of the method in \cite{sheng2021deep} was lower than that of the proposed method. However, except for one point cloud, its rate-distortion performance was worse.

\subsection{Comparison with the state-of-the-art}

In this section, we first compare GQE-Net to the multi-scale graph attention network (MSGAT) \cite{sheng2022attribute}, which was the first neural network used for enhancing the color quality of point clouds. Note that MSGAT trained $12$ separate models for the Y, Cb, and Cr components with bit rates ranging from R01 to R04. This means that MSGAT is only suitable for point clouds at specific bit rates, limiting its applicability. For a comprehensive comparison, we considered two versions of GQE-Net. The first one, which we call GQE-Net\_1, consists of a single model for all bit rates, which was trained on the WPCSD training set for the corresponding color channel. The second one, which we call GQE-Net\_2, consists of several models, where each model was trained on the training set in \cite{sheng2022attribute} for one target bit rate and color channel. In both cases, the training and test point clouds were compressed using G-PCC TMC13v12.0 with the lifting transform according to the configuration adopted in \cite{sheng2022attribute}. Table VIII gives the BD-rates of MSGAT, GQE-Net\_1, and GQE-Net\_2 on the test set in \cite{sheng2022attribute}. The results confirm the superiority of our approach. 

Next, we compare the proposed method and a Wiener filter-based point cloud color quality enhancement method \cite{xing2022gpcc}, in which the optimal coefficients of a Wiener filter are calculated in the encoder and transmitted to the decoder. We used G-PCC TMC13v14.0 with the lifting transform for compression. Table IX shows the BD-rates and the time complexity of the encoding and decoding for Cat\_A and Cat\_B point clouds. We can observe that the proposed method performed better no matter which category of point clouds was used.

\begin{table}[!ht]
\scriptsize
\renewcommand{\arraystretch}{1}
\captionsetup{font={small}}
\caption{BD-rate comparison between GQE-Net and V-PCC color smoothing \cite{color_smoothing}.}
\setlength{\belowcaptionskip}{-1.4em}
\centering
{\begin{tabular}{ccccccc}
\toprule
\centering
\multirow{2}{*}{\textbf{Sequence}}& \multicolumn{3}{c}{\textbf{Smoothing}}  & \multicolumn{3}{c}{\textbf{GQE-Net}}    \\
  &  \textbf{Y [\%]} & \textbf{Cb [\%]} & \textbf{Cr [\%]} & \textbf{Y [\%]} & \textbf{Cb [\%]} & \textbf{Cr [\%]}  \\
\midrule
basketball & 0.5 & -0.5 & -0.5  & \textbf{-4.9}  & \textbf{-10.3} &  \textbf{-11.2}  \\
dancer & \textbf{-10.8} & 1.3 & 12.2   & -5.6 & \textbf{-11.4} & \textbf{-12.6} \\
exercise & \textbf{-7.2} & 2.6 & -1.0   & -4.8 &  \textbf{-13.7} & \textbf{-15.0} \\
longdress & 1.5 & 0.1 & 0.4   & \textbf{-5.2} &  \textbf{-6.4} &  \textbf{-7.4} \\
loot & 1.4 & -1.6 & -1.6  & \textbf{-6.1}  & \textbf{-19.6} & \textbf{-18.4} \\
model & 1.1 & -0.4 & -0.5  &  \textbf{-4.9} &  \textbf{-9.0}& \textbf{-10.3} \\
queen & 1.3 & -0.8  & -0.8 &  \textbf{-15.1} & \textbf{-11.0} &  \textbf{-9.0} \\
redandblack & 1.3 & -0.6 & 1.7 & \textbf{-6.8} & \textbf{-11.3} & \textbf{-8.7} \\
soldier & 2.1 & -0.6 & 1.7  & \textbf{-8.2} & \textbf{-16.1} & \textbf{-15.3} \\
Andrew & 15.3 & 3.8 & 1.9  &\textbf{-6.3} & \textbf{-5.8} & \textbf{-1.9} \\
David & 2.3 & 0.5 & -1.8  & \textbf{-3.5} & \textbf{-17.0} & \textbf{-14.5} \\
Phil & 18.9 & 4.0 & 5.0 & \textbf{-6.8} & \textbf{-16.5} & \textbf{-15.8} \\
Ricardo & 10.3 & -0.9 & 1.7 & \textbf{-9.8} & \textbf{-11.1} & \textbf{-18.0} \\
Sarah & 18.6 & 1.2 & 0.9 & \textbf{-5.6} &  \textbf{-15.6} & \textbf{-15.5} \\
\midrule

\textbf{Average} & \textcolor{red}{4.0} & \textcolor{red}{0.5} & \textcolor{red}{1.2}  & \textcolor{red}{\textbf{-6.5}} & \textcolor{red}{\textbf{-12.4}} & \textcolor{red}{\textbf{-11.5}} \\ 
\bottomrule
\end{tabular}}
\end{table}

In Table IX, we provide the processing time of G-PCC, the additional processing time of Wiener filtering, and the additional processing time of the proposed method in different stages (patch generation, GQE-Net, and patch fusion). We can see that the time complexity of the proposed method was higher than that of G-PCC and the Wiener filter-based method. It is a common problem that neural network-based methods consume more computational resources.

Finally, we compared the performance of GQE-Net to that of V-PCC TMC2v18.0's low-complexity color smoothing method \cite{color_smoothing}. Table X compares the BD-rates. As can be seen, our method was more stable and provided clear BD-rate gains over the V-PCC test model color smoothing method.

\section{Conclusion}
We proposed GQE-Net, a graph-based quality enhancement network to restore the true colors of distorted point clouds. GQE-Net includes two novel modules: PSGA and FR. PSGA, which consists of three single-head attention layers, extracts, integrates and reinforces attention features. FR efficiently uses local information through normal fusion and distance-guided feature weighting. GQE-Net can be applied to point clouds encoded at various quantization parameters, making it flexible in practical applications. GQE-Net achieved BD-PSNR gains of $0.43$ dB, $0.25$ dB, $0.36$ dB with corresponding BD-rates of -$14.0\%$, -$9.3\%$ and -$14.5\%$ for the Luma, Cb, and Cr components of dense point clouds (Cat\_A) coded by G-PCC with the lifting transform. For sparse point clouds (Cat\_B), the BD-rate gains were -$1.3\%$, -$4.0\%$ and -$2.8\%$. We also showed that GQE-Net is effective for other coding configurations, such as G-PCC RAHT and V-PCC. 

As future work, we will focus on improving our method for sparse point clouds. We will also build a lightweight version of our model to reduce its time complexity. 

\bibliographystyle{IEEEbib}
\bibliography{IEEEexample}

\begin{IEEEbiography}
[{\includegraphics[width=1in,clip,keepaspectratio]{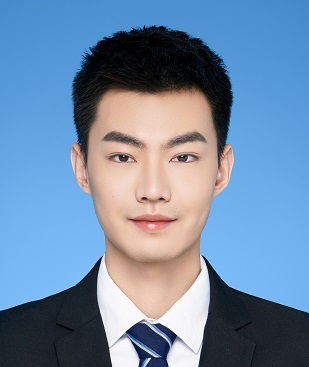}}] 		                                     
{Jinrui Xing} received the B.E. degree in automation with the Department of Control Science and Engineering from Shandong University, Ji'nan, China, in 2021. He is now pursuing the M.E. degree in control science and engineering from Shandong University, Ji'nan, China. His research interests include 3D point cloud compression and post processing.
\end{IEEEbiography}

\begin{IEEEbiography}
[{\includegraphics[width=1in,height=1.25in,clip,keepaspectratio]{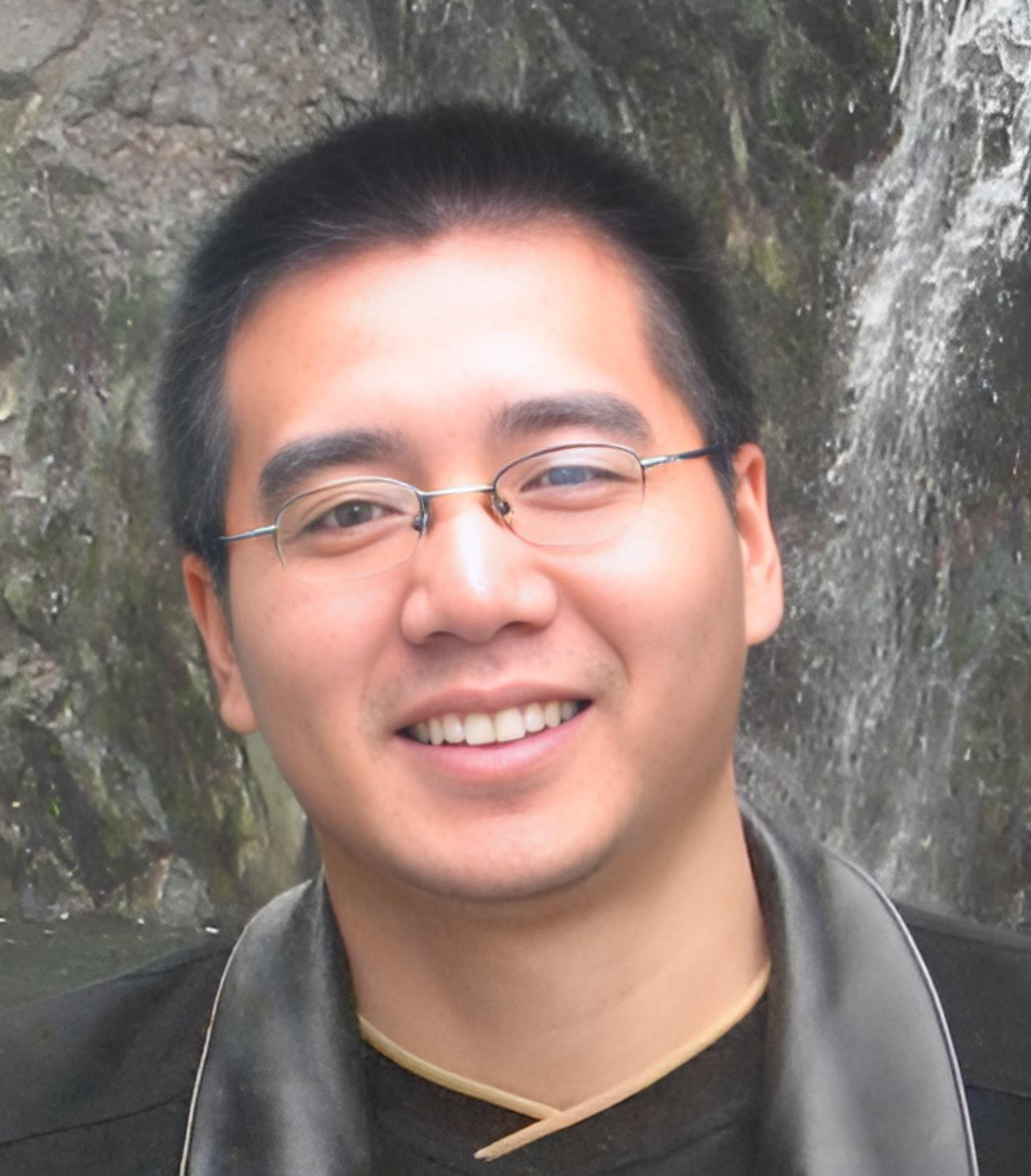}}] 		                                     
{Hui Yuan} (Senior Member, IEEE) received the B.E. and Ph.D. degrees in telecommunication engineering from Xidian University, Xi’an, China, in 2006 and 2011, respectively. In April 2011, he joined Shandong University, Ji’nan, China, as a Lecturer (April 2011–December 2014), an Associate Professor (January 2015-August 2016), and a Professor (September 2016). From January 2013 to December 2014, and from November 2017 to February 2018, he worked as a Postdoctoral Fellow (Granted by the Hong Kong Scholar Project) and a Research Fellow, respectively, with the Department of Computer Science, City University of Hong Kong. From November 2020 to November 2021, he worked as a Marie Curie Fellow (Granted by the Marie Skłodowska-Curie Actions Individual Fellowship under Horizon2020 Europe) with the School of Engineering and Sustainable Development, De Montfort University, Leicester, U.K. From October 2021 to November 2021, he also worked as a visiting researcher (secondment of the Marie Skłodowska-Curie Individual Fellowships) with the Computer Vision and Graphics group, Fraunhofer Heinrich-Hertz-Institut (HHI), Germany. His current research interests include 3D visual coding and communication. He served as an Associate Editor for IET Image Processing (from 2023), an Area Chair for IEEE ICME 2023, ICME 2022, ICME 2021, IEEE ICME 2020, and IEEE VCIP 2020, and Senior Area Chair for PRCV 2023. He also serves as a member of IEEE CTSoc Audio/Video Systems and Signal Processing Technical Committee (AVS TC), IEEE CASSoc Visual Signal Processing and Communication Technical Committee (VSPC TC), and APSIPA Image, Video, and Multimedia Technical Committee (IVM TC). His research interest is 3D visual media coding, processing, and communication.
\end{IEEEbiography}

\vspace{0mm}

\begin{IEEEbiography}
[{\includegraphics[width=1in,height=1.25in,clip,keepaspectratio]{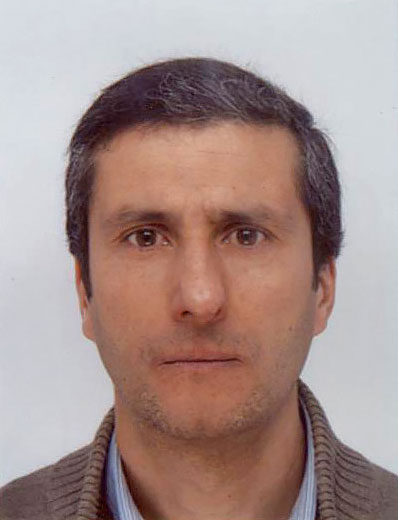}}] 		{Raouf Hamzaoui} (Senior Member, IEEE) received the MSc degree in mathematics from the University of Montreal, Canada, in 1993, the Dr.rer.nat. degree from the University of Freiburg, Germany, in 1997 and the Habilitation degree in computer science from the University of Konstanz, Germany, in 2004. He was an Assistant Professor with the Department of Computer Science of the University of Leipzig, Germany and with the Department of Computer and Information Science of the University of Konstanz. In September 2006, he joined DMU where he is a Professor in Media Technology and Head of the Signal Processing and Communications Systems Group in the Institute of Engineering Sciences. He was a member of the Editorial Board of the IEEE Transactions on Multimedia and IEEE Transactions on Circuits and Systems for Video Technology. He has published more than 100 research papers in books, journals, and conferences. His research has been funded by the EU, DFG, Royal Society, and industry and received best paper awards (ICME 2002, PV’07, CONTENT 2010, MESM’2012, UIC-2019).
\end{IEEEbiography}

\begin{IEEEbiography}
[{\includegraphics[width=1in,height=1.25in,clip,keepaspectratio]{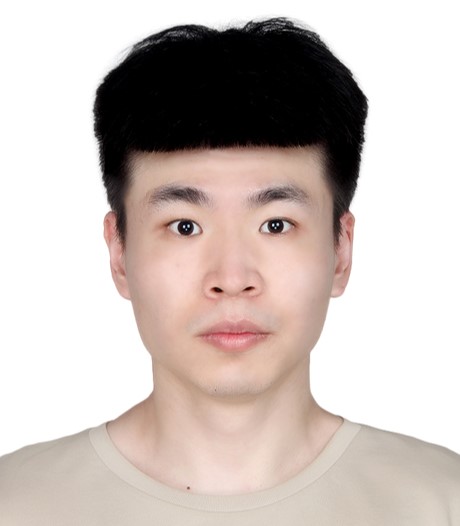}}] 		                                     
{Hao Liu} received the B.E. degree in telecommunication engineering from Shandong Agricultural University, Taian, China, in 2017, and the Ph.D. degree in information science and engineering from Shandong University, Qingdao, China, in 2022. From July 2022, he works a Lecturer with the School of Computer Science and Control Engineering, Yantai University. His research interests include 3D point cloud compression and processing.
\end{IEEEbiography}

\begin{IEEEbiography}
[{\includegraphics[width=1in,height=1.25in,clip,keepaspectratio]{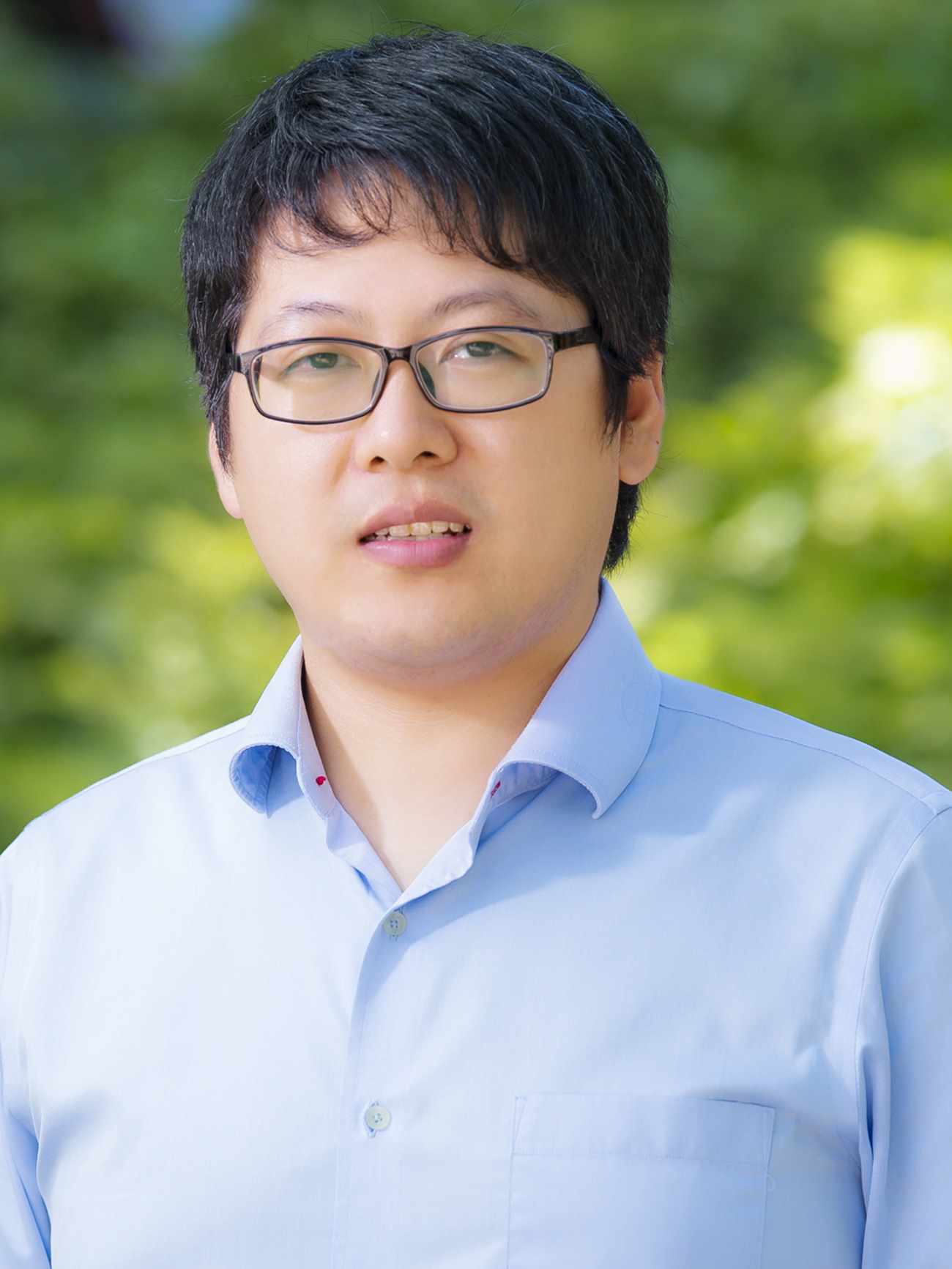}}] 		                                     
{Junhui Hou} is an Associate Professor with the Department of Computer Science, City University of Hong Kong. He holds a B.Eng. degree in information engineering (Talented Students Program) from the South China University of Technology, Guangzhou, China (2009), an M.Eng. degree in signal and information processing from Northwestern Polytechnical University, Xi’an, China (2012), and a Ph.D. degree from the School of Electrical and Electronic Engineering, Nanyang Technological University, Singapore (2016). His research interests are multi-dimensional visual computing.

Dr. Hou received the Early Career Award (3/381) from the Hong Kong Research Grants Council in 2018. He is an elected member of IEEE MSATC, VSPC-TC, and MMSP-TC. He is currently serving as an Associate Editor for IEEE Transactions on Visualization and Computer Graphics, IEEE Transactions on Circuits and Systems for Video Technology, IEEE Transactions on Image Processing, Signal Processing: Image Communication, and The Visual Computer.
\end{IEEEbiography}

\vspace{45mm}

\end{document}